\documentclass{aa}
\usepackage{txfonts}
\usepackage{epsfig}
\usepackage{subfigure}
\usepackage{multirow}

\usepackage{times}
\usepackage{natbib}
\usepackage{longtable,lscape}
\bibpunct{(}{)}{;}{a}{}{,}

\usepackage[bookmarks=false]{hyperref}

\def\xmm{{\it XMM-Newton }}
\def\rosat{{\it ROSAT}}
\def\asca{{\it ASCA}}

\def\xmm{{\it XMM-Newton}}
\def\chandra{{\it Chandra}}

\begin{document}

\title{CAIXA: a catalogue of AGN in the XMM-\textit{Newton} archive \\ III. Excess 
Variance Analysis}
\author{Gabriele Ponti\inst{1}, Iossif Papadakis\inst{2,3}, Stefano Bianchi\inst{4}, 
Matteo Guainazzi\inst{5}, Giorgio Matt \inst{4}, Phil Uttley\inst{1} and 
Fonseca Bonilla, Nuria\inst{5}}

\offprints{Gabriele Ponti\\ \email{ponti@iasfbo.inaf.it}}

\institute{School of Physics and Astronomy, University of Southampton, Highfield, Southampton SO17 1BJ
\and Department of Physics and Institute of Theoretical \& Computational Physics, University of Crete, 71003 Heraklion, Greece
\and IESL, Foundation for Research and Technology, 71110 Heraklion, Greece
\and Dipartimento di Fisica, Universit\`a degli Studi Roma Tre, via della Vasca Navale 84, 00146 Roma, Italy
\and XMM-Newton Science Operations Center, European Space Astronomy Center, ESA, Apartado 50727, E-28080 Madrid, Spain}

\date{Received / Accepted}

\authorrunning{G. Ponti et al.}

\abstract
{We report on the results of the first \xmm\ systematic ``excess variance"  study of all the radio quiet, X-ray un-obscured AGN. The entire sample consist of 161 sources observed by \xmm\ for more than 10 ks in pointed observations, which is the largest sample used so far to study AGN X-ray variability on time scales less than a day. }
{Recently it has been suggested that the same engine might be at work in the core of every Black Hole (BH) accreting object. In this hypothesis, the same variability should be observed in all AGN, once rescaled by the M$_{\rm BH}$ (M$_{\rm BH}$) and accretion rate ($\dot{m}$).}
{We systematically compute the excess variance for all AGN, on different time-scales (10, 20, 40 and 80 ks) and in different energy bands (0.3-0.7, 0.7-2 and 2-10 keV).} 
{We observe a highly significant and tight ($\sim0.7$ dex) correlation between $\sigma^2_{\rm rms}$ and M$_{\rm BH}$. The subsample of reverberation mapped AGN shows an even smaller scatter (only a factor of 2-3) comparable to the one induced by the M$_{\rm BH}$ uncertainties. This implies that X-ray variability can be used as an accurate tool to measure M$_{\rm BH}$ and this method is more accurate than the ones based on single epoch optical spectra. This allows us to measure M$_{\rm BH}$ for 65 AGN and estimate lower limits for the remaining 96 AGN.
On the other hand, the $\sigma^2_{\rm rms}$ vs. accretion rate dependence is weaker than expected based on the PSD break frequency scaling. This strongly suggests that both the PSD high frequency break and the normalisation depend on accretion rate in such a way that they almost completely counterbalance each other (PSD$_{\rm amp} \propto\dot{m}^{-0.8}$).
A highly significant correlation between $\sigma^2_{\rm rms}$ and 2-10 keV spectral index is observed. The highly significant correlations between $\sigma^2_{\rm rms}$ and both the L$_{\rm Bol}$ and the FWHM$_{\rm H\beta}$ are consistent with being just by-products of the $\sigma^2_{\rm rms}$ vs. M$_{\rm BH}$ relation. The soft and medium $\sigma^2_{\rm rms}$ is very well correlated with the hard $\sigma^2_{\rm rms}$, with no deviations from a linear one to one correlation. This suggests that the additional soft components (i.e. soft excess, warm absorber) add a minor contribution to the total variability. Once the variability is rescaled for M$_{\rm BH}$ and $\dot{m}$, no significant difference between narrow-line and broad-line Seyfert 1 is observed.}
{The results are in agreement with a picture where, to first approximation, all local AGN have the same variability properties once rescaled for M$_{\rm BH}$ and $\dot{m}$.}


\maketitle

\section{Introduction}

Rapid variability is one of the major observational properties of accretion onto Black Holes (BH). Since the early observations, it was realised that the fast variability observed in AGN implied that the primary source had to be both very compact and able to emit with a large luminosity, thus requiring an extremely efficient engine, providing the first compelling argument for the presence of BH in the heart of AGN (e.g. Lynden-Bell 1969; Rees 1984). Early studies of the AGN X--ray variability properties suggested that more luminous sources are ``less" variable (Barr \& Mushotzky 1986). 
Later on, the {\it EXOSAT} long looks allowed the first study of the observed variations with the use of Fourier analysis methods, like the power spectral density function (PSD), for the brightest AGN.  Lawrence \& Papadakis (1993) and Green, McHardy \& Lehto (1993) found that the majority of the {\it EXOSAT} AGN light curves exhibited ``red noise" variations, i.e.  their PSD were well fitted by power-laws with slopes steeper than 1. They also found that the PSD amplitude at a fixed frequency  was anti-correlated with the source luminosity,  confirming previous results. 

The availability of well sampled, very long (up to $\sim 10-15$ years in some cases) {\it RXTE} light curves, and their use together with high signal to noise, continuous {\it XMM-Newton} light curves has revolutionised our view of AGN X--ray variability in the last 15 years. For a dozen objects, the combined use of these light curves allowed the accurate determination of their PSD over a broad range of time scales. The past suggestions that the steep PSD at high frequencies had to flatten below a certain time scale (see e.g. Papadakis \& McHardy, 1995) were soon confirmed: all the best quality PSD are best fitted by a steep power law of slope $\sim -2$ which ``breaks" to a flatter ($\sim -1$) slope at frequencies below the so-called  ``break frequency" $\nu_{\rm b}$ (e.g. Edelson \& Nandra, 1999; Uttley et al. 2002; Markowitz et al. 2004; Kelly et al. 2010). In some AGN the $-1$ slope part could be measured for more than 3 decades in frequency, with no lower frequency breaks being detected, thus indicating a behaviour similar to Galactic X--ray black hole binary candidates (BHB) in their so-called ``soft state" (e.g. McHardy et al, 2004). In at least one case, namely Ark 564, two frequency breaks have been detected (Papadakis et al 2002). In fact, the PSD of this object may be better fitted by ``Lorentzian" functions (as opposed to "broken" power-laws; McHardy et al. 2007), similar to the functions that are widely used in the parametrisation of the BHB PSD. 

The detailed PSD analysis of the high quality {\it RXTE} and {\it XMM} light curves revealed a remarkable similarity between the AGN and BHB power-spectra, although the time scales sampled in the two classes of objects were vastly different. This result implied a deep link between all BH accreting objects, reinforcing initial suggestions that both systems host the same engine, and that the same emission and variability mechanisms operate in them. The difference in time-scales could be explained by the difference in the mass of the compact object in the center of these objects. The results from these first studies made it clear that the observed AGN X--ray variations are most probably determined by BH mass (M$_{\rm BH}$) rather than luminosity (Hayashida et al. 1998; Czerny et al. 2001; Uttley et al. 2002), and even suggested that  X-ray variability  might be used as a tool to estimate M$_{\rm BH}$ in these objects. 

McHardy et al. (2006) were the first to demonstrate clearly that AGN are scaled-up versions of BHB. Using PSD results for almost a dozen objects, spanning a range of $\sim 8$ orders of magnitude in M$_{\rm BH}$ and $\sim 3$ orders of magnitude in accretion rate (hereinafter we will use the term accretion rate as a sinonimus of Eddington ratio),  they demonstrated that the PSD ``break time scales" increase proportionally with M$_{\rm BH}$, and decrease with increasing accretion rate.  Koerding et al. (2007) showed that the three physical parameters, namely M$_{\rm BH}$, accretion rate and break frequency, are intimately related, both in AGN and BHB in the soft state, determining a ``variability plane" of accreting BHs. Interestingly, if a constant offset is introduced, this plane can also be extended to the ``hard state" objects as well. Casella et al. (2008) even used this ``variability plane" to estimate the M$_{\rm BH}$ of Ultra Luminous X-ray sources (ULXs).

Although the PSD analysis is a powerful tool to characterise the variability properties of AGN and BHB, it nevertheless requires long, uninterrupted observations from especially tailored monitoring campaigns to fully exploit its potential. Such observations are available at the moment for no more than two dozen AGN. However, the last few years, shorter, high signal-to-noise X--ray observations for tens of objects have populated the archive of X--ray observatories such as {\it Chandra}, and, perhaps even more importantly {\it XMM-Newton} (due to its high sensitivity, and broad energy band pass). A convenient analysis tool for such short data sets is the so-called ``normalized excess variance", $\sigma^2_{\rm rms}$ (Nandra et al. 1997). Although it does not offer the same wealth of information like the PSD analysis, it can certainly be used to confirm the PSD results using larger data samples. In fact, the availability of larger samples can also allow the discovery of new correlations between the X--ray variability amplitude and other AGN physical parameters. 

The first AGN excess variance surveys were performed using  \asca\ light curves. The results confirmed the variability vs. luminosity anti-correlation (Nandra et al. 1997; George et al. 2000) but also indicated, for the first time, that the variability amplitude (i.e. the excess variance) correlates with the X--ray spectral index, and anti-correlates with the FWHM of the H$\beta$ line (Turner et al. 1999). Similar studies also indicated, for the first time, that the so called ``Narrow Line Seyfert 1" galaxies were systematically ``more variable" than the classical AGN of equal luminosity (Leighly 1999). Later on, the variability vs. luminosity anti-correlation was also confirmed on long time scales (Markowitz \& Edelson 2001), and it was soon suggested that the excess variance vs. luminosity relation might be just a by-product of a more ``fundamental" relation, that of the ``variability vs. M$_{\rm BH}$ relation" (Lu \& Yu, 2001; Bian \& Zhao, 2003; Papadakis, 2004). These suggestions were put forward even before the McHardy et al (2006) and Koerding et al (2007) M$_{\rm BH}$ (and accretion rate) scaling relations were published. These relations, together with the standard accretion theory of $\alpha$-disks (Shakura \& Sunyaev 1973), which predicts that all the disc characteristic time scales should depend linearly on M$_{\rm BH}$ (e.g. Treves, Maraschi \& Abramowicz 1988), reinforce the hypothesis of the ``variability vs. M$_{\rm BH}$" relation being the fundamental relation which determines all the other observed relations, although alternative suggestions are still being considered (Liu \& Zhang 2008). 

The largest collection so far of excess variance measurements in local AGN has been presented by O'Neill et al. (2005). These authors used all the available 40~ks long \asca\ light curves to calculate the excess variance of 46 AGN. They found a strong anti-correlation of variability with M$_{\rm BH}$ and with luminosity. However, the latter correlation disappeared once the $\sigma^2_{\rm rms}$ vs. M$_{\rm BH}$ correlation was taken into account. O'Neill et al. (2005) found a weaker correlation ($\sim96$ \%) between $\sigma^2_{\rm rms}$ and 2-10 keV spectral index, $\Gamma$, than observed previously (Turner et al. 1999). Miniutti et al. (2009) confirmed the variability vs. M$_{\rm BH}$ trend extending the variability estimation to smaller M$_{\rm BH}$ AGN. Finally, Zhou et al. (2010) tried to calibrate accurately, and estimate the intrinsic scatter of the $\sigma^2_{\rm rms}$ vs. M$_{\rm BH}$ relation, using high quality {\it XMM-Newton} light curves of AGN with M$_{\rm BH}$ measured through the ``reverberation mapping" technique. They found that the intrinsic scatter of this relation is even smaller than the one implied by the M$_{\rm BH}$ uncertainties. Moreover, no dependence of the variability on either accretion rate or spectral index was observed.

In this paper we present the results from the excess variance measurements of a sample of 161 AGN. This is about three times larger than the \asca\ sample of O'Neill et al. (2005) and, at present, the largest sample in which the short time variability (less than a day) has been systematically investigated. In addition, we estimated the excess variance in various energy bands, and on different time scales. The first aim of our work is to investigate whether the results obtained from the detailed PSD analysis of good quality light curves of a few AGN are applicable to the ``majority" of the X--ray studied AGN as well. Note that the McHardy et al (2006) and Koerding et al (2007) scaling relations were based on the accurate PSD results for a relatively small number of objects (for example there were just 10 AGN in the McHardy et al. 2006 sample). The excess variance, being the integral of the PSD over the frequency window sampled in the light curve that is used to compute it, can be a powerful tool that can be used to investigate the applicability of these relations to a much larger sample of AGN. This can be done with the excess variance vs. M$_{\rm BH}$ and excess variance vs. accretion rate relations, which we study in detail in this work, putting particular emphasis on the comparison between the PSD model predictions and the observed relations. 

The second major aim of our work is to investigate how accurately we can ``weigh" the central M$_{\rm BH}$ in AGN with $\sigma^2_{\rm rms}$, and we provide accurate recipes to measure M$_{\rm BH}$ from X-ray variability using light curves of various lengths. Moreover, we present the results from the study of the variability amplitude with the X--ray spectral index and FWHM H$\beta$. Our results indicate that the same variability mechanism operates in all AGN, and that significant differences in the variability amplitude of NLS1 and ``typical" broad line Seyferts (BLS1) can be fully understood once the observed variability is properly ``normalized" to the M$_{\rm BH}$ and accretion rate of these objects. 

\section{The catalogue: CAIXAvar}

XMM-\textit{Newton} provides high statistics, low background, uninterrupted
light-curves, allowing the calculation of the excess variance in large samples
of AGN. We study here a new catalogue, CAIXAvar, that is a sub-catalogue of the
CAIXA sample presented by Bianchi et al. 2009a,b. CAIXA consists of all the
radio-quiet X-ray unobscured (N$_{H}<2\times10^{22}$ cm$^{-2}$) AGN observed by
\xmm\ in targeted observations. We selected a sub-catalogue of CAIXA, creating
CAIXAvar, which includes datasets of sources with the following
characteristics: 1) cleaned exposure times larger than 10 ks; and 2) at least
20 counts in the (rest-frame) 2-10 keV band for each time bin of 250 s.
Compared to CAIXA (Bianchi et al. 2009a,b), we expand the sample including all
observations whose data are public as of June 2010 and consider here also
multiple observations of the same object. There are 161 sources in CAIXAvar that 
fulfil these criteria (with 260 observations), 125 and 158 of which have a
measurement of the M$_{\rm BH}$ and FWHM of H$_{\beta}$, respectively.  
All sources have at least one observation with exposure longer than 10 ks, 
and there are 89, 56 and 32 sources with at least one observation 
with exposure longer than 20, 40 and 80 ks, respectively. 

We estimate the bolometric luminosities (L$_{\rm Bol}$) 
through four different recipes. First, we collected the values reported 
by Woo \& Urry 2002, whose measure of the L$_{\rm Bol}$ is derived by integrating the source 
spectral energy distribution. 
The other three methods use the mean X-ray luminosity of the longest \xmm\ observation. 
We used the constant X-ray bolometric correction by Elvis et al. (1994), 
the luminosity-dependent one provided by Marconi et al.
(2004) and, finally, the BH-mass dependent bolometric correction by Vasudevan et al. 2007. We checked and confirmed that the results presented here are not affected by the bolometric correction we use. Thus, in this work we report the results when we used the Marconi et al. (2004) bolometric correction to estimate L$_{\rm Bol}$ for all objects in the sample. This correction is applicable to all the sources in CAIXAvar, allowing us to use the same recipe uniformly throughout the sample. We also use these values to compute the ratio L$_{\rm Bol}/$L$_{\rm Edd}$ (for the sources with available M$_{\rm BH}$ estimates) which we will refer to as the 
``accretion rate" ($\dot m$) of the source.

M$_{\rm BH}$ from reverberation mapping are preferentially used, then
stellar velocity dispersion, measurements using the relation between broad-line
region radius and optical luminosity (primarily based on H$\sc {\beta}$, but
also on Mg {\sc II} whenever the former was not available) and finally the
relations for the narrow line region. The measurements of the stellar velocity
dispersion are collected from the HyperLeda database ({\it
http://leda.univ-lyon1.fr/}) and the M$_{\rm BH}$ estimated using the relation:
$Log(M_{BH}/M_{\odot}) = 8.12 + 4.24~Log(\sigma_*/200~km~s^{-1})$; see Gultekin
et al. (2009). However, as Greene \& Ho (2006) and Greene et al. (2008)
suggest, the M$_{\rm BH}$ vs. stellar velocity dispersion might deviate at low 
M$_{\rm BH}$ from this relation, thus for Log(M$_{\rm BH}$)$<$6 the values 
are directly taken from Greene \& Ho (2006). 

\begin{figure}
\begin{center}
\epsfig{file=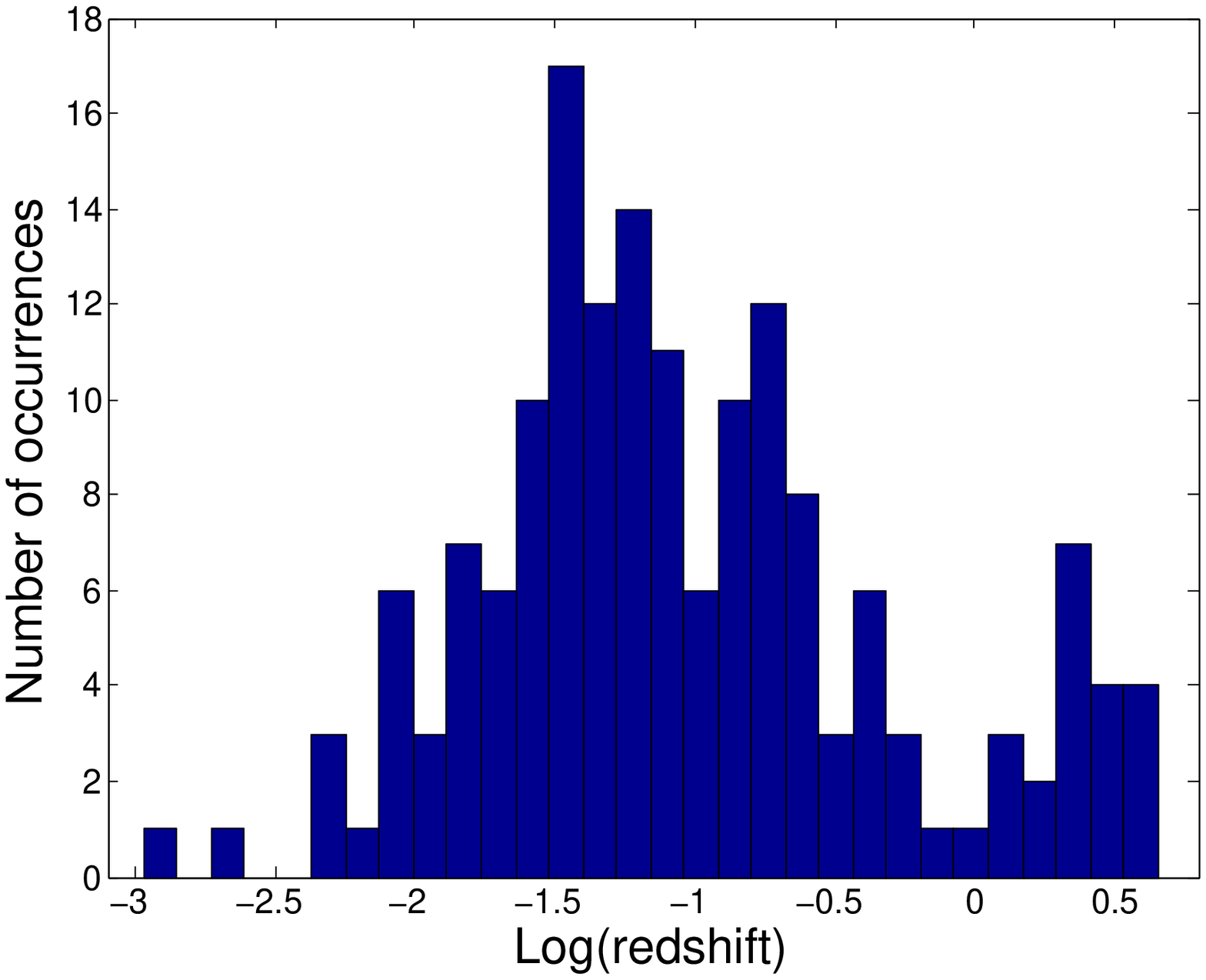,width=0.24\textwidth}
\epsfig{file=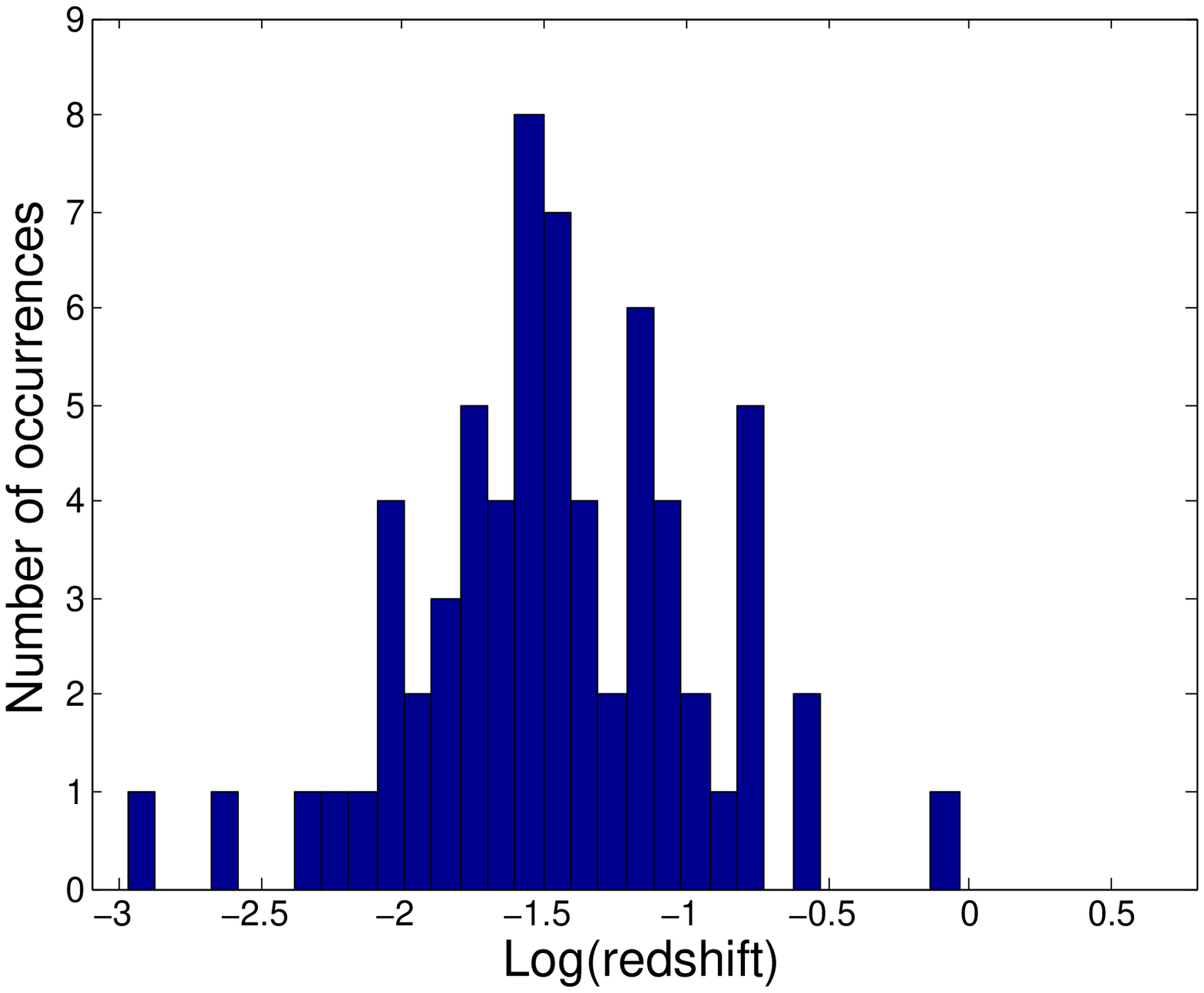,width=0.24\textwidth}
\end{center}
\caption{{\it (Left panel)} Redshift distribution of the all the 161 sources in CAIXAvar. Although the vast majority of AGN belong to the local Universe z$<0.2$, 20 sources in CAIXAvar have z$>1$, with 4 objects with redshifts as high as 4.  {\it (Right panel)} Redshift distributions of all the sources with at least 1 variable segment (see bold values in Tab. 2) used hereinafter to study the correlations. Only 3 objects, with at least a variable segment, have a redshift higher than 0.2 and only one have z $\simeq0.9$. }
\label{z} 
\end{figure}

Table \ref{multiWL} summarises the multi-wavelength data collected for the
sources in CAIXAvar. 
\begin{table*}
\begin{center}
\footnotesize
\begin{tabular}{ | c c c c c c c c c c |}
\hline                      
Sample             & Name    & z          & Log(M$_{\rm BH})$       & Log(L$_{\rm Bol,Woo}$) & Log(L$_{\rm Bol,Elv}$) & Log(L$_{\rm Bol,Mar}$) & Log(L$_{\rm Bol,Vas}$) & FWHM H$_{\beta}$ & $\Gamma$ \\ 
                           &                &             & (M$_{\odot}$)                 & (erg s$^{-1}$)                     & (erg s$^{-1}$)                 & (erg s$^{-1}$)                  & (erg s$^{-1}$)                   & (km s$^{-1}$)            &                      \\
\hline                      
R+C                  &MRK335& 0.0258 & 7.15$\pm$0.11             & 44.69                                  &  44.3561                           & 44.0919                            & 44.5322                             &          1620                  &  1.53           \\
\hline                      
\end{tabular}

\caption{List of all Multi-wavelength data for each source. The first column indicates to which samples the source belongs (R for the Rev sample 
and C for the CAIXAvar sample). The entire table is published in the on line version of the paper, only.
}
\label{multiWL}
\end{center}
\end{table*} 
The left panel of Figure \ref{z} shows the redshift distribution of all the 161 AGN in the CAIXAvar sample. The vast majority of these AGN are local with redshift lower than 0.2. However, CAIXAvar contains also 20 AGN with redshift higher than 1 and in particular 4 AGN with z as high as 4. The right panel of Fig. \ref{z} shows the distribution of the sources of CAIXAvar with at least one variable segment. The sources with at least one variable segment (which we use to study the variability and correlations of the CAIXAvar sample with different physical parameters) is primarily made of local AGN. In fact, there are only 3 objects with redshift higher than 0.2 and only one with z $>0.9$ with at least one variable segment. 

\subsection{Reverberation sample (Rev)} \label{rev}

In order to keep the scatter on the relations with the M$_{\rm BH}$ as small as
possible we also selected a side sample (``Rev", hereafter) of AGN with M$_{\rm BH}$ 
measured through reverberation only. Peterson et al. (2004) measured the M$_{\rm BH}$ 
for 37 AGN, 8 of which have either: no; or still proprietary; or too short; \xmm\
observations. The remaining 29 sources are analysed regardless of the radio
loudness and absorption (after individually checking that these do not play a
major contribution to the measured 2-10 keV variability). All the reverberation
M$_{\rm BH}$ are from Peterson et al. (2004) apart from PG1351+442, whose 
M$_{\rm BH}$ estimate is taken from Kaspi et al. (2000). The M$_{\rm BH}$ 
estimate for  the 6 objects (one of which, Mrk290, is not in the Peterson et al. 2004 sample)
re-analysed by Denney et al. (2010) is taken from that paper.  Finally, the M$_{\rm BH}$ 
estimate for Mrk766 (which is not included in Peterson et al. 2004) is taken from Bentz et al. (2009). 

Due to the poor quality of the data, Peterson et al. (2004) mark as less
accurate the measurements of the M$_{\rm BH}$ of IC4329A and PG1211+143. Following
Markowitz et al. (2009) we use for IC4329A a value of M$_{BH}=2.17\times10^8$
M$_{\odot}$, obtained through the stellar velocity dispersion
($\sigma_*=218-231$ km s$^{-1}$; Oliva et al. 1999) that is in agreement with
all the other methods used to estimate the M$_{\rm BH}$. PG1211+143, instead, has no
stellar velocity dispersion measurement. We note that this value is almost one
order of magnitude higher than the reverberation mapping estimate given by
Kaspi et al. (2000; Log(M$_{BH})=7.37$) and more than one order of magnitude
higher compared to the M$_{\rm BH}$ estimates from single epoch spectra (H$_{\beta}$;
Log(M$_{\rm BH})=6.72\pm0.43$; C{\sc IV} $6.95\pm0.40$ and from BLR estimate
$6.84\pm0.29$; Kelly \& Bechtold 2007). In the end, we decided to use the
Peterson et al. (2004) measurement.  We also added NGC4395 in the Rev sample
because  an HST campaign allowed the reverberation mapping measurement of
$M_{BH} = 3.6\times10^5 M_{\odot}$ for this object (Peterson et al. 2005).  For
conformity we use this value even if we note that it might be an overestimate
of the true mass. In fact the low bulge velocity dispersion ($\sigma_*<30$ km
s$^{-1}$; Filippenko \& Ho 2003) deviates from the extrapolation of the
M-$\sigma$ relation when this estimate of the M$_{\rm BH}$ is adopted (but see
Greene \& Ho 2006). The final Rev sample consists of 32 AGN (11 more AGN 
than in the Zhou et al. 2010 sample), 6 of which are either absorbed or radio 
loud objects, hence they are  not present in CAIXAvar.

\section{Observations and data reduction}

The analysis products have been obtained starting from the {\sc ODF} files and
after reprocessing with SASv6.9. The source and background regions have been
selected with the same procedure as Bianchi et al. 2009a,b apart from the
observations in Small Window mode for which the background has been selected
from a source-free region on the same chip as the source (source and background
have the same extraction radius). The screening for flaring of particle
background was performed in two steps. The first cut has been performed in the
same way as Bianchi et al. (2009a,b), via an iterative process that leads to a
maximisation of the signal-to-noise (see Piconcelli et al. 2005). We also
filtered out every timebin during which the 10-15 keV light curve of the entire
field of view (but the source region) had more than 30 and 1250 counts during
the 250 s bins for the Small Window and Full Frame mode, respectively. This
prevents small fluctuations of the particle background from making an
appreciable contribution to the excess variance measurement. Objects
requiring long cuts have been individually screened, and an \textit{ad hoc}
background selection has been performed on a case-by-case basis.

Light curves have been constructed in the energy bands: 0.3--0.7, 0.7--2, 2--10
and 0.3--10 keV, by selecting events with {\sc flag$==$0}, {\sc pattern$<=$4}
and {\sc xmmea\_ep} . The energy band limits  refer to rest-frame energies
(thus we were unable to compute the low energy excess variances for the 15 
objects with a redshift larger than about 2).  Light curves in all bands are 
corrected with the {\sc SAS} task {\sc epiclccorr}. 
Every light curve was calculated with 250 s bins, and divided into segments 
of 10, 20, 40 or 80 ks. During the estimation of
the excess variance (see below), we rejected the time bins with fractional
exposure lower than 0.35, regardless of the observation mode (in Small Window
mode the EPIC pn camera is active for only about $\sim70$ \% of the time, thus
we reject only time bins with background flare lasting more than half of the
active time). 

\subsection{Normalised excess variance computation}

Following Nandra et al. (1997), Turner et al. (1999), Vaughan et al. (2003) and
Ponti et al. (2004), we compute, for each light curve segment, the normalised excess
variance with the formula: 

\begin{equation}
\sigma^2_{\rm rms}=\frac{1}{N\mu^2}\sum^N_{i=1}[(X_i-\mu)^2-\sigma^2_i]
\end{equation}

where $N$ is the number of good time bins in the segment, $\mu$ is the
unweighted arithmetic mean of the counting rates within the segment, $X_i$ and
$\sigma_i$ are the counting rates and uncertainties, respectively, in each bin
(hereinafter, we refer to $\sigma^2_{\rm rms}$ simply as "excess variance" of the 
light curve).
We rejected all the segments with less than 50 \% of good time bins. When more
than one valid segment is available, the excess variance has been determined
computing the unweighted mean of all the individual estimates. This has the
potential to reduce the large uncertainty owing to the stochastic nature of the
excess variance. The method of the excess variance uncertainty estimation 
and the impact of different source redshifts is described in detail in the 
Appendix A and B. 

\section{The excess variances}

\begin{table*}
\begin{center}
\footnotesize
\begin{tabular}{ | cccc cccc |}
\hline                      
Sample             & Name    & Log(M$_{\rm BH})$       & Log(L$_{\rm Bol,Mar}$) & $\sigma^2_{\rm rms,80,2-10keV}$ &  $\sigma^2_{\rm rms,40,2-10keV}$ &  $\sigma^2_{\rm rms,20,2-10keV}$  &  $\sigma^2_{\rm rms,10,2-10keV}$ \\
                           &                & (M$_{\odot}$)                 & (erg s$^{-1}$)                  & & & & \\
\hline                      
R+C                  &MRK335 & 7.15$\pm$0.12              & 44.09                            & $2.6^{+1.9}_{-0.9}\times10^{-2}$    & $2.3^{+1.1}_{-0.5}\times10^{-2}$     & $1.5\pm0.2\times10^{-2}$                  & $1.02\pm0.16\times10^{-2}$    \\
\hline                      
\end{tabular}
\caption{List of all the $\sigma^2_{\rm rms,2-10keV}$ computed, in the 2-10 keV band, with 10, 20, 40 and 80 ks intervals. This is an update of the multi-wavelength table published in Bianchi et al. (2009). The entire table is published in the on line version of the paper, only.}
\label{exCAIXA}
\end{center}
\end{table*} 
Table \ref{exCAIXA} shows the excess variance estimates and the source parameters for the CAIXAvar and Rev samples, respectively, in the 2--10 keV band.

In several cases, due to the fact that we were quite conservative in assigning 
the 90\% confidence limits on our estimates, and especially for shorter 
intervals (10-20 ks) and less variable AGN (which are those objects with the 
largest M$_{\rm BH}$), the lower limit on $\sigma^2_{\rm rms}$ implies an intrinsic 
excess variance less than zero. In this case, we consider our measurement 
as a ``non-detection", and we simply list in the respective tables the 90\% upper 
limit of our estimate. 

\subsection{Method of analysis}

In the following sections we study the correlations between the excess variance in various energy bands, and also between $\sigma^2_{\rm rms}$ and other source parameters. We used various methods (such as a linear bisector and/or  a Y/X linear model fit to the data in the log-log space, methods based on the points scatter, etc) to measure the correlations between two parameters.  We observed that, despite some slight differences,  all the methods yielded similar results. We thus decided to apply a censored fit of a linear model with the bisector method to all the data sets we considered (see  Appendix A of Bianchi et al. 2009), mainly for consistency with previous results (e.g. Bianchi et al. 2009a,b). All model fits were performed in the log-log space, and in all cases, the excess variance is considered to be the ``dependent" variable, and the other parameters the ``independent" variable. We performed MonteCarlo simulations to compute  the uncertainties on the best-fit parameters and the correlation significance as in Bianchi et al. (2009).

Finally, following Nandra et al. (1997) and O'Neill et al. (2005), in our model fits we considered only the excess variances of the objects with at least one segment with significant variations (we also plot only these objects in all figures below). To identify these ``variable" segments we performed a $\chi^2$ test to each one of them (for each source and any interval length, 10, 20, 40 and 80 ks), accepting as evidence for "significant" variability a null hypothesis probability of less than 1\%. We measured at least a "variable" interval in 25, 42, 54 and 47 sources in the 80, 40, 20 and 10 ks intervals, respectively. In total, there are 65 different sources (i.e. 40~\% of the 161 sources in the CAIXAvar sample) with at least one variable segment in at least one time interval. Table 2 lists in boldface the $\sigma^2_{\rm rms}$ measurements for these sources, which we use in the studies we present here. For the remaining 96 sources we do not detect significant variability during any segment and the corresponding upper limits are reported in Table 2. Not surprisingly, it turned out that most objects with upper limits on $\sigma^2_{\rm rms}$ (in the case of ``non-detections") did not have a single ``variable" segment. This fact implies that we did not consider most of the objects with upper limits on their $\sigma^2_{\rm rms}$ in the model-fits. This is not a serious problem as in most cases  the upper limits are so large that their presence in the correlation plot do not add any physically constraining power. 
Nevertheless, we always checked ``a posteriori" that none of these discarded upper limits is observed below or close to the best fit relations (thus carrying potentially important information).

\section{Comparison of $\sigma^2_{\rm rms}$ in various energy bands}

\begin{figure*}
\begin{center}
\epsfig{file=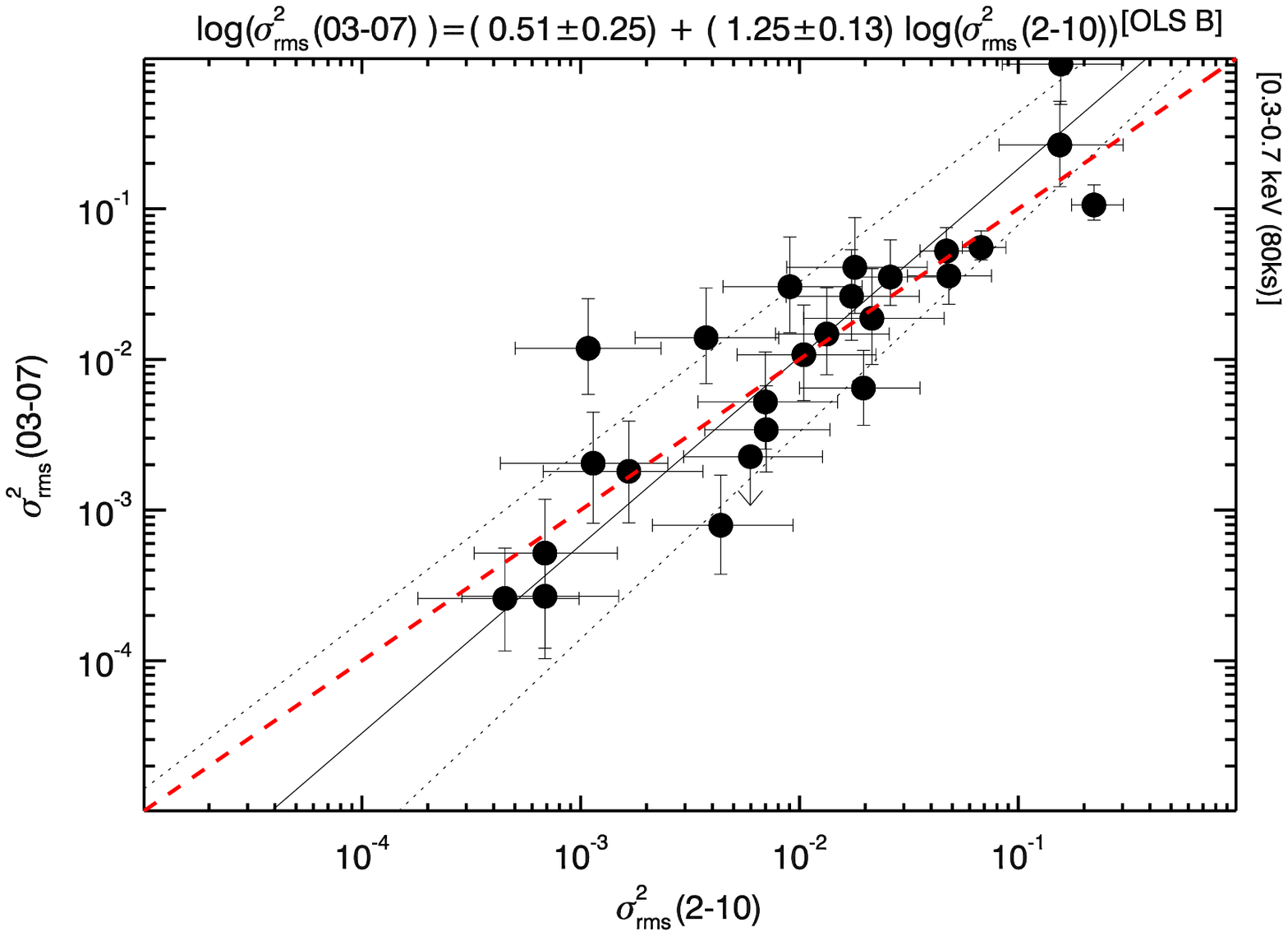, width=0.49\textwidth}
\epsfig{file=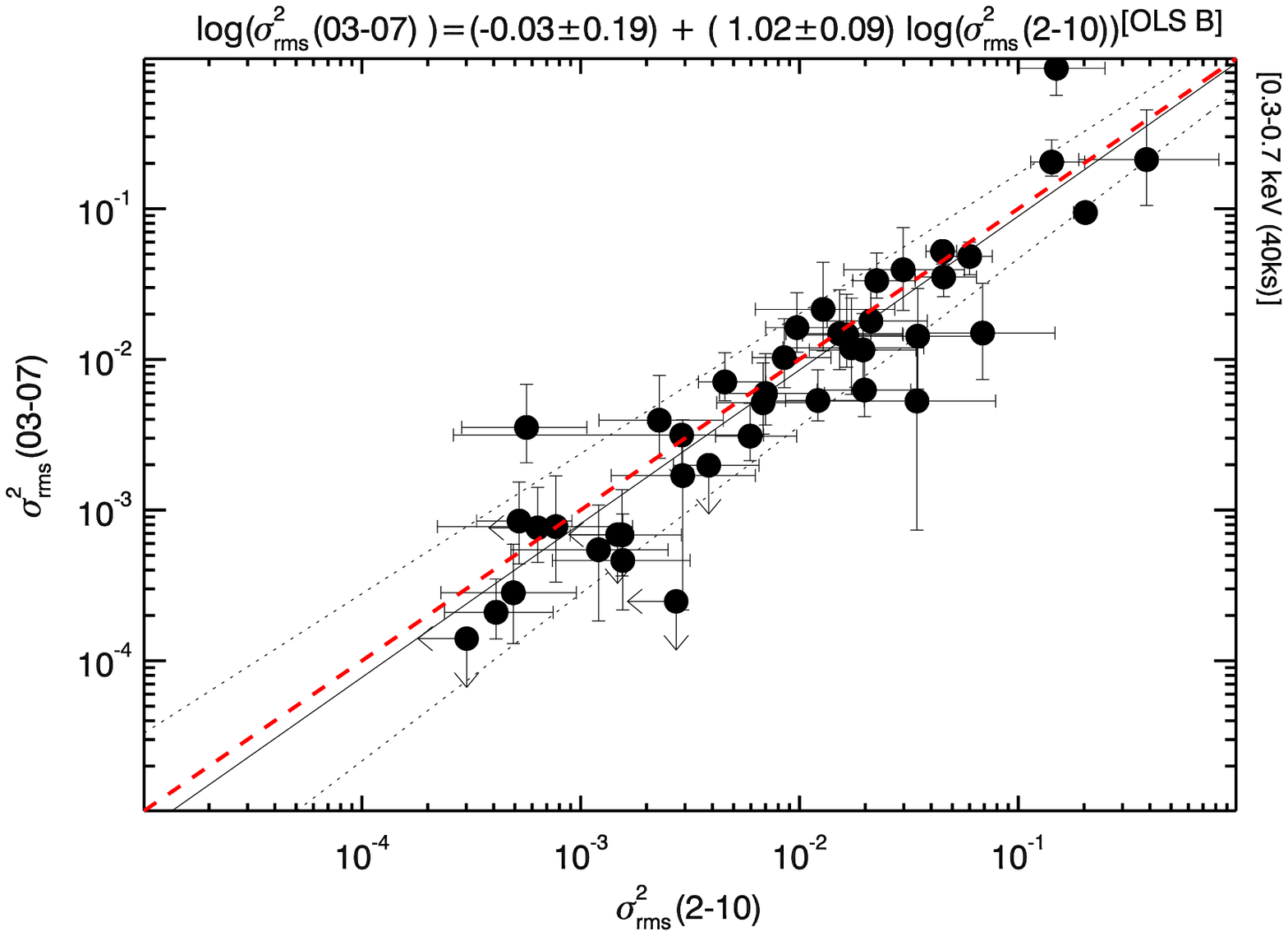, width=0.49\textwidth}
\epsfig{file=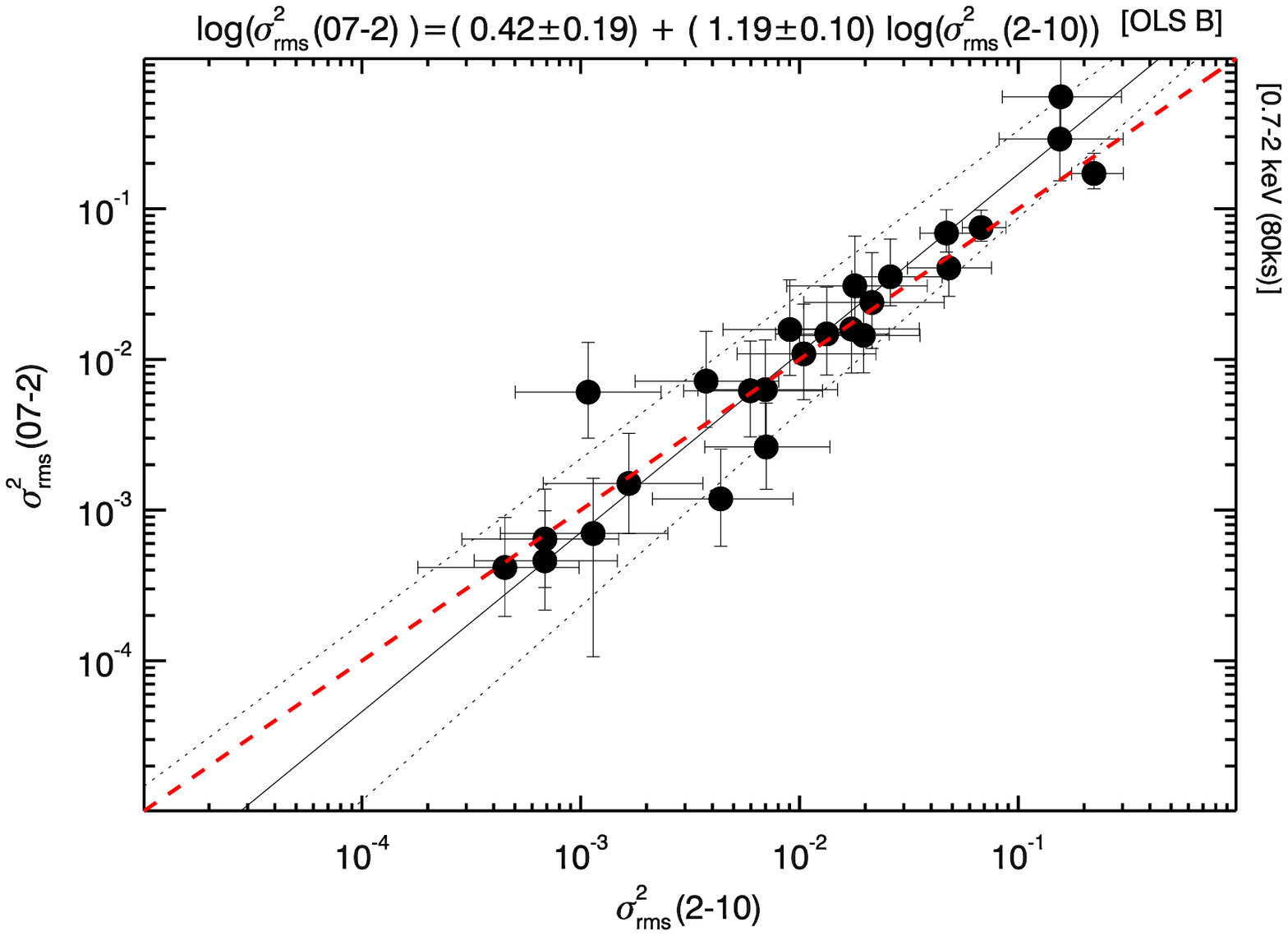, width=0.49\textwidth}
\epsfig{file=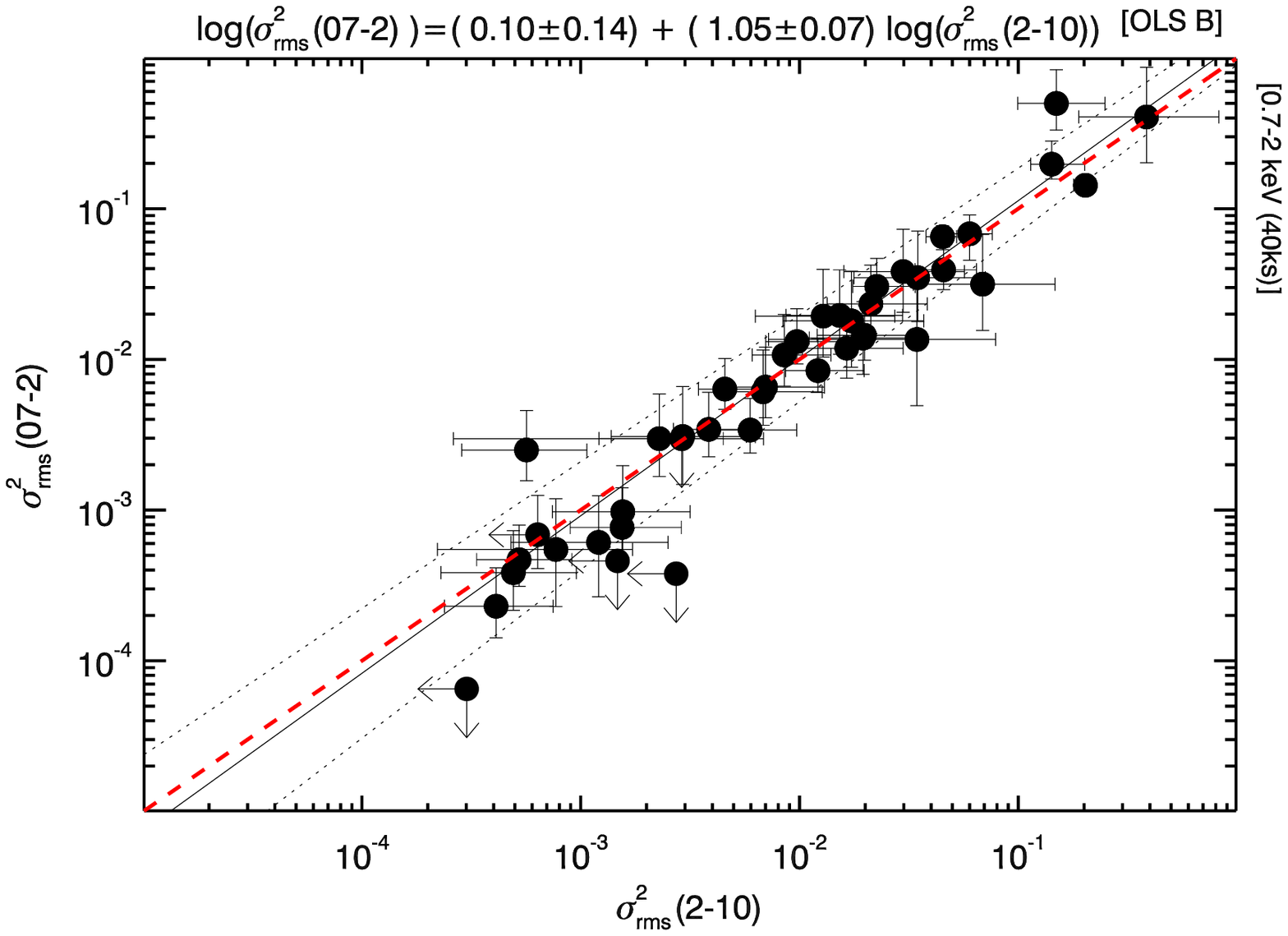, width=0.49\textwidth}
\end{center}
\caption{\label{exvsex} {\it Upper panels} Soft (0.3-0.7 keV)  vs. hard (2-10 keV) $\sigma^2_{\rm rms}$, computed within the 80 ks (left), and 40 ks (right) intervals. {\it Lower panels}. Similar plots for $\sigma^2_{\rm rms}$ in the 0.7-2 vs. $\sigma^2_{\rm rms}$ in the 2-10 keV band. The best fit curves are plotted  with solid lines and the combined 1-$\sigma$ error on the slope and normalisation with dotted lines. The red dashed lines represents the one to one relations expected in the case of achromatic variations. On top of each panel (in this and all similar subsequent figures) we report the best fit models (see also Tab. \ref{relation}). }
\end{figure*}

The three energy bands (0.3--0.7, 0.7--2 and 2--10 keV) we consider in this work are dominated by different spectral components (i.e. warm absorber and the so-called  soft excess are more pronounced in the softer energy bands). Their presence can introduce additional spectral variability in these bands. Consequently, a non-linear relation between the measured $\sigma^2_{\rm rms}$ in these bands may not be surprising. Even if the best-fit relation is linear when we consider the $\sigma^2_{\rm rms}$ in one band vs. $\sigma^2_{\rm rms}$ in another, a departure from a slope of $\sim 1$, or a best-fit normalisation different than unity, could be expected.

The upper left and right panels of Figure \ref{exvsex} show a plot of  $\sigma^2_{\rm rms, 0.3-0.7 keV}$ vs. $\sigma^2_{\rm rms, 2-10 keV}$ (for the CAIXAvar sample), when computed within time intervals of 80 ks and 40 ks, respectively. Similarly, the lower panels show the same plots for the 0.7-2 vs.  2-10 keV bands. Clearly the excess variance in the soft bands are well correlated with $\sigma^2_{\rm rms, 2-10 keV}$. Our best-fit results are listed in Tab. \ref{relation}. The best-fit models are also plotted in Fig. \ref{exvsex} (solid lines). The dotted lines represent the combined 1-$\sigma$ error on the best-fit slope and  normalisation values, while the red dashed line shows the one to  one relationship.

The best-fit slope and normalisation estimates are consistent, within the errors, with $1$ and $0$, respectively, in the case of the excess variance estimates which are calculated using the 40, 20 and 10 ks light curves. This result implies that, on these ``short" time scales, the main driver for the X-ray variability in these objects is the continuum normalisation variations, while variations of the other spectral components must be of a much smaller amplitude. 
In the case of the 80 ks based $\sigma^2_{\rm rms}$ estimates, the best-fit results suggest a slope steeper than $1$ and a normalisation larger than zero (in the log-log space). A steeper slope implies that the more variable AGN tend to sit above the 1 to 1 relation, while less variable objects below. Consequently, the PSD shape and/or amplitude may not  scale from one energy band to the other in the same way for all objects and, as a consequence, we observe a not null normalisation once the relation is extrapolated to 0. However, the significance of this result is rather low (at the $\sim 2\sigma$ level) . 

The very good correlation between the excess variance measurements in the three bands could allow us to integrate the signal over the largest possible energy band (i.e. 0.3--10 keV) in order to gain better precision on the variability measurements. However  we present below the results from the correlation of the  2-10 keV band excess variance measurements with other source parameters, for comparison with previous work, and in order to allow a future extension of this work using longer timescales data from the RXTE PCA observations.

\section{Correlations between the excess variance and other source parameters}
\label{cor}

To investigate potential physical mechanisms which drive the X--ray variability in local AGN on short time scales (i.e. less than a day) we search for correlations between $\sigma^2_{\rm rms}$ (i.e. the source variability amplitude) of the variable sources in CAIXAvar and various  physical parameters of the sources such as: M$_{\rm BH}$, accretion rate, luminosity, width of the H$_{\beta}$ line, and X--ray spectral slope. We use the excess variance measurements from the 10, 20, 40 and 80 ks segments when we study the variability vs. M$_{\rm BH}$ or $\dot{m}$ relation. In all other cases, for brevity reasons, we use the $\sigma^2_{\rm rms}$ measurements from the 20 and/or 40 ks intervals only. 

CAIXAvar is the largest sample used so far to study the AGN X-ray variability on short time scales (less than a day). Large samples of AGN have been used to study the variability properties of higher redshift AGN, and on longer time scales, albeit with light curves which are heavily undersampled. For example, Almaini et al. (2000) and Manners et al. (2002) have used \rosat\ data to study the variability properties of 86 and 156 AGN, respectively. Both Paolillo et al. (2004) and Papadakis et al. (2008), thanks to the study of variability of \chandra\ and \xmm\ deep field observations, trace a higher redshift AGN population. Paolillo et al. (2004) study a sample of 430 Chandra sources from the deep field south (74 of which result to be variable) and Papadakis et al. (2008) 66 AGN from the \xmm\ Lockman Hole observations. Both these works study AGN variability, but only on time scales longer than $\sim1$ day. Finally, Vagnetti et al. (2011) used a different technique, the structure function, to study the variability of a large sample (412) of AGN from the \xmm\ Serendipitous Source Catalogue. 

\subsection{The $\sigma^2_{\rm rms}$ vs. M$_{\rm BH}$ relation}
\label{mbh}

The presence of a correlation between X-ray variability and M$_{\rm BH}$ is already well established (Lu \& Yu 2001; Bian \& Zhao 2003; Papadakis 2004; O'Neill et al. 2005;  Nikolajuk et al. 2006; 2009; Zhou et al. 2007; Miniutti et al. 2009; Zhou et al. 2010). First, we study this correlation using the results of the Rev sample,  for which every object has a well determined M$_{\rm BH}$. This allows us to keep under control the scatter introduced by the mass uncertainties, that are estimated to be of the order of a factor of 3 or smaller, for this method (Peterson et al. 2004). Then, we extend our study to the full CAIXAvar sample, using the sources in this sample with a M$_{\rm BH}$ estimate.

\subsubsection{Results on the Rev sample}

\begin{figure*}
\begin{center}
\epsfig{file=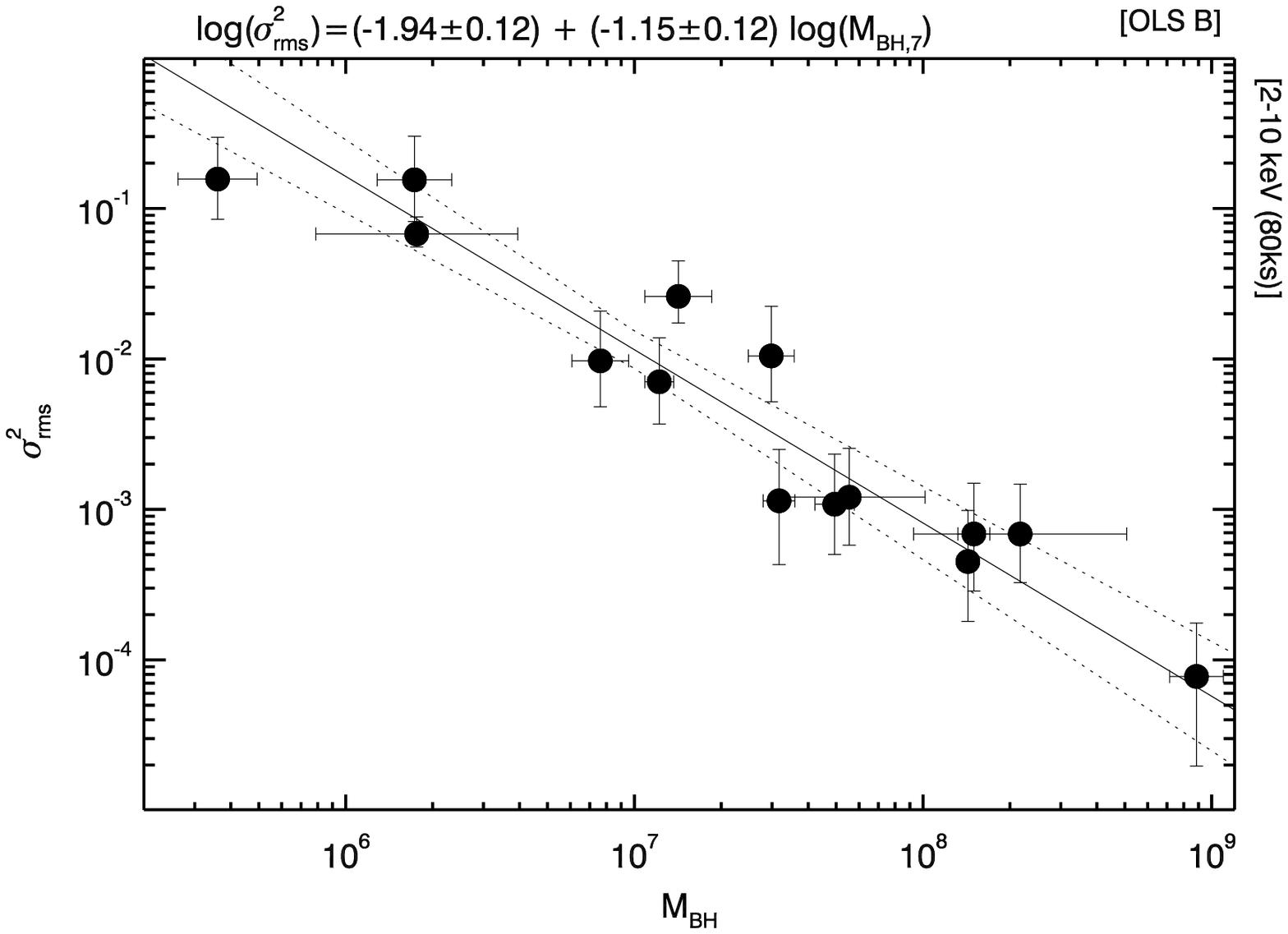, width=0.49\textwidth}
\epsfig{file=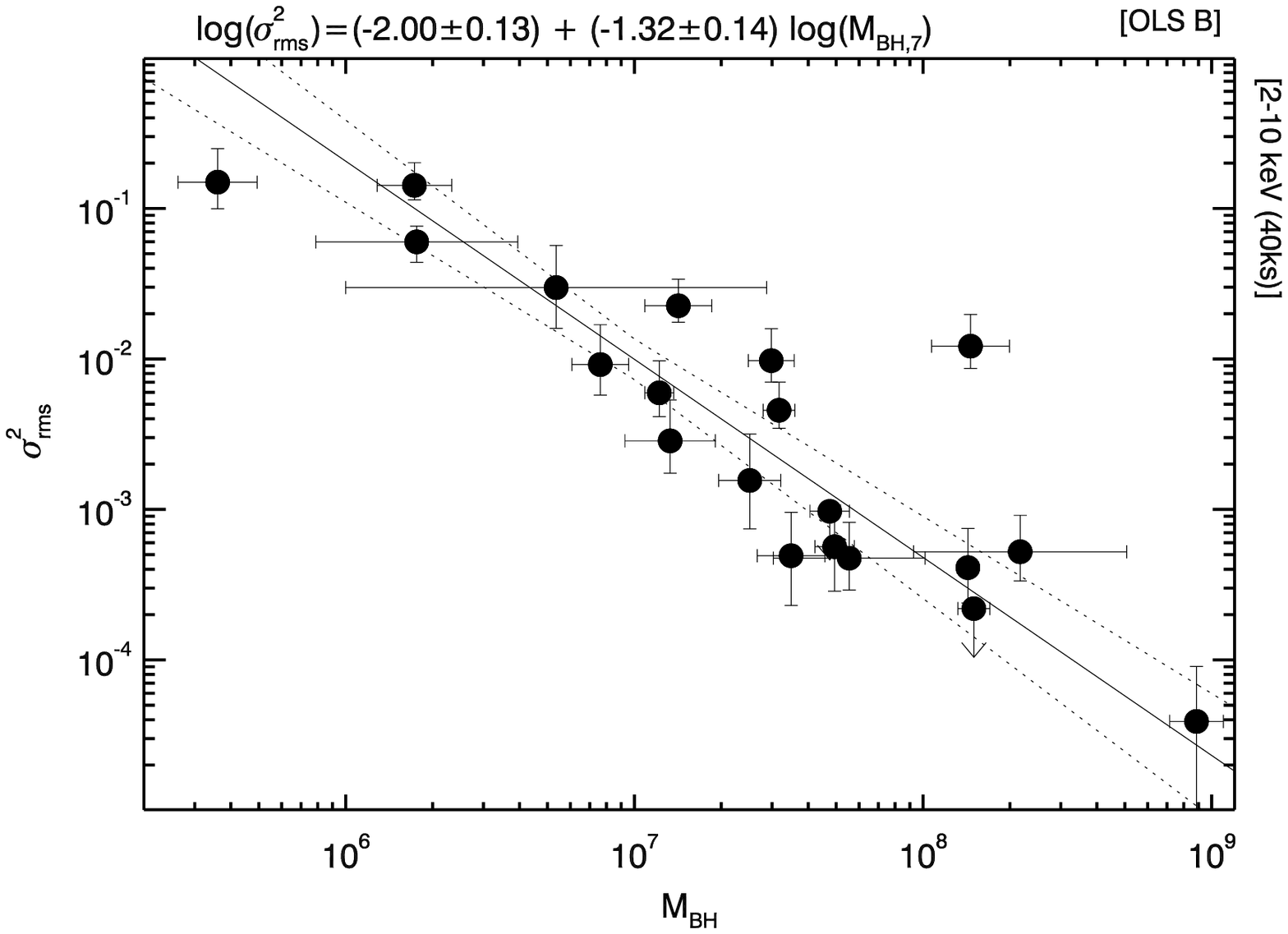, width=0.49\textwidth}
\epsfig{file=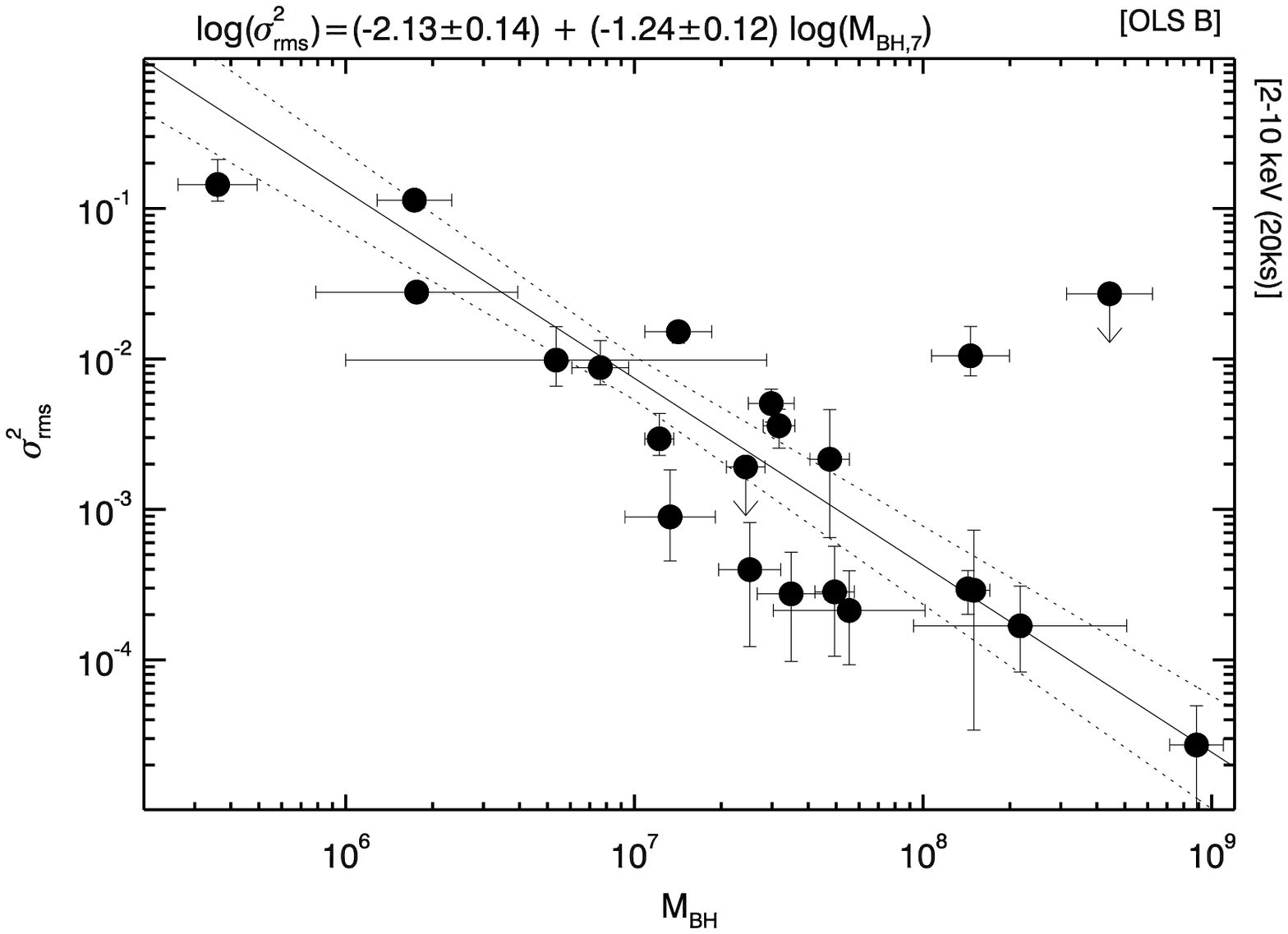, width=0.49\textwidth}
\epsfig{file=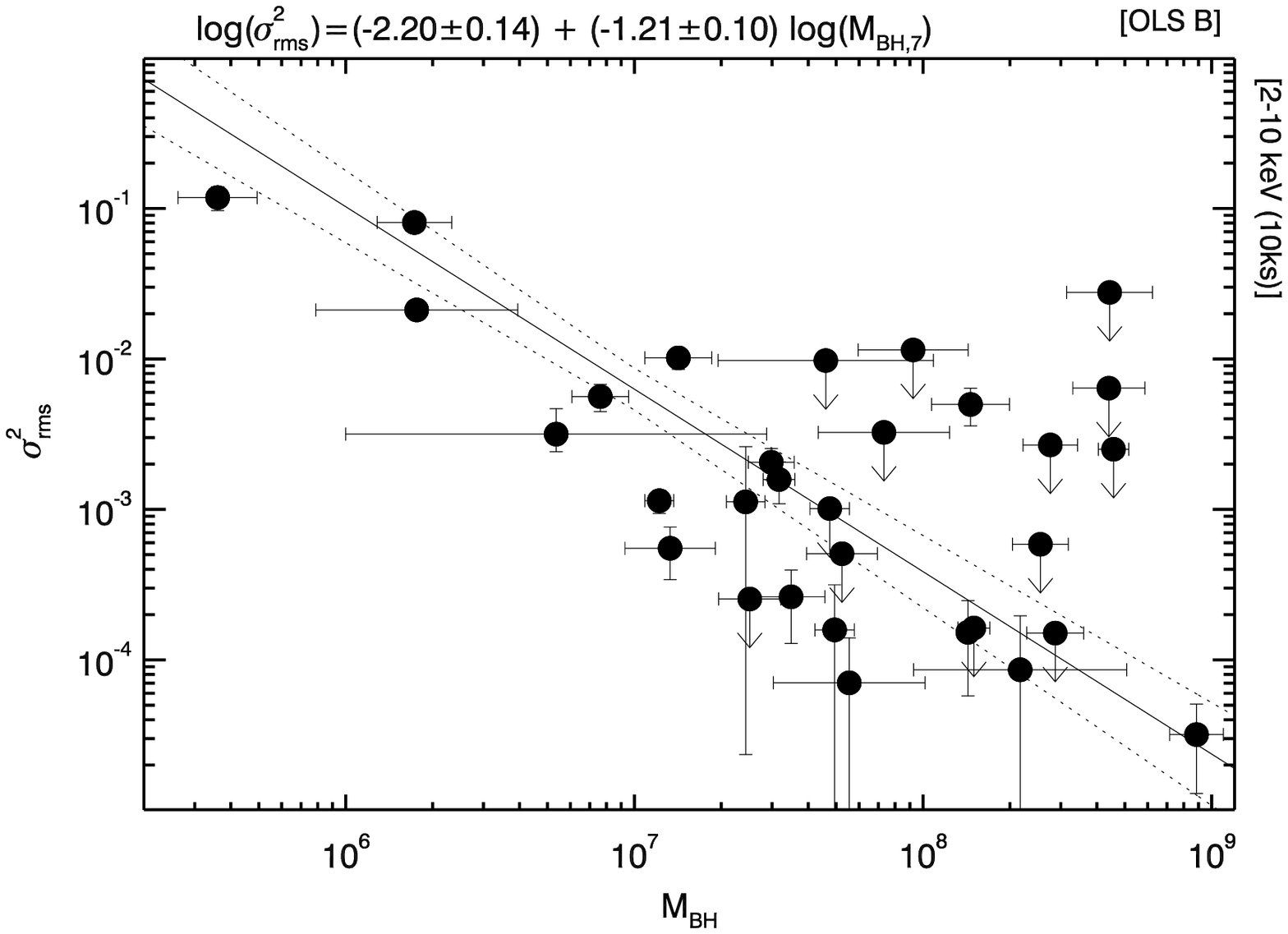, width=0.49\textwidth}
\end{center}
\caption{{\it Upper left}, {\it upper right}, \textit{lower left} and {\it lower right} panels show the $\sigma^2_{\rm rms,80}$, $\sigma^2_{\rm rms,40}$, $\sigma^2_{\rm rms,20}$ and $\sigma^2_{\rm rms,10}$ vs. M$_{\rm BH}$ for the Rev sample. The best fit relationships (see Tab. \ref{relationRev}) are plotted with solid lines and the combined 1-$\sigma$ error on the slope and normalisation with dotted lines. M$_{\rm BH,7}$ indicates M$_{\rm BH}$ in units of 10$^7$ M$_{\odot}$. }
\label{exvsMBH210} 
\end{figure*}

\begin{table*}
\begin{center}
\footnotesize
\begin{tabular}{ |l |  ccl |  ccl  | ccl |}
\hline                      
Relation & & 80 ks & & & 40 ks &  \\
              & Norm & Slope & prob & Norm & Slope & prob \\
\hline
Rev & & & & & & \\
$\sigma^2_{\rm rms}$ vs. $M_{\rm BH}$     & $-1.94\pm0.12$ & $-1.15\pm0.12$ & 99.998 & $-2.00\pm0.13$ & $-1.32\pm0.14$ & 99.998 \\
\hline
\hline
Relation & & 20 ks & & & 10 ks & \\
              & Norm & Slope & prob & Norm & Slope & prob \\
\hline
Rev & & & & & &  \\
$\sigma^2_{\rm rms}$ vs. $M_{\rm BH}$     & $-2.13\pm0.14$ & $-1.24\pm0.12$ & 99.79 & $-2.20\pm0.14$ & $-1.21\pm0.10$ & 99.6  \\
\hline                      
\end{tabular}
\caption{List of all best fit relations of the Rev sample as well as their probabilities.}
\label{relationRev}
\end{center}
\end{table*} 

The four panels  of Figure \ref{exvsMBH210} show the $\sigma^2_{\rm rms,80,40,20,10}$ vs. M$_{\rm BH}$ plot for the Rev sample, respectively  ($\sigma^2_{\rm rms,80}, \sigma^2_{\rm rms,40}$, $\sigma^2_{\rm rms,20}$ and $\sigma^2_{\rm rms,10}$ denote the excess variance measurements from the 80, 40, 20 and 10 ks segments, respectively). A highly significant anti-correlation is observed in all cases (see Tab. \ref{relationRev}). Solid lines in these panels indicate the best-fit relations, and the best-fit results are listed in Tab. \ref{relationRev}. We note that, as expected, the normalisation of the relation increases for longer intervals. As the light curve segment duration increases, the expected variance is expected to increase  as well, as in effect it is like integrating the PSD over a larger frequency window, hence  we do expect to measure a larger variability amplitude.

To quantify the scatter of the data around the best-fit line we compute the quantity $\sigma_{\rm scatter}$ as follows: 
$\sigma_{\rm scatter}=  \sqrt{\sum_{i=1}^{N} \{Log(\sigma^2_{\rm rms,i})-Log[f(M_{BH,i})]\}^2 /N}$; 
where $\sigma^2_{\rm rms, i}$ and $f(M_{BH,i})$ are the measured and best-fit excess variances for a source with M$_{\rm BH}$ of $M_{BH}$, respectively, and $N$ is the total number of sources, excluding sources with upper limits on their $\sigma^2_{\rm rms}$. We also exclude PG1211+143; this source is the only clear outlier in the middle and right panels of Fig.~3, but also has the largest uncertainty associated with its M$_{\rm BH}$ measurement (see \S \ref{rev}). 

The $\sigma_{\rm scatter, 80, 40}$ and $\sigma_{\rm scatter, 20}$ values are 0.44, 0.49 and 0.47, respectively, and they imply a scatter of a factor of 3 in linear space. Interestingly this is comparable with the  uncertainty associated with the reverberation mapping M$_{\rm BH}$ measurements. This result implies that the variability M$_{\rm BH}$ estimates, using the best-fit lines with the parameters listed in Tab. \ref{relationRev}, are at least as accurate as the reverberation mapping ones. 

\subsubsection{The full sample: CAIXAvar}
\label{rmsvsMBH}

\begin{figure*}
\begin{center}
\epsfig{file=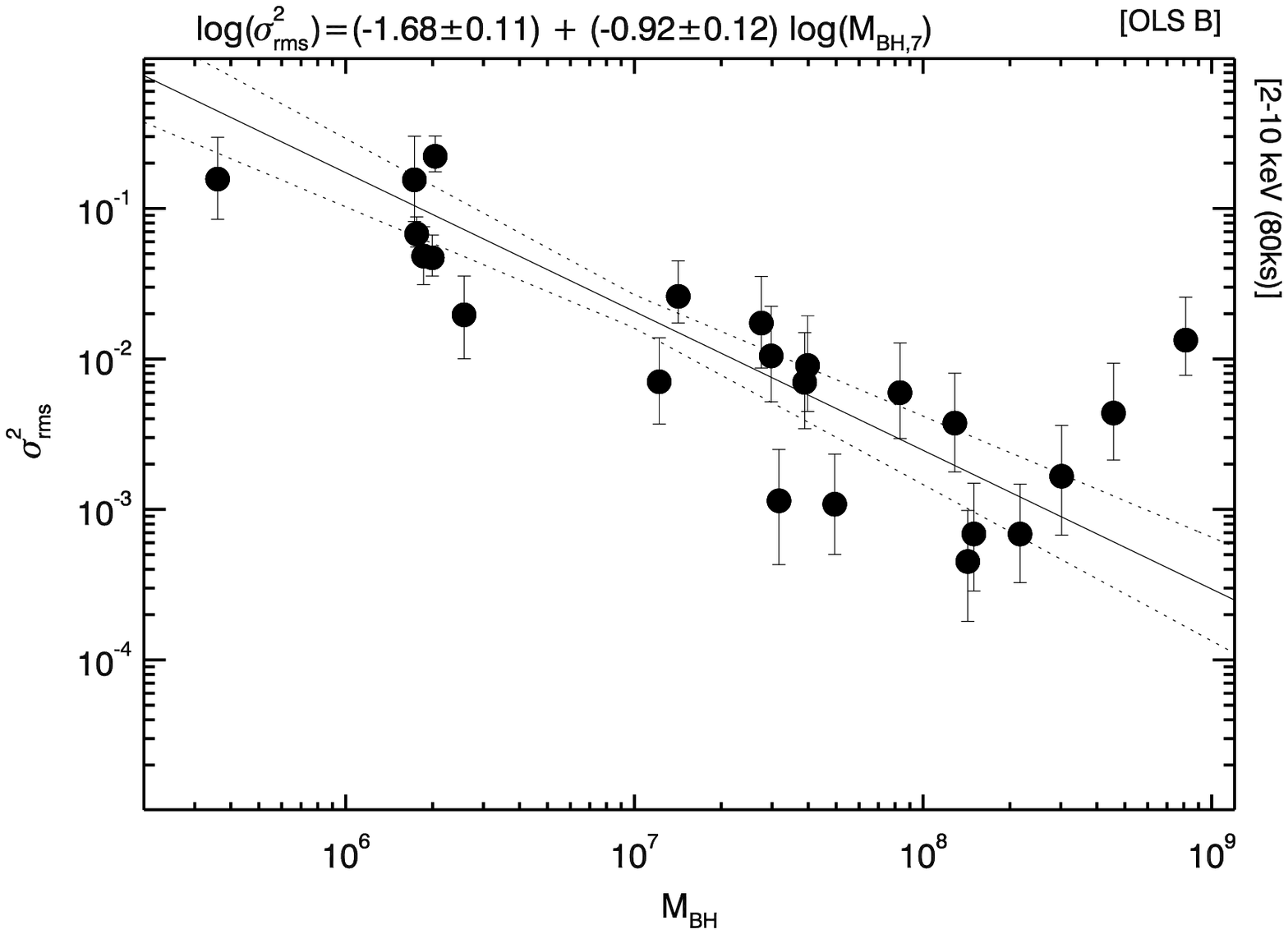,width=0.49\textwidth}
\epsfig{file=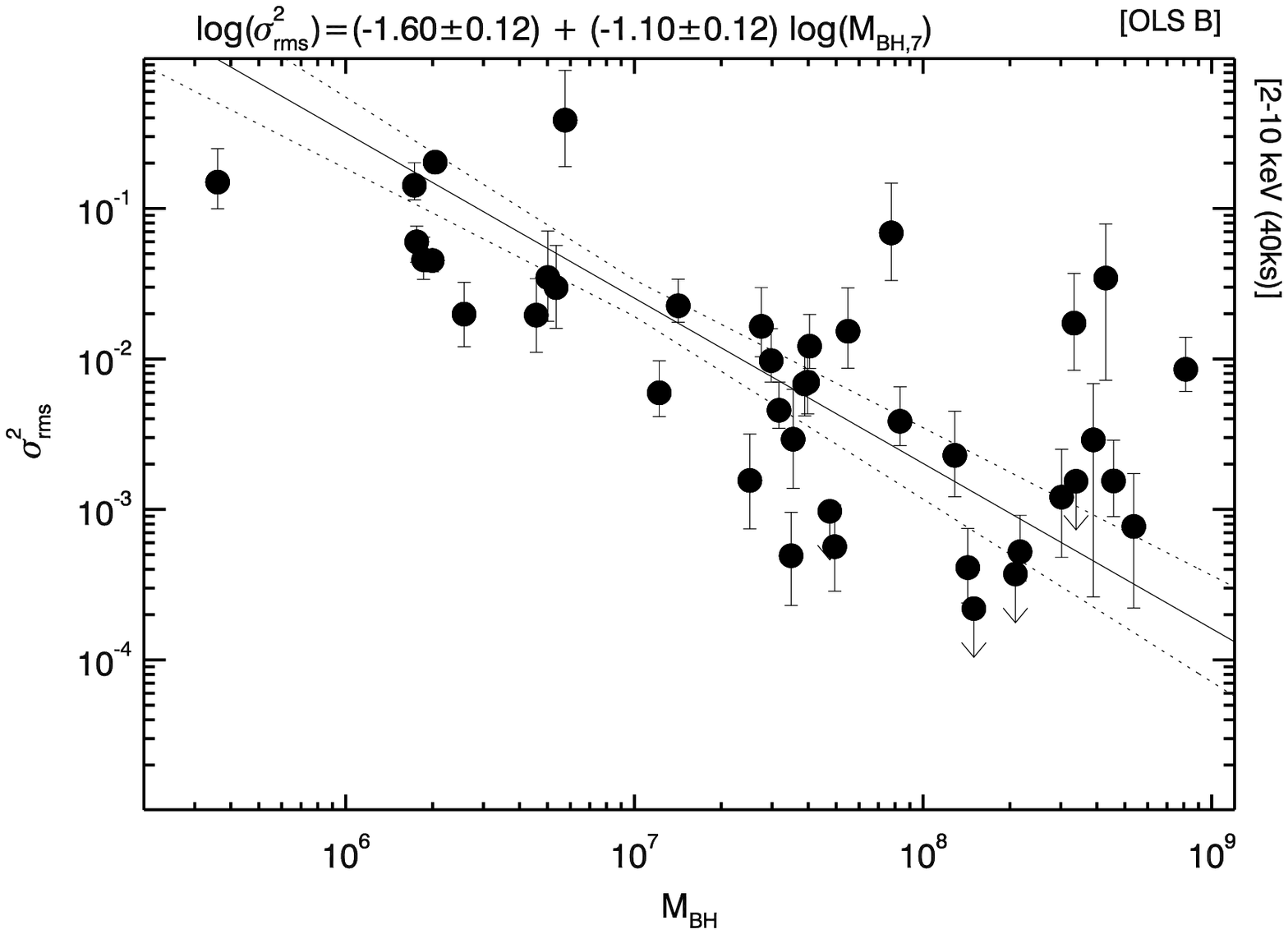, width=0.49\textwidth}
\epsfig{file=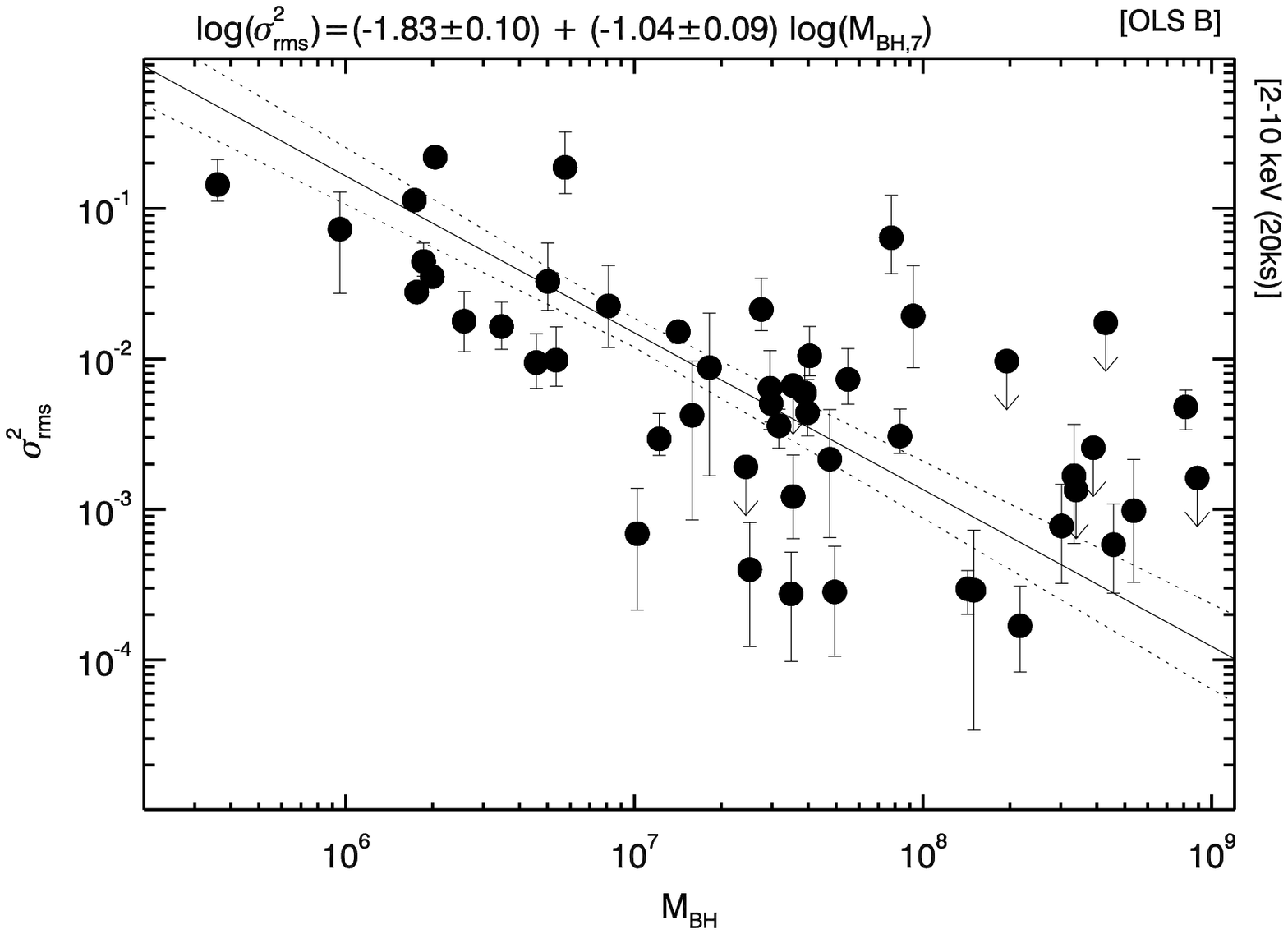,width=0.49\textwidth}
\epsfig{file=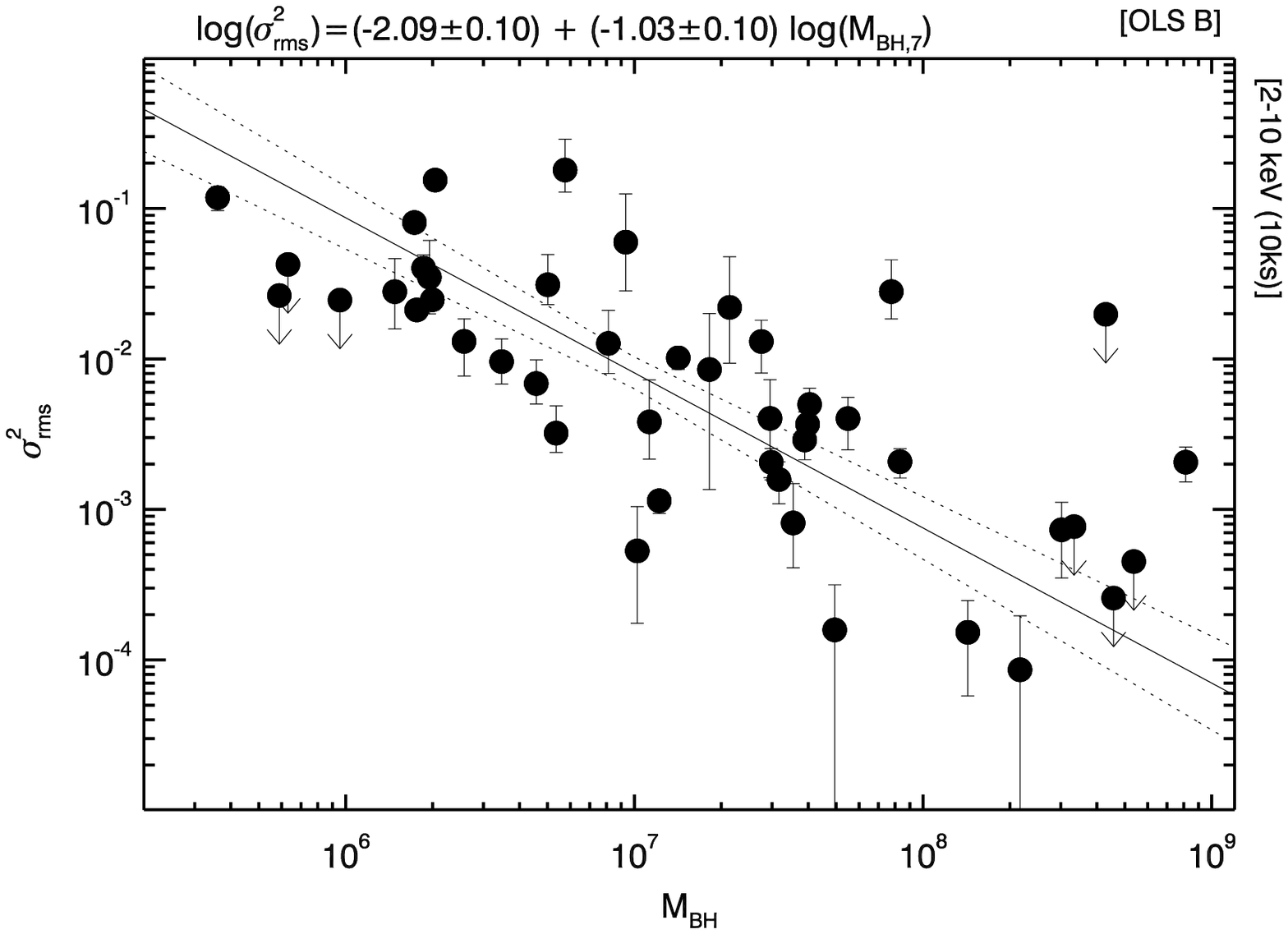,width=0.49\textwidth}
\end{center}
\caption{ {\it Upper left}, {\it upper right}, \textit{lower left} and {\it lower right} panels show the $\sigma^2_{\rm rms,80}$, $\sigma^2_{\rm rms,40}$, $\sigma^2_{\rm rms,20}$ and $\sigma^2_{\rm rms,10}$ vs. M$_{\rm BH}$ for the CAIXAvar sample. The best fit relationships (see Tab. \ref{relation}) and combined 1-$\sigma$ error are plotted as in Fig. \ref{exvsMBH210}. }
\label{exvsMBHca210} 
\end{figure*}

The four panels of Fig. \ref{exvsMBHca210} show the $\sigma^2_{\rm rms,80,40,20,10}$ vs. M$_{\rm BH}$ plots for the CAIXAvar sample, respectively.  As with the Rev sample plots, the excess variance is highly correlated with M$_{\rm BH}$ (with probability as high as 99.999 \%, see Tab. \ref{relation}). Moreover, the best fit lines over the entire CAIXAvar sample (solid lines in these panels) have slopes consistent within the errors with $-1$ (see Tab. \ref{relation}), and with the value obtained for the Rev sample. The normalisation is systematically higher in the case of the CAIXAvar sample. But even the largest difference of 0.4, in the case of the $\sigma^2_{\rm rms,40}$ vs. M$_{\rm BH}$ plots, is significant at just the $\sim 2.2\sigma$ level. 

The scatter of the data around the best-fit lines in the  CAIXAvar sample is larger than the scatter in the Rev sample. We found that $\sigma_{\rm scatter, 80, 40, 20, 10}=0.62$, 0.73, 0.72 and 0.68. These correspond to an average scatter by a factor of $\sim 5$ in linear space. 
The easiest explanation is that the M$_{\rm BH}$ estimates for the objects in the CAIXAvar sample have an uncertainty larger than the uncertainty of the reverberation M$_{\rm BH}$ estimates. On the other hand, it is possible that the sources in the CAIXAvar and Rev samples do not sample exactly the same AGN population. 

\begin{figure} 
\begin{center}
\epsfig{file=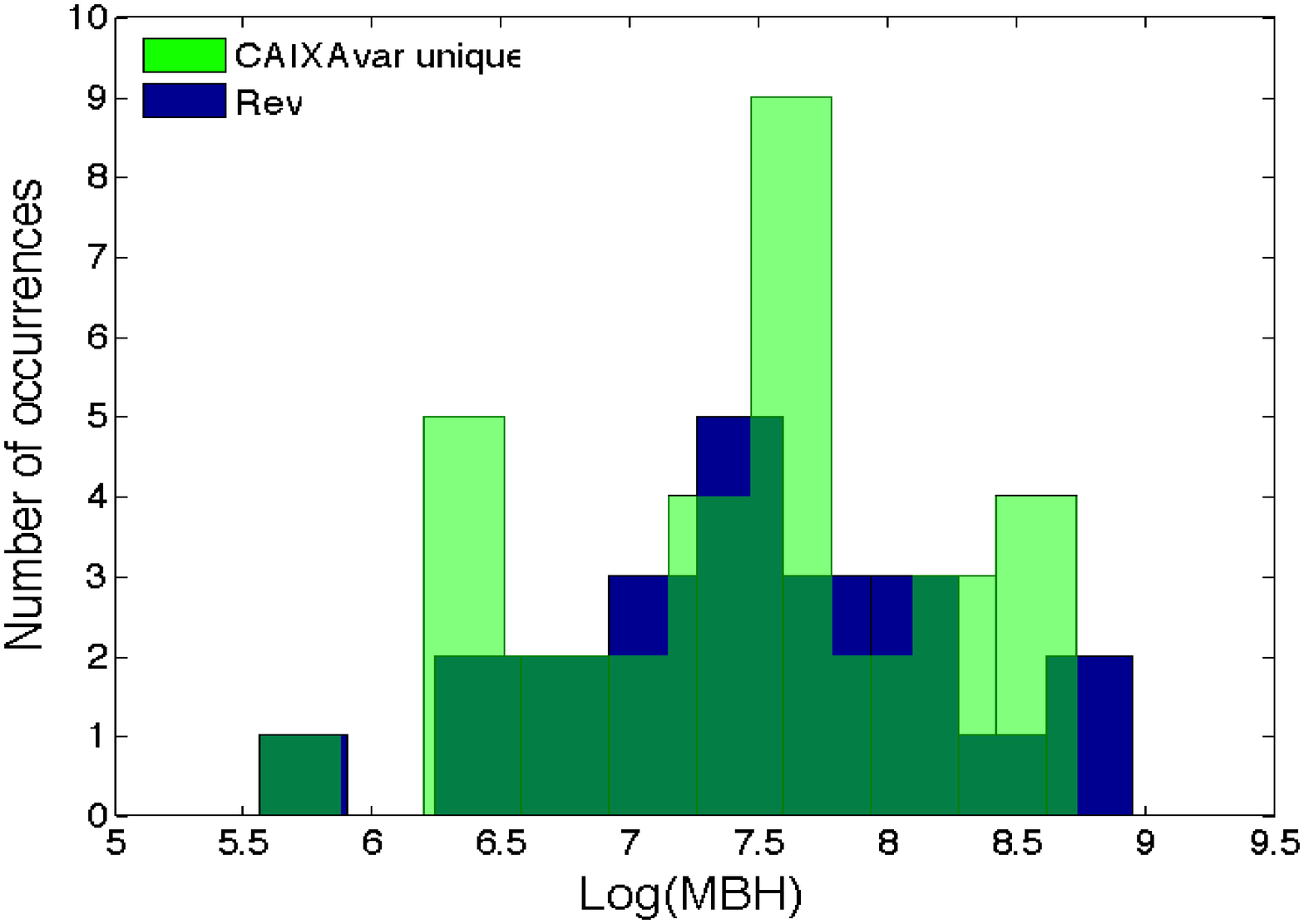,width=5.2cm,height=4.4cm,angle=-90}
\epsfig{file=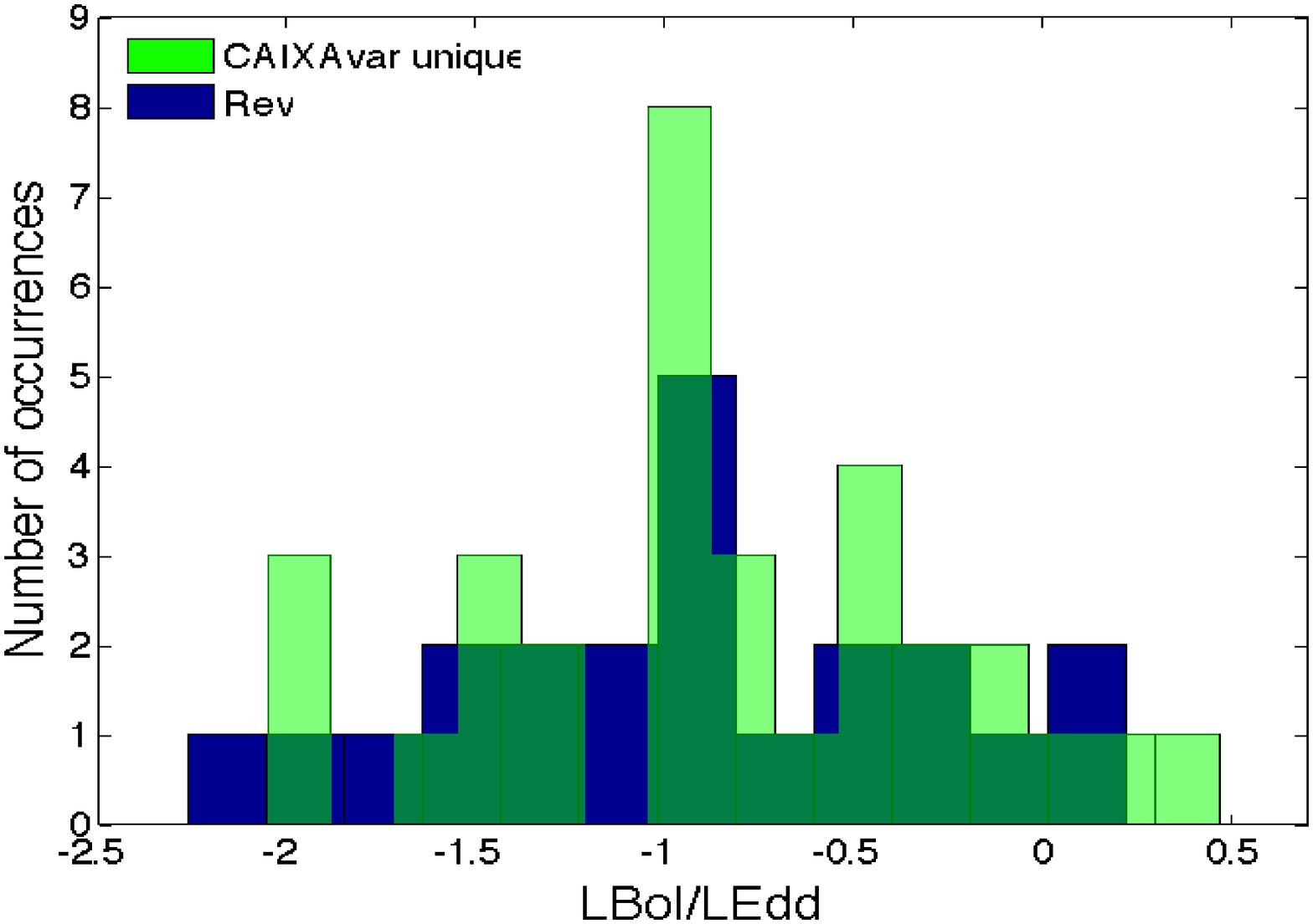,width=5.2cm,height=4.4cm,angle=-90}
\end{center}
\caption{
{\it (Left panel)} The M$_{\rm BH}$ distribution of the sources which are variable in 20 ks intervals
and belong to the Rev sample (blue histogram), and to the CAIXAvar sample only (light green histogram).  
{\it (Right panel)} The accretion rate distribution of the same sources. }
\label{histosample} 
\end{figure}

To investigate this possibility, we used the $\sigma^2_{\rm rms}$ for the 20~ks intervals. Fifty sources in the CAIXAvar sample are significantly variable on this time scale and also have M$_{\rm BH}$ estimates. Thirty two of these objects are unique to CAIXAvar. Similarly, 21 sources in the Rev sample show significant variations in at least one of their 20 ks segments. Figure \ref{histosample} shows the M$_{\rm BH}$ and accretion rate distribution (left and right panel, respectively) of the 21 Rev sample sources which are variable within 20 ks (in blue) and of the 32 unique CAIXAvar sources, variable on the same time scales (light green). Application of the KS test indicates that M$_{\rm BH}$ and accretion rate distribution of the Rev and CAIXAvar samples are identical (a KS-test indicate that the two distributions are drawn from the same population at more than 90 \% probability).

This result implies that both the CAIXAvar and Rev samples are representative of the same AGN population. Therefore, the observed higher scatter of the $\sigma^2_{\rm rms}$ measurements around the best-fit models in the case of the CAIXAvar sample is primarily due to the less accurate M$_{\rm BH}$ measurements of the non--reverberation sources.  

\subsubsection{Comparison with the results from PSD analysis of the AGN light curves}

All the best fit slopes of the correlation between $\sigma^2_{\rm rms}$ and M$_{\rm BH}$ for the Rev and CAIXAvar samples are consistent (within $\sim 2\sigma$)  with the $-1$ value (Tab. \ref{relationRev} and Tab. \ref{relation}). This is the expected slope in case of an ubiquitous PSD shape with a $-2$ slope in the frequency range over which the excess variance is computed. In fact, we can predict the expected $\sigma^2_{\rm rms} - $M$_{\rm BH}$ relation for the AGN in the two samples following Gonzalez-Martin et al. (2011). 

\begin{figure}
\begin{center}
\epsfig{file=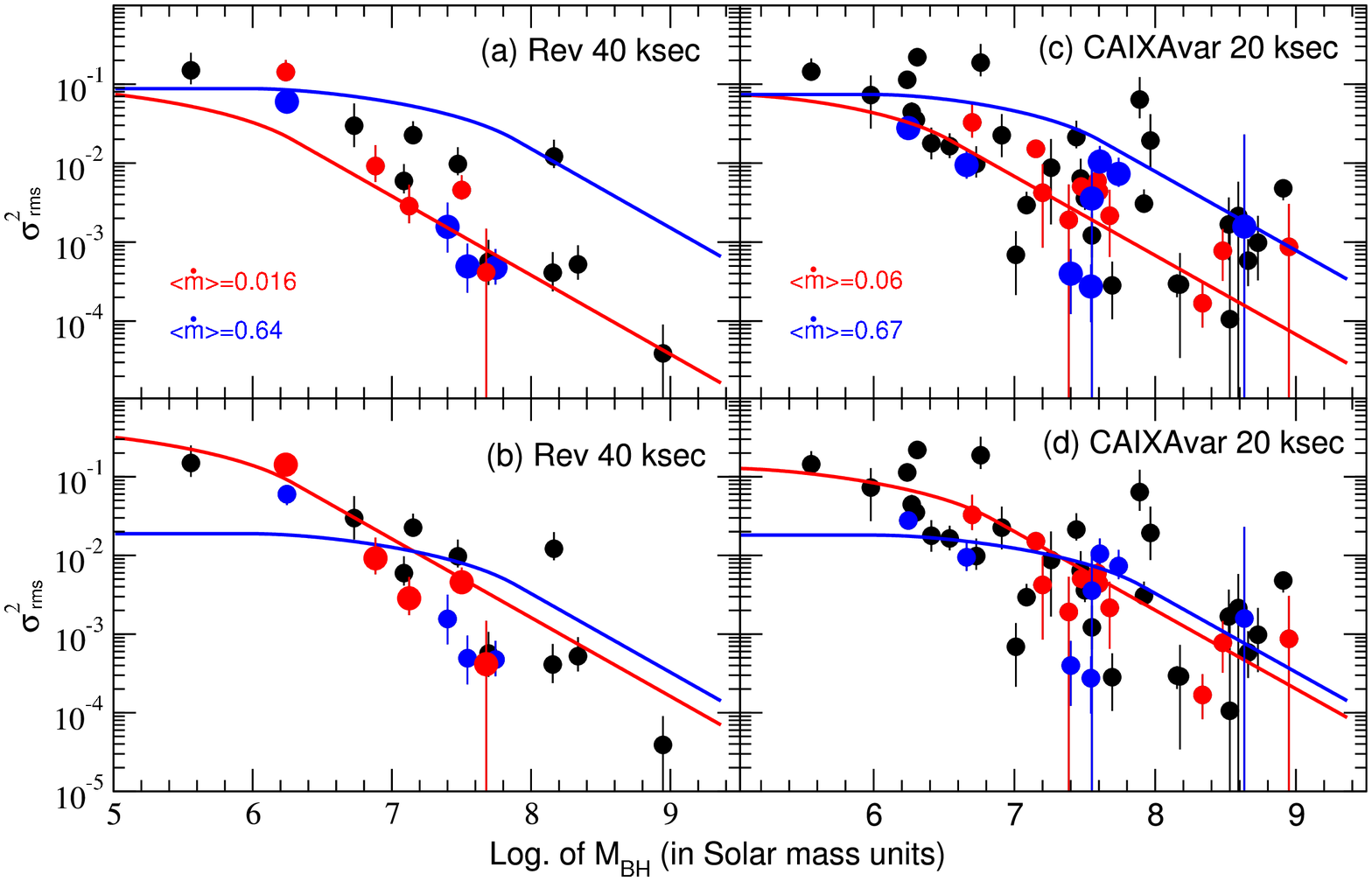,width=0.49\textwidth,bb=30 40 762 510,clip=}
\end{center}
\caption{{\it (Upper left):} $\sigma^2_{\rm rms, 40}$ vs. M$_{\rm BH}$ for the Rev sample. Red and blue filled circles indicate sources with $\dot {m}$ in the ranges (0.0098 -- 0.035) and (0.42 -- 0.86), respectively, thus with average accretion rates of $<\dot{m}> = 0.016$ and $0.64$, respectively. While the black filled circles show the $\sigma^2_{\rm rms, 40}$ for the sources with different accretion rates. The red and blue lines indicate the expected Model A relations for the two accretion rates. {\it (Upper right):} CAIXAvar (20 ks) and Model A predictions (color code as before). {\it Lower left and right:} Model B predictions.}
\label{mbhTheo} 
\end{figure}

Panels (a) and (b) in Fig. \ref{mbhTheo} show the $\sigma^2_{\rm rms, 40}$ vs. M$_{\rm BH}$ plots for the Rev sample objects. Red and blue filled circles indicate sources with accretion rate, $\dot {m}$, in the range (0.0098 -- 0.035) and (0.42 -- 0.86), respectively. The average accretion rate, $<\dot{m}>$,  of the objects in these two groups is 0.016 and 0.64, respectively. The red and blue lines in panel (a) of Fig. \ref{mbhTheo} indicate the expected $\sigma^2_{\rm rms, 40}$ vs. M$_{\rm BH}$ relation according to the ``case A" model of Gonzalez-Martin et al. (2011)\footnote{The  ``case A" model of Gonzalez-Martin et al. (2011) is based on the results from recent detailed PSD analysis of long X--ray light curves of a few AGN. It  assumes that the AGN PSD has a $-1$ slope up to a break frequency, $\nu_{\rm br}$, above which the slope steepens to $-2$. This break frequency decreases with increasing M$_{\rm BH}$ and increases proportionally with the accretion rate, as in McHardy et al. (2006), while the PSD amplitude, defined as the  product of the PSD value at the break frequency times the break frequency itself,  is the same for all objects, and equal to 0.02.}, i.e. using their equations (7) in the case when $\dot{m}=0.016$ (red line), and $\dot{m}=0.64$ (blue line). We refer to this case as ``Model A" hereafter. 
\begin{figure}
\begin{center}
\epsfig{file=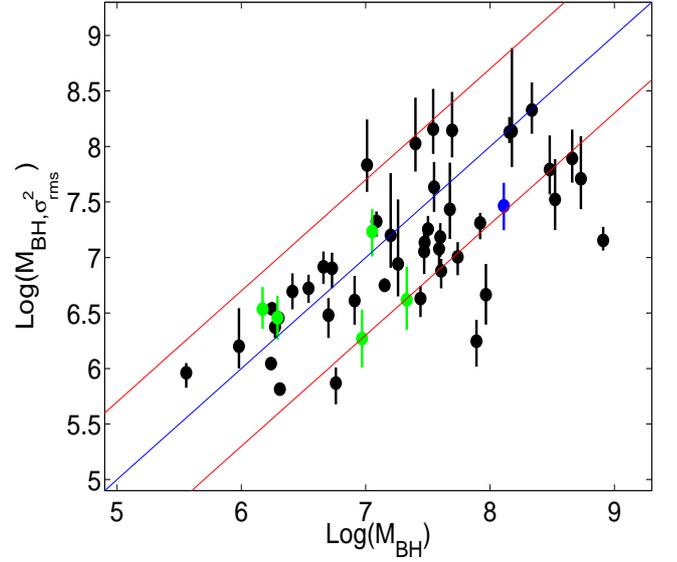,width=8.5cm,height=7.5cm}
\end{center}
\caption{M$_{\rm BH}$ measured from the $\sigma^2_{\rm rms}$ vs. M$_{\rm BH}$ best fit relation of the Rev sample, as a function of the M$_{\rm BH}$ estimated in other ways. Black, blue and green points show the results from $\sigma^2_{\rm rms,20,40,10}$, respectively. The blue line, in Fig. \ref{MBHv} corresponds to the one-to-one relation between the two M$_{\rm BH}$ estimates, and the red lines indicate a dispersion of a factor of 5, along the x-axis.}
\label{MBHv}
\end{figure}

Although the slope of the model lines is $-1$, hence they describe well the decreasing trend of $\sigma^2_{\rm rms}$ with increasing M$_{\rm BH}$, this model also predicts that, at a given M$_{\rm BH}$, $\sigma^2_{\rm rms}$ should increase considerably with $\dot m$, contrary to what is observed. The blue and red points in Fig. \ref{mbhTheo} suggest that, at a given M$_{\rm BH}$, both the high and low accretion rate objects have the same excess variance. In fact, it is because of this effect that the scatter of the Rev sample ``$\sigma^2_{\rm rms, 40}$ vs. M$_{\rm BH}$" relation is so small.

The discrepancy between the model predictions and the observed $\sigma^2_{\rm rms, 40}$ vs. M$_{\rm BH}$ relations becomes even more evident when we consider the CAIXAvar $\sigma^2_{\rm rms, 20}$ vs. M$_{\rm BH}$ relation (shown in panels (c) and (d) of Fig. \ref{mbhTheo}). In these plots we have considered all the positive CAIXAvar $\sigma^2_{\rm rms,20}$ values (irrespective of the magnitude of their error). Red and blue circles in these panels indicate the CAIXAvar objects with $\dot{m}$ in the range (0.035 -- 0.076) and (0.42 -- 0.96), respectively. The red and blue lines in panel (c) indicate the Model A curves for $\dot{m}=0.06$ and 0.67 (the mean accretion rate for the objects indicated with the red and blue circles, respectively). The model A prediction slightly underestimates the observed relation for the low accretion rate objects. A PSD amplitude larger than 0.02 (i.e. the value we have adopted in this work), could improve the agreement between the Model A predictions and the data for the low accretion rate AGN in CAIXAvar. In this case, however, the disagreement between the Model A relation and the data for the high accretion rate AGN (blue line and blue circles, respectively) would increase even more. 

\begin{figure*}
\begin{center}
\epsfig{file=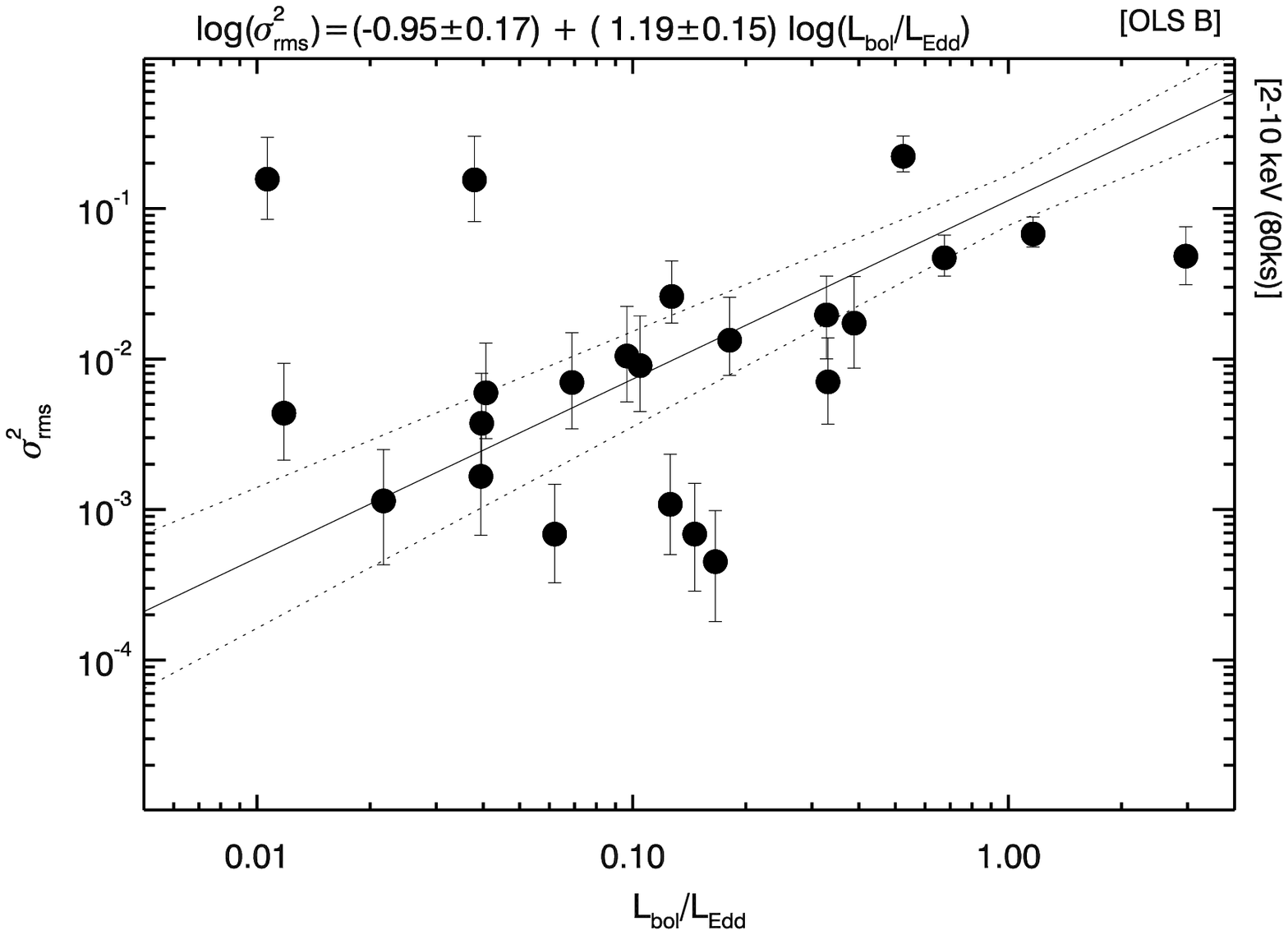,width=0.49\textwidth}
\epsfig{file=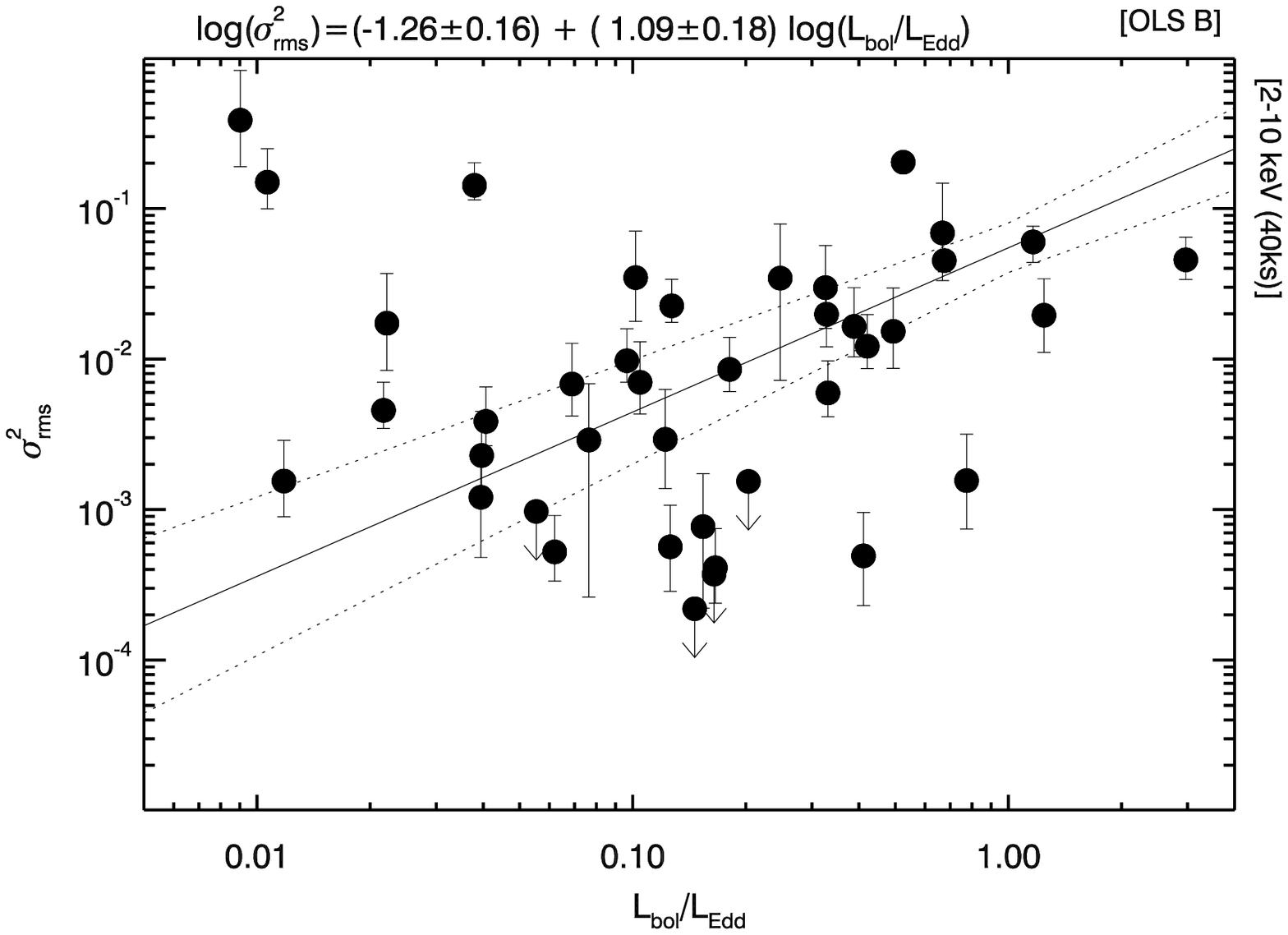,width=0.49\textwidth}
\epsfig{file=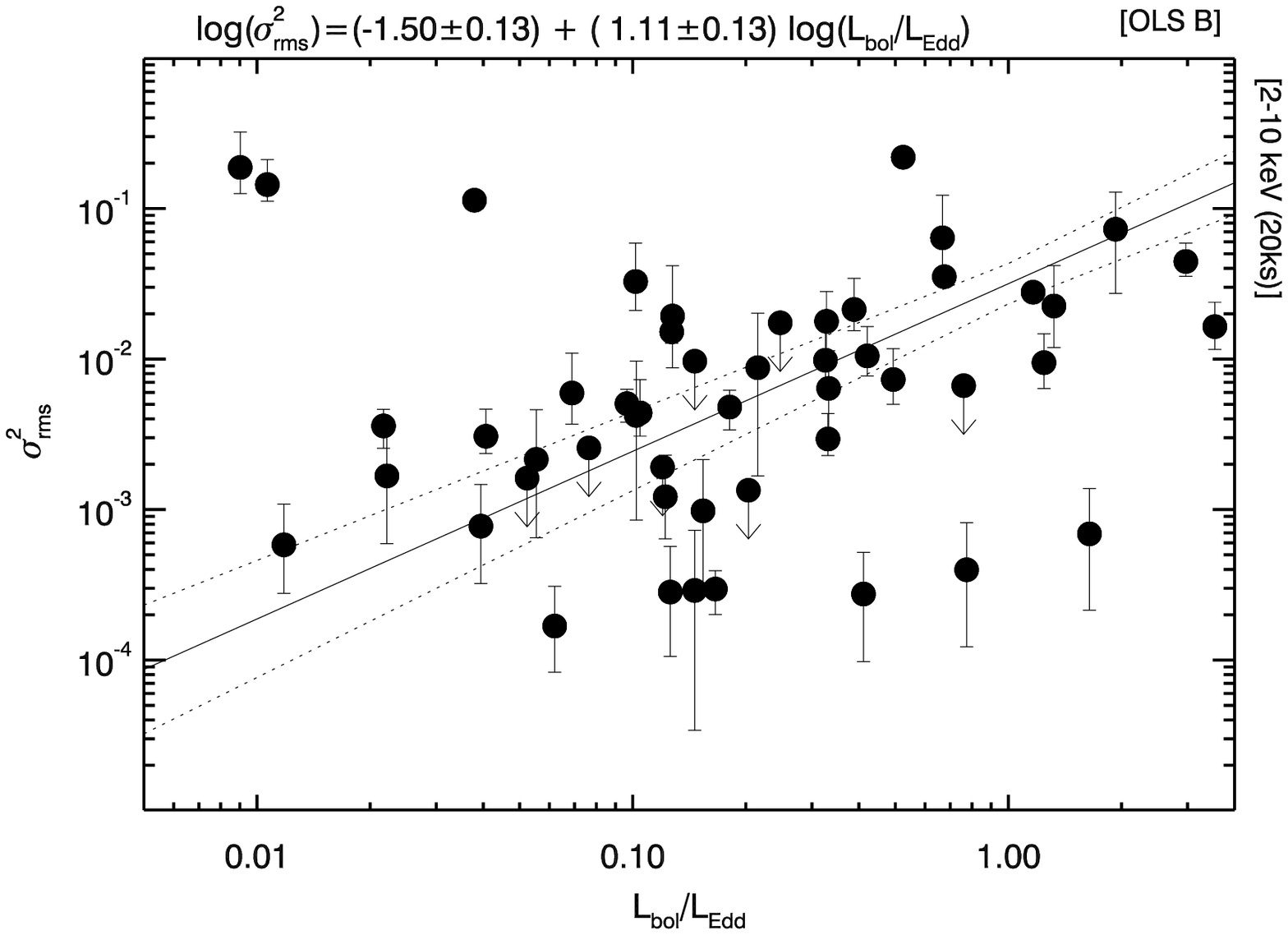,width=0.49\textwidth}
\epsfig{file=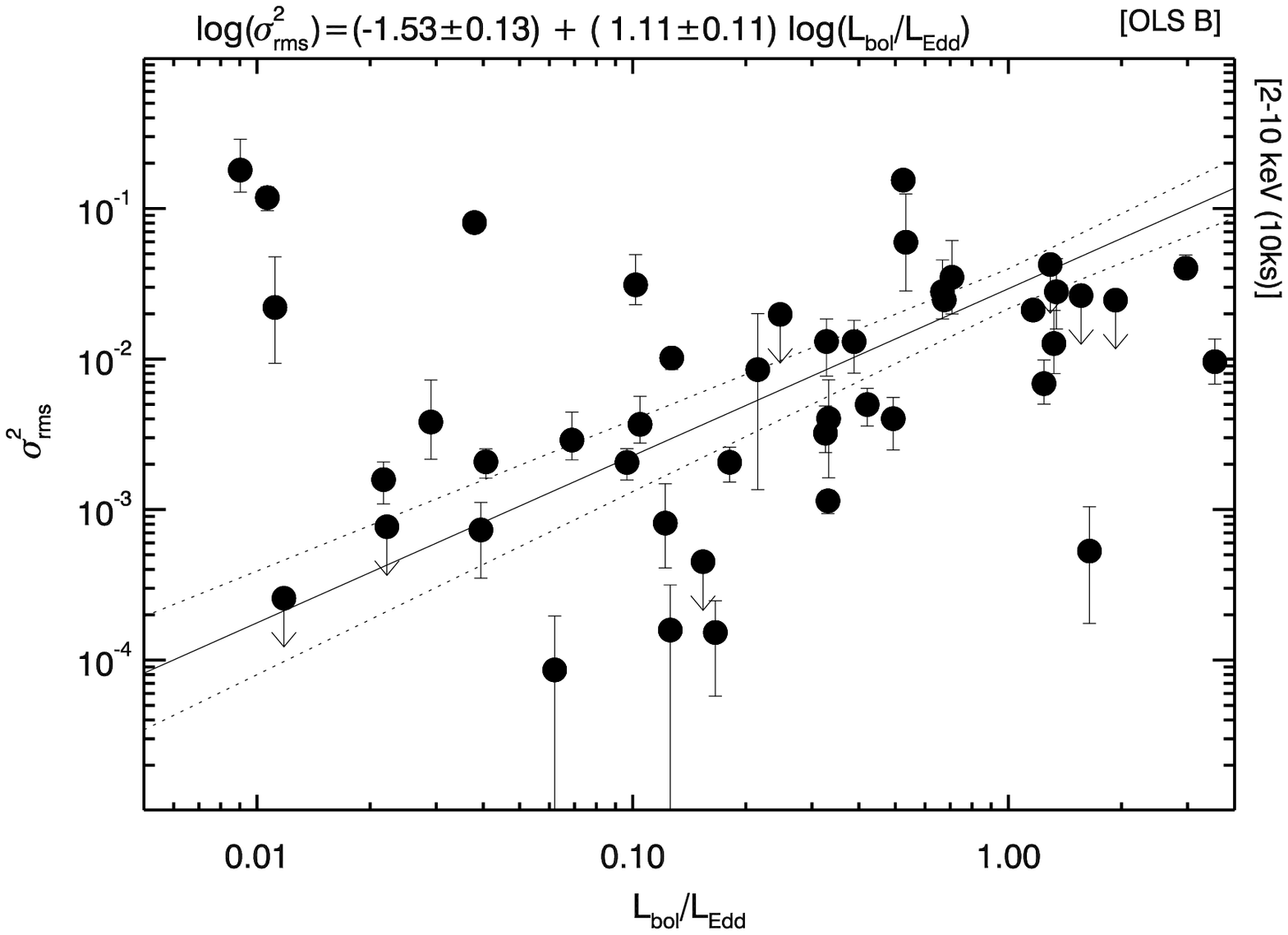,width=0.49\textwidth}
\end{center}
\caption{
{\it Upper left}, {\it upper right} and \textit{lower left} panels show the $\sigma^2_{\rm rms,80}, 
\sigma^2_{\rm rms,40}, \sigma^2_{\rm rms,20}$ and $\sigma^2_{\rm rms,10}$ vs. accretion rate 
relation for the CAIXvar sample. The best fit relationships (see Tab. \ref{relation}) and combined 1-$\sigma$ error are plotted as in Fig. \ref{exvsMBH210}.
}
\label{accretion} 
\end{figure*}
This discrepancy seems to suggest that one of the two assumptions in Model A is wrong. Thus, that the PSD either does not depend on accretion rate or its normalisation is not the same in all AGN. However, the PSD break frequency does depend on the accretion rate (McHardy et al. 2006; Koerding et al. 2007). This led us to investigate the possibility that the PSD amplitude is {\it not} the same in all AGN, but depends on $\dot m$ instead. A decreasing PSD$_{\rm amp}$ with increasing $\dot m$ could counterbalance the expected increase of the excess variance with increasing $\dot{m}$ (which is mainly due to the subsequent  increase of the break frequency). For that reason, we fitted the ($\sigma^2_{\rm rms, 40}$, M$_{\rm BH}$) data of the Rev sample, using again the ``case A" model of Gonzalez-Martin et al. (2011) but assuming that PSD$_{\rm amp}=A\dot{m}^{-\beta}$. The best-fit results are as follows: $\chi^2=46.7$ for 17 degrees of freedom, $A=0.003^{+0.002}_{-0.001}$, and $\beta=0.8\pm 0.15$ (errors indicate the 90\% confidence region for a single interesting parameter). The red and blue lines in the bottom panels of Fig. \ref{mbhTheo} indicate the predicted $\sigma^2_{\rm rms, 40}$ vs. M$_{\rm BH}$ relations for the Rev and CAIXAvar low and high accretion rate objects, using the best-fit parameter values we mentioned above (hereafter we refer to this case as ``Model B"). The agreement between the Model B predictions and both low-$\dot{m}$ and high-$\dot{m}$ Rev sample objects has been improved, although for the high accretion rate objects a significant discrepancy between the model curve and the data (see the bottom left panel of Fig. \ref{mbhTheo}) is still present (this being the reason for the large best-fit $\chi^2$ value). However, when we consider the CAIXAvar $\sigma^2_{\rm rms} -$M$_{\rm BH}$ relation, the model B curves agree very well with both the high and low accretion rate objects (panel (d) in Fig. \ref{mbhTheo}). 

Furthermore, the Model B lines can also explain the fact that the observed $\sigma^2_{\rm rms}$ vs. M$_{\rm BH}$ plots are well fitted by a single line, with the spread of the points around the best-fit line being very small (at least in the Rev sample case). At M$_{\rm BH}$ higher than $\sim 10^7$M$_{\odot}$, Model B predicts similar excess variances, irrespective of the accretion rate of the objects. However, when M$_{\rm BH}< 10^7$M$_{\odot}$, we expect a difference up to $\sim 5$ in the observed $\sigma^2_{\rm rms}$ values of low and high accretion rate objects, with higher accretion objects showing $\it smaller$ excess variance. 
\begin{table}
\begin{center}
\begin{tabular}{ | l c c |}
\hline                      
Name & Log(M$_{\rm BH}$)$_{\rm opt}$  & Log(M$_{\rm BH}$)$_{\rm var}$ \\  
\hline                      
MRK335 & $7.15\pm0.12$ & 6.75$_{-0.05}^{+0.06}$ \\
\hline                      
\end{tabular}
\small
\caption{List of AGN for which M$_{\rm BH}$ has been measured through variability (Log(M$_{\rm BH}$)$_{\rm var}$) and compared to M$_{\rm BH}$ estimated in otherwhise. }
\label{MBHvar}
\end{center}
\end{table} 

\subsubsection{Weighing AGN black holes using excess variance measurements}

In Section 6.1.1. we commented that the variability M$_{\rm BH}$ estimates, using the best-fit lines for the Rev sample, should be as accurate as the reverberation mapping ones. It seems then appropriate to use the best-fit relations for the Rev sample (with the parameter values as listed in Table 1) and the excess variance measurements of the CAIXAvar sources, to estimate their M$_{\rm BH}$. These estimates should have smaller uncertainty than the uncertainty of the present estimates which are mainly based on single epoch spectra. Table \ref{MBHvar} lists the variability M$_{\rm BH}$ estimates for 55, out of the 161 AGN, in CAIXAvar (for 6 of these AGN no other M$_{\rm BH}$ estimate was found in literature). For the remaining objects, we are able to provide upper limits on their M$_{\rm BH}$, as follows. 

We first measured M$_{\rm BH}$ from $\sigma^2_{\rm rms,20}$ for all the sources with at least one variable 20 ks interval (black points in Fig. \ref{MBHv}). There are 54 such AGN in CAIXAvar (47 measurements and 7 upper limits). For each one of these objects, we estimate M$_{\rm BH}$ using their $\sigma^2_{\rm rms,20}$ and the best-fit $\sigma^2_{\rm rms,20}$ vs. M$_{\rm BH}$ relation for the Rev sample. The uncertainties on M$_{\rm BH}$ are estimated propagating the $\sigma^2_{\rm rms,20}$ uncertainty, only. Whenever the resulting uncertainty is smaller than 0.4 dex (estimated to be the uncertainty of the Rev best-fit relation itself), the latter must be used. We then followed the same procedure and used the $\sigma^2_{\rm rms,40}$ and $\sigma^2_{\rm rms,10}$, to estimate the M$_{\rm BH}$ for further 2 objects (one of which is an upper limit) and 9 more AGN (two of which are upper limits), respectively (blue and green points in Fig. \ref{MBHv}). 

In this way, we have been able to measure the M$_{\rm BH}$ for a total of 55 AGN (out of 161 AGN in CAIXAvar) and provide stringent (90\%) upper limits for 10 more objects. Out of these 65 AGN, 44 have neither reverberation mapping nor stellar velocity dispersion M$_{\rm BH}$ estimates. The remaining AGN, for which we detect no variability in any of the intervals analysed, are mainly weak sources with short exposures. However, for some of these sources the data are of good quality, thus indicating that the sources have a small variability amplitude on the time scales considered here. Nevertheless, even for these objects, we used their 90\% upper limit on $\sigma^2_{\rm rms}$ (which is positive in at least one interval), and the method outlined above, to provide a {\it lower} limit on their M$_{\rm BH}$. These limits are also listed in Table \ref{MBHvar}. 

Figure \ref{MBHv} shows a plot of the variability M$_{\rm BH}$ measurements as a function of the M$_{\rm BH}$ estimated in other ways. Black, blue and green points show the results from $\sigma^2_{\rm rms,20,40,10}$, respectively. The blue line, in Fig. \ref{MBHv} corresponds to the one-to-one relation between the two M$_{\rm BH}$ estimates, and the red lines indicate a dispersion of a factor of 5, along the x-axis. The vast majority of the objects lie within the red lines. This result suggests that, since the uncertainty associated with the current M$_{\rm BH}$ estimates for most of the CAIXAvar AGN  are of the order of $\sim 5$ (see discussion in Section 6.1.2 above), the variability estimates we list in this work should be considerably less uncertain, as otherwise, we would expect more objects to lie outside the region defined by the red lines in Fig. \ref{MBHv}. 
\begin{figure} [t]
\begin{center}
\epsfig{file=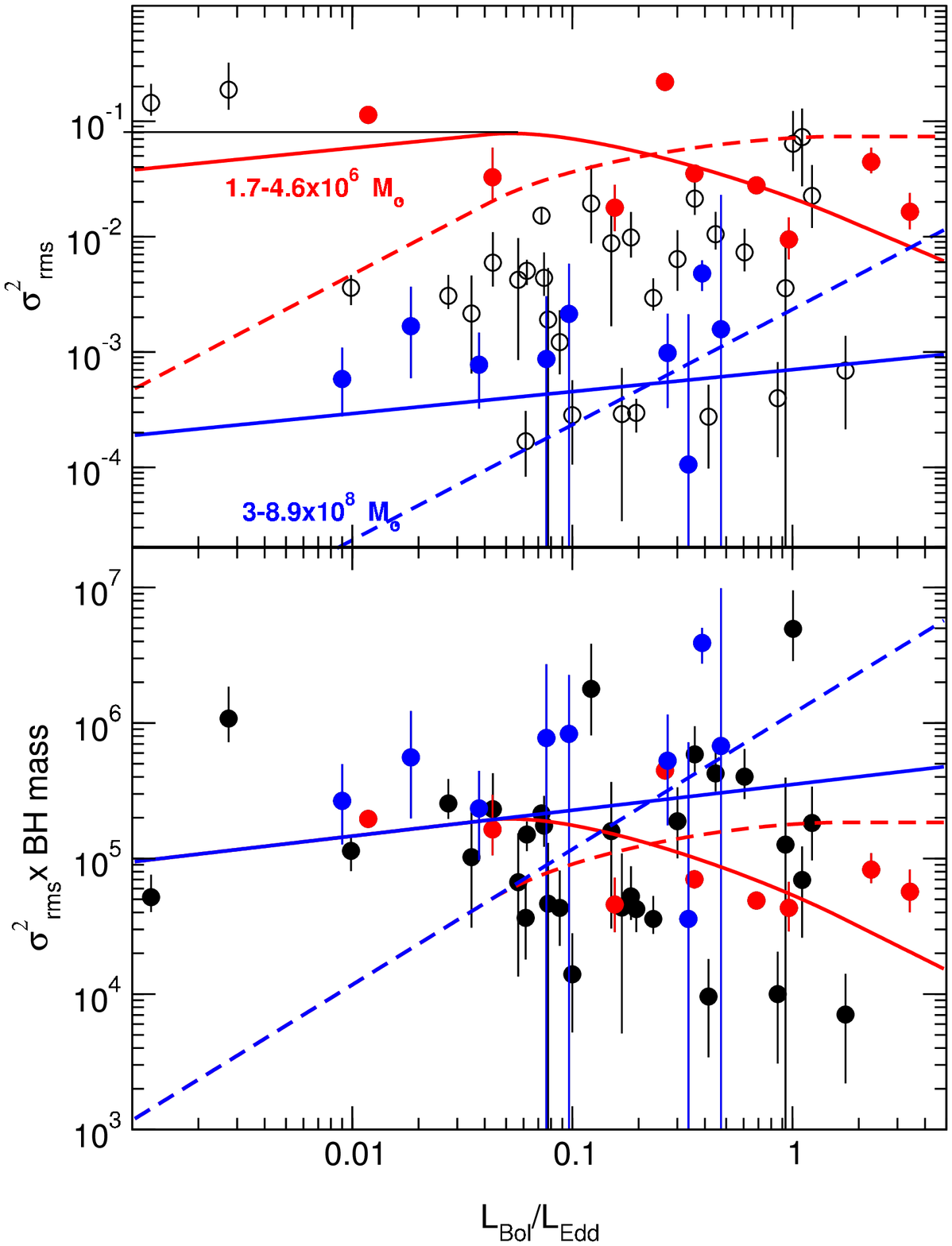,width=0.49\textwidth,bb=10 120 535 770, clip=}
\end{center}
\caption{The $\sigma^2_{\rm rms,20}\times$M$_{\rm BH}$ vs. $\dot{m}$ relation for the CAIXAvar (20 ks) sample. Red and blue circles indicate sources with M$_{\rm BH}$ in the range 1.7--4.6$\times 10^6$M$_{\odot}$ and 3--8.9$\times10^8$M$_{\odot}$. Red and blue lines indicate the expected relations for mean M$_{\rm BH}$ of $2.5\times10^6$ and $5\times10^8$M$_{\odot}$, respectively. Dashed and solid lines show the Model A and Model B predictions.}
\label{accr_all} 
\end{figure}

\subsection{$\sigma^2_{\rm rms}$ vs. accretion rate}
\label{acc}

Figure \ref{accretion} shows the variability $\sigma^2_{\rm rms,80}, \sigma^2_{\rm rms,40}, \sigma^2_{\rm rms,20}$ and $\sigma^2_{\rm rms,10}$ vs. accretion rate relation for the CAIXAvar sample. The best-fit parameters for each plot in Fig.~\ref{accretion} are listed in Table \ref{relation}, together with the probability of significant correlations.

The plots in Fig.~\ref{accretion} suggest the presence of a trend where higher accretion rate sources appear to be more variable. But there is also a significant  scatter around the best-fit relation (shown with a solid line in the same plots). Not surprisingly, given the large scatter of the points around the best-fit lines, the probability of a significant correlation between excess variance and accretion rate is small (even smaller than 90\% in almost all cases). Therefore, strictly speaking, our results suggest that there is no significant correlation between the variability amplitude and the accretion rate in CAIXAvar. 

The scatter of the points around the best-fit lines in Figure \ref{accretion} may be caused by the strong variability vs. M$_{\rm BH}$ correlation in the CAIXAvar sample. If less massive BH are intrinsically more variable than higher mass BH, they should  populate the higher part of the variability vs. accretion rate plot at any given accretion rate. Thus, even if there exists a variability vs. accretion rate correlation, we should expect  a significant scatter in the plots shown in Fig. \ref{accretion}, simply because of the large M$_{\rm BH}$ range of the objects in our sample. 

In order to eliminate the M$_{\rm BH}$ dependence of $\sigma^2_{\rm rms}$, we considered the product $\sigma^2_{\rm rms,20}\times$M$_{\rm BH}$. Given the fact that the best-fit slope of the $\sigma^2_{\rm rms,20}$ vs. M$_{\rm BH}$ relation is consistent with $-1$ (see  Table \ref{relation}), this product should result in a quantity independent of M$_{\rm BH}$. The bottom panel in Fig. \ref{accr_all} shows a plot of the $\sigma^2_{\rm rms,20}\times$M$_{\rm BH}$ vs. accretion rate relation for all AGN in CAIXAvar (irrespective of the amplitude of their excess variance uncertainty). Clearly, a large scatter in this plot is introduced by the uncertainties on the M$_{\rm BH}$, and of the bolometric luminosity estimates for each object (as before, L$_{\rm Bol}$ is  estimated using the prescription of Marconi et al. 2004), which should increase the uncertainty of the L$_{\rm Bol}/$L$_{\rm Edd}$ values. In any case though, and in agreement  with previous studies (O'Neill et al. 2005; Gierlinski et al. 2008; Nikolajuk et al. 2009; Zhou et al. 2010), no correlation between $\sigma^2_{\rm rms}\times$M$_{\rm BH}$ and accretion rate appears. We therefore conclude that $\sigma^2_{\rm rms}$ does not appear to correlate with accretion rate.

To investigate this issue further, the red and blue circles in both panels of Fig. \ref{accr_all} indicate the $\sigma^2_{\rm rms}$ vs. $\dot{m}$ (top panel) and $\sigma^2_{\rm rms}\times$M$_{\rm BH}$ vs. $\dot{m}$ data (bottom panel) for AGN in the CAIXAvar sample with M$_{\rm BH}$ in the range 1.7--4.6$\times 10^6$~M$_{\odot}$ and 3--8.9$\times10^8$~M$_{\odot}$, respectively. The solid (dashed) red and blue lines indicate the Model B (Model A) curves for an AGN with a M$_{\rm BH}$ of $2.5\times10^6$ and $5\times10^8$M$_{\odot}$ (these values are equal to the  mean M$_{\rm BH}$ of the points indicated with the red and blue circles in Fig. \ref{mbhTheo}). 

\begin{figure*} 
\begin{center}
\epsfig{file=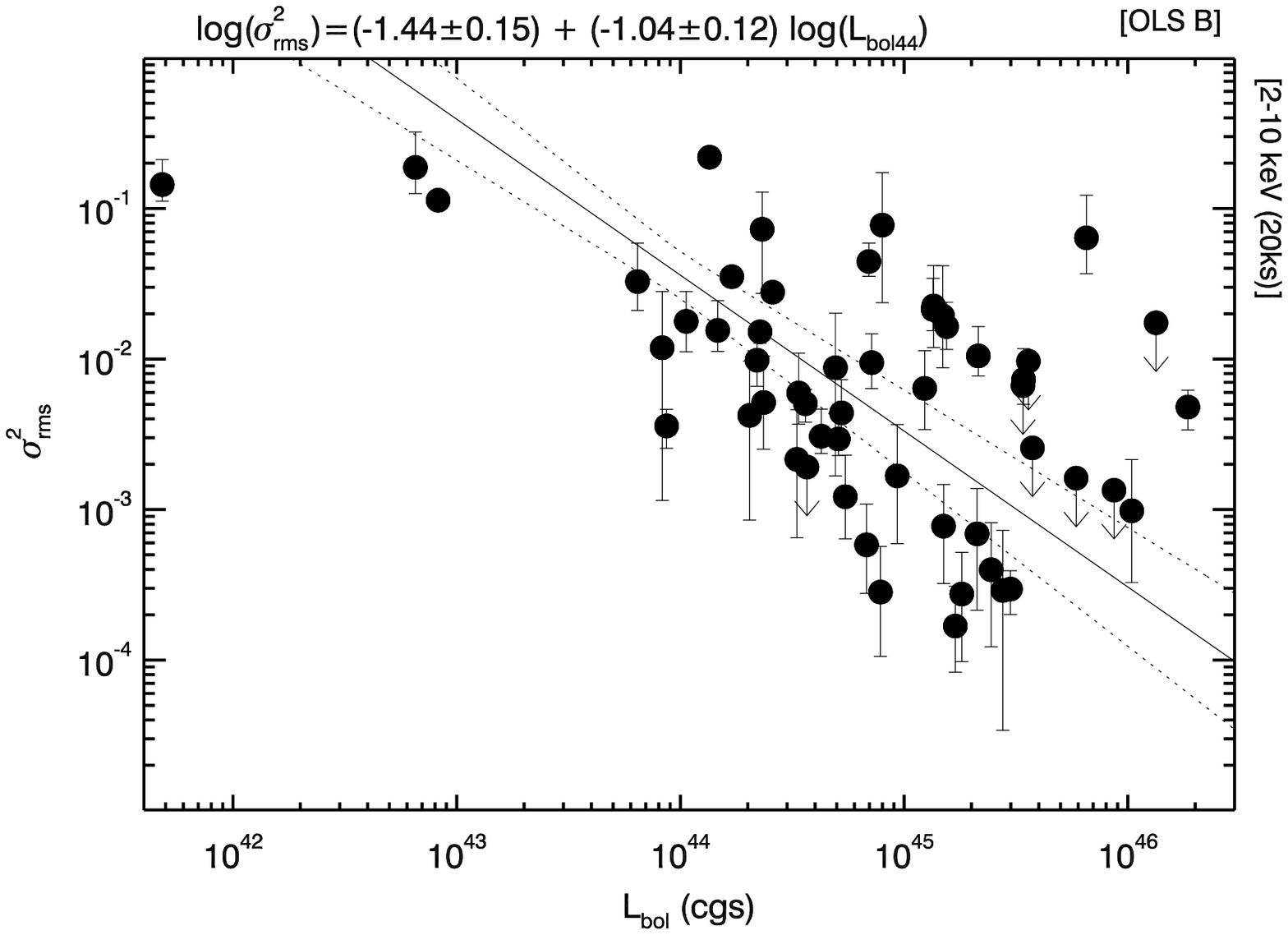,width=0.49\textwidth}
\epsfig{file=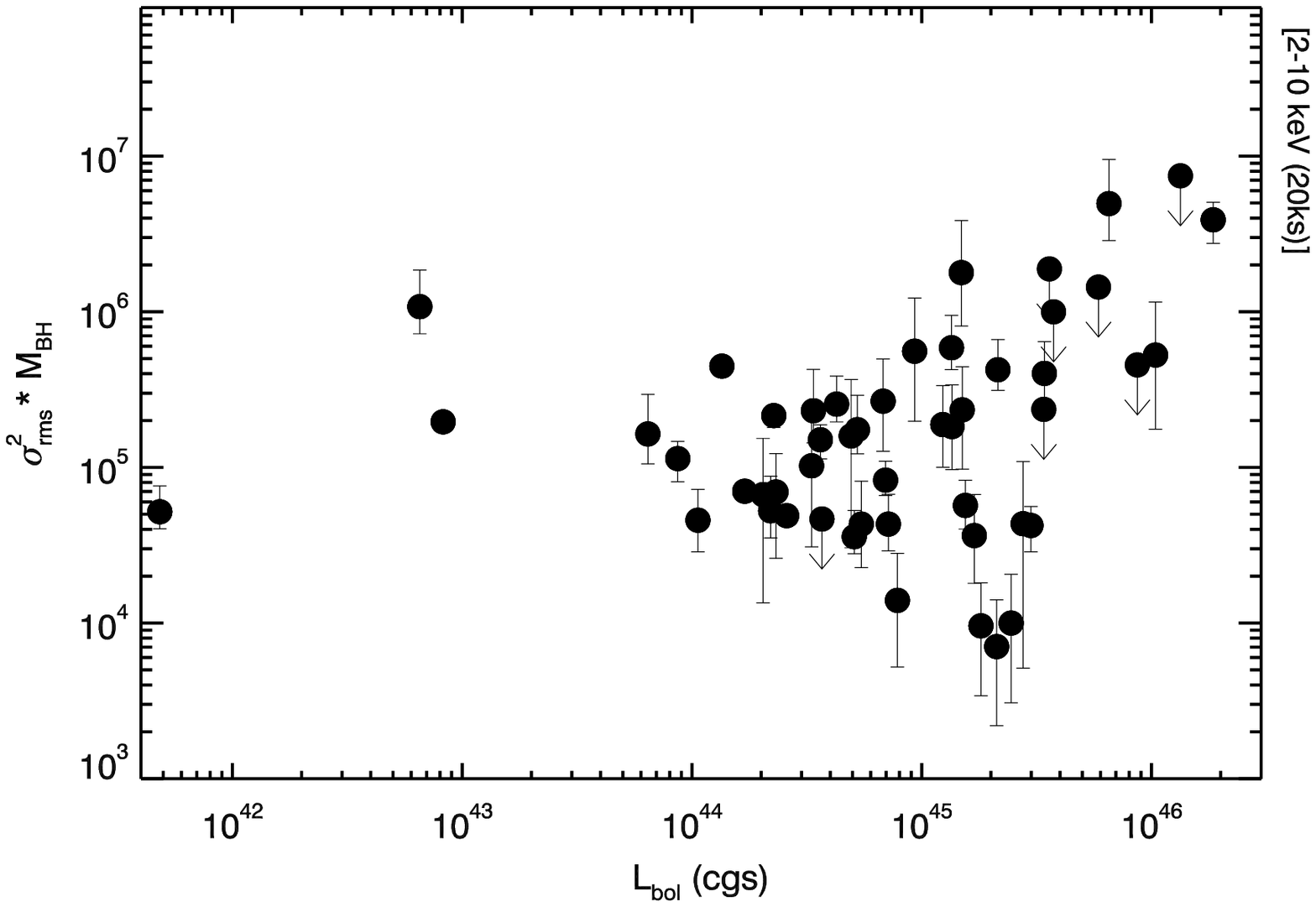,width=0.49\textwidth}
\end{center}
\caption{The $\sigma^2_{\rm rms,20}$ vs. bolometric luminosity, and the  $\sigma^2_{\rm rms,20}\times{\rm M_{\rm BH}}$ vs. bolometric luminosity plots for the CAIXAvar sample (left and right panels, respectively). 
L$_{\rm bol,44}$ indicates the bolometric luminosity in units of 10$^{44}$ erg s$^{-1}$. }
\label{LBOL} 
\end{figure*}

The top panel in Fig. \ref{accr_all} shows that, because of the $\sigma^2_{\rm rms}$ vs. M$_{\rm BH}$ dependence, neither Model A nor Model B predict a universal $\sigma^2_{\rm rms}-\dot m$ relation for AGN. We do expect a significant scatter in the $\sigma^2_{\rm rms}$ vs. $\dot{m}$ plot, because, at any given accretion rate, the excess variance should be different for AGN with different M$_{\rm BH}$. In fact, it should increase strongly with decreasing M$_{\rm BH}$, due to the ``PSD break time scale -- M$_{\rm BH}$" relation of McHardy et al (2006). On the other hand, Model A predicts a $\sigma^2_{\rm rms}\propto \dot{m}$ relation, while Model B predicts a much shallower relation of the form $\sigma^2_{\rm rms}\propto \dot{m}^{0.2}$, due to the decrease of the PSD amplitude with increasing accretion rate. The McHardy et al (2006) relations predict that the PSD break-time scale becomes smaller than 20 ks (thus entering in the frequency window in which the $\sigma^2_{\rm rms}$ is computed) for an AGN with M$_{\rm BH}=2.5\times10^6$ M$_{\odot}$ (red curves in the top panel of Fig. \ref{accr_all}) when $\dot m \sim 0.05$. As a result, the Model A, with constant PSD$_{\rm amp}$, predicts that the excess variance of the small M$_{\rm BH}$ should increase linearly with $\dot{m}$ until $\dot{m}=0.05$, then bend until the PSD break time scale becomes even smaller than 250 s. At that point the excess variance should flatten, becoming constant. While the more massive AGN should always have the PSD break time scale longer than 20~ks, thus no flattening is expected. When we consider the predictions of Model B, on top of this trend with accretion rate, we also have to consider the anti-correlation between PSD$_{\rm amp}$ and $\dot{m}$. In Model B, the PSD normalisation vs. accretion rate relation will reduce the expected trends predicted by the Model A relation by 0.8 (PSD$_{\rm amp}\propto\dot{m}^{-0.8}$), producing, i.e. for the small M$_{\rm BH}$, a flatter relation at low $\dot{m}$ and a decreasing variability at high $\dot{m}$.

The main effect of multiplying the excess variance by the M$_{\rm BH}$ is to eliminate of the M$_{\rm BH}$ dependence in the $\sigma^2_{\rm rms}$vs. $\dot m$ relation. In fact, as expected, the predicted values of $\sigma^2_{\rm rms}\times$M$_{\rm BH}$, for different M$_{\rm BH}$ do overlap both in Model A and B scenarios, at least when the PSD break time scale is not entering in the frequency window in which $\sigma^2_{\rm rms}$ is computed (i.e. lower than 20 ks).

The Model A (dashed curves) predictions in the bottom panel of Fig. \ref{accr_all} clearly fail to reproduce the behaviour of both the low and high $\dot m$ AGN. The low $\dot{m}$ sources show $\sigma^2_{\rm rms}\times$M$_{\rm BH}$ values which are significantly larger than the Model A curve. At high $\dot m$, although the model correctly predicts a flattening of the relation for the small M$_{\rm BH}$, it systematically overestimates the $\sigma^2_{\rm rms}\times$M$_{\rm BH}$ values. As before, a PSD$_{\rm amp}$ lower than 0.02 could result in model curves which are consistent with high $\dot{m}$ small M$_{\rm BH}$ AGN, however, in this case, the disagreement between Model A curve and the low $\dot{m}$ AGN (red dashed line) will get even worse. On the other hand, the data for all the low $\dot m$ objects are now consistent (within 2$\sigma$) with the Model B curve (solid line), i.e. the solid blue line shown in the bottom panel of Fig. \ref{accr_all}. The solid red line (in the same panel) is also in agreement with the data for the low M$_{\rm BH}$ objects. 

Therefore, the apparent lack of correlation between $\sigma^2_{\rm rms}$ (when ``normalized" to M$_{\rm BH}$) and $\dot m$ strongly supports the hypothesis  that the PSD amplitude in AGN decreases with increasing accretion rate. We conclude that the $\sigma^2_{\rm rms} $vs. $\dot{m}$ relation for CAIXAvar is consistent with the McHardy et al. (2006) $\nu_{\rm br}\propto \dot{m}$ correlation, but {\it only if} the PSD normalisation decreases with increasing accretion rate.

\subsection{$\sigma^2_{\rm rms}$ vs. luminosity}

The left panel of Fig. \ref{LBOL} shows the plot of  $\sigma^2_{\rm rms,20}$ vs. L$_{\rm bol}$ for the CAIXAvar sample. Table \ref{relation} lists the best linear fit (in the log-log space) in the case of the $\sigma^2_{\rm rms,80,40,20,10}$ vs. L$_{\rm bol}$ relations, as well as the probability of a significant correlation between these two quantities. Variability and luminosity are well correlated with correlation probabilities higher than 99.9 \%.The best-fit slopes are also consistent with $-1$, in all cases. 

The right panel in Fig. \ref{LBOL} shows the  $\sigma^2_{\rm rms,20}\times$M$_{\rm BH}$ vs. L$_{\rm Bol}$ relation. Contrary to the data plotted in the left panel of the same figure, no significant correlation is observed any more between $\sigma^2_{\rm rms}\times$M$_{\rm BH}$ and luminosity (see also Tab.~\ref{relation}). This result suggests that the apparent correlation between $\sigma^2_{\rm rms}$ and L$_{\rm Bol}$ is driven mainly by the $\sigma^2_{\rm rms}$ vs. M$_{\rm BH}$ correlation. This result is also consistent with the suggestion that the variability vs. luminosity relation in AGN is a byproduct of the variability vs. M$_{\rm BH}$ relation in the same objects (see e.g. Papadakis 2004). 

\begin{figure}
\begin{center}
\epsfig{file=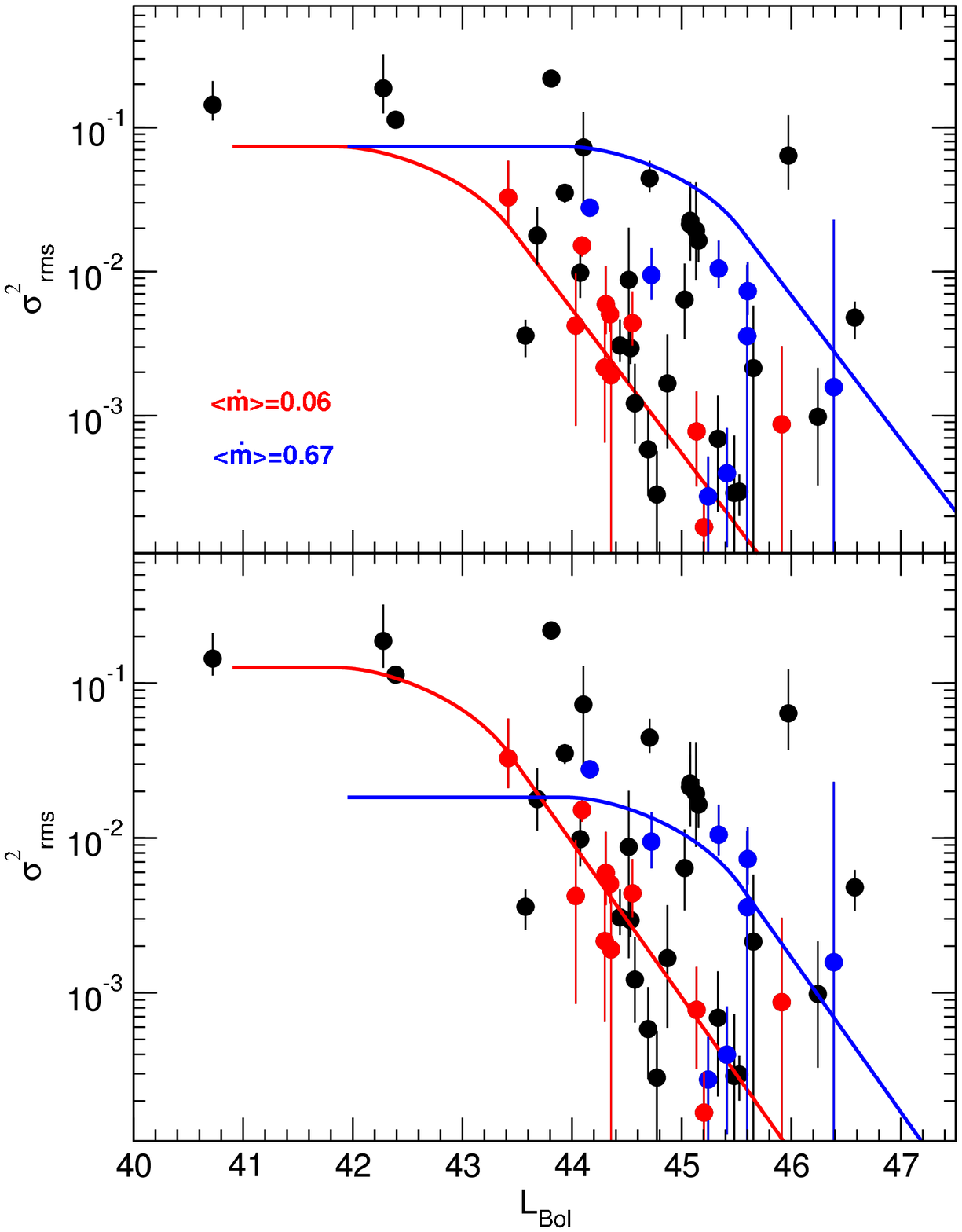,width=0.49\textwidth,bb=10 120 535 770, clip=}
\end{center}
\caption{{\it Top panel:} The $\sigma^2_{\rm rms,20}$ vs. $L_{Bol}$ relation for the CAIXAvar (20 ks) sample. Red and blue circles indicate sources with mean $\dot{m}=0.06$ and $0.67$, respectively. The solid lines show the Model A predictions for different accretion rates. {\it Bottom panel:} Model B predictions.}
\label{LBOLTheo} 
\end{figure}

To investigate this issue further, we computed the expected Model A and Model B $\sigma^2_{\rm rms}$ vs. L$_{\rm Bol}$ relation for various accretion rates. The red and blue lines in the top and bottom panels of Fig. \ref{LBOLTheo} indicate the Model A (top panel) and Model B (bottom panel) relation for AGN with $\dot m=0.06$ and $\dot m=0.67$, respectively. Black circles in both panels indicate all the $\sigma^2_{\rm rms}$ vs. L$_{\rm Bol}$ data for the CAIXAvar sample, and red/blue circles indicate the data for sources with an accretion rate between (0.0098 -- 0.035) and (0.42 -- 0.86), respectively (as we did in Section 6.3.1). In effect, the model lines in this figure are identical to the model lines plotted in panels (c) and (d) of Fig. \ref{mbhTheo}, except that, for each M$_{\rm BH}$, we have estimated L$_{\rm Edd}$, and hence L$_{\rm Bol}$, through the relation: L$_{\rm Bol}=\dot{m}$L$_{\rm Edd}$ (note that, due to the multiplication of L$_{\rm Edd}$ with $\dot m$, even the Model B curves are well separated in this case).

Although the agreement between Model A curve prediction with the low $\dot m$ objects is rather good, this is not the case with the model prediction for the high accretion rate objects (all points lie below the blue line in the top panel of Fig. \ref{LBOLTheo}). On the other hand, the Model B curves agree very well with both the low and the high $\dot m$ data points. We therefore conclude that the apparent variability vs. luminosity relation in the CAIXAvar sample is a byproduct of the variability vs. M$_{\rm BH}$ relation, and it can be explained well if the PSD$_{\rm amp}$ decreases with accretion rate as we discussed in Section 6.3.1. 

\subsection{$\sigma^2_{\rm rms}$ vs. H$\beta$}

\begin{figure*}
\begin{center}
\epsfig{file=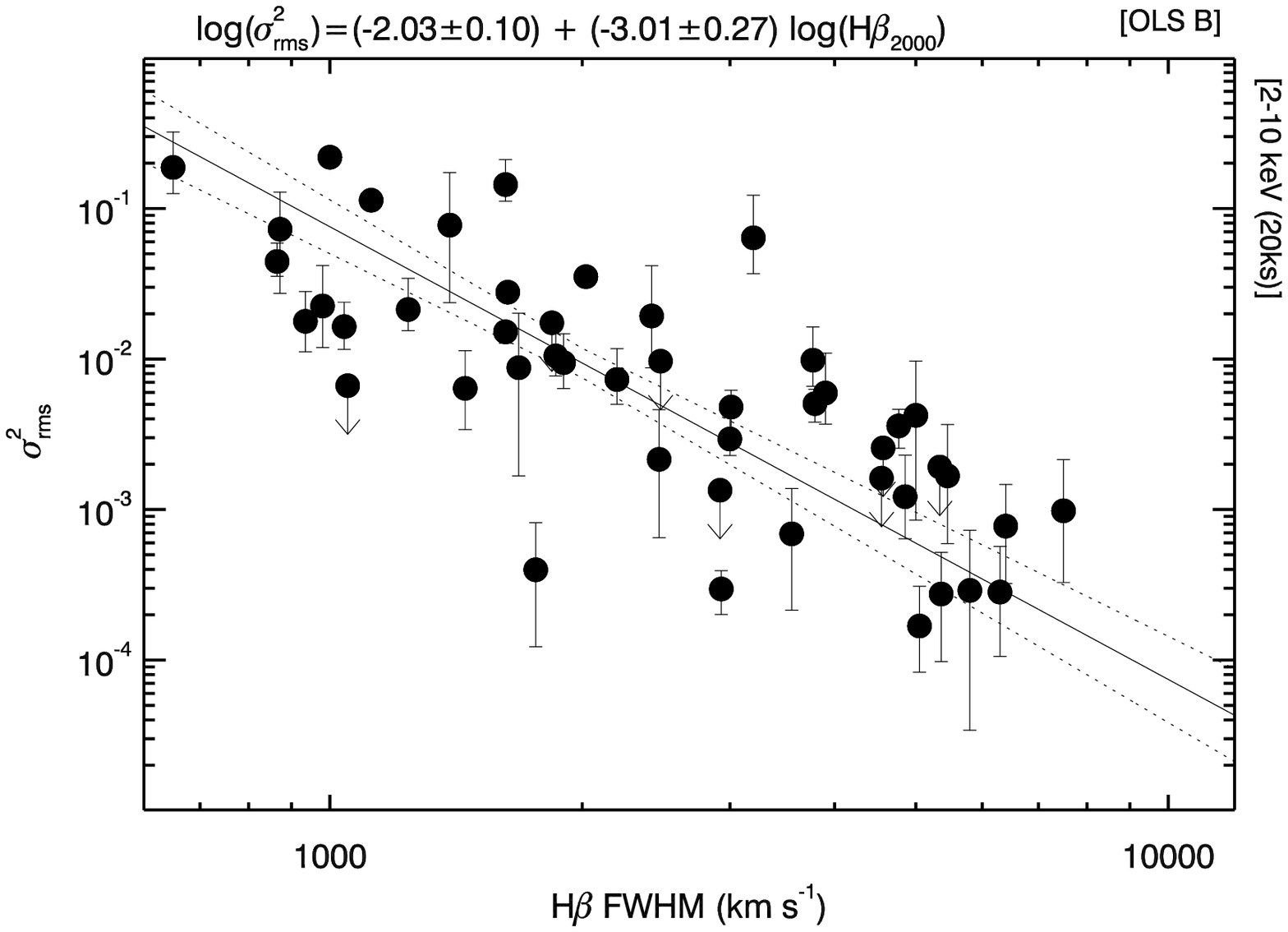,width=0.49\textwidth}
\epsfig{file=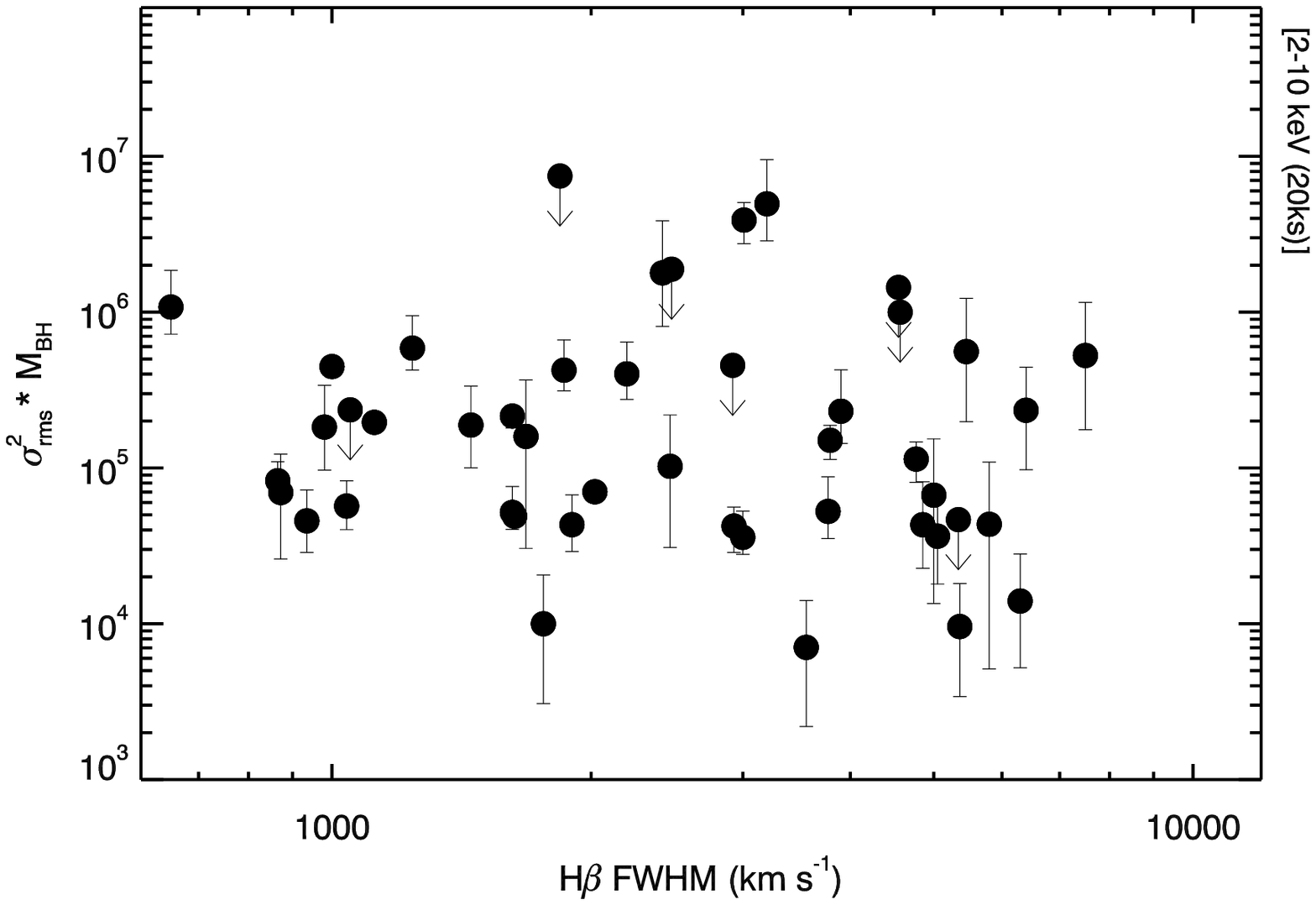,width=0.49\textwidth}
\end{center}
\caption{{\it (Left panel)} $\sigma^2_{\rm rms,20}$ vs. FWHM H$_{\beta}$.
A strong correlation is present. {\it (Right panel)} $\sigma^2_{\rm rms,20}\times{\rm 
M_{\rm BH}}$ vs. FWHM H$_{\beta}$. The correlation disappears. H$\beta_{\rm 2000}$ indicates H$\beta$ in units of 2000 km s$^{-1}$.}
\label{FWHM} 
\end{figure*}

The left panel of Fig. \ref{FWHM} shows the $\sigma^2_{\rm rms,20}$ vs. the H$_{\beta}$ Full Width at Half Maximum (FWHM) relation. As observed by McHardy et al. (2006), a strong anti-correlation is present in CAIXAvar (the best-fit results, as well as the probability of a correlation between variability amplitude and H$_{\beta}$ FWHM are listed in Table \ref{relation}). However, it is well know that H$_{\beta}$ FWHM strongly depends on M$_{\rm BH}$. In fact, Bianchi et al. 2009 show a very strong correlation between M$_{\rm BH}$ and  H$_{\beta}$ for the objects in the original CAIXA sample. This is not surprising, in fact the width of the H$_{\beta}$ line is often used to estimate M$_{\rm BH}$ in AGN.  Therefore, the anti-correlation between variability and H$_{\beta}$ FWHM that we observe, is almost certainly a byproduct of the variability vs. M$_{\rm BH}$ relation. The right panel of Fig. \ref{FWHM} shows that the variability vs. width of the H$_{\beta}$ line correlation disappears when the dependance on the M$_{\rm BH}$ on excess variance is taken off by multiplying $\sigma^2_{\rm rms}$ with the M$_{\rm BH}$. This is true also for the excess variance measurements from the 80, and 40 ks segments (see Table \ref{relation}). 

\subsection{$\sigma^2_{\rm rms}$ vs. spectral index $\Gamma$}
\label{secGamma}

\begin{figure*}
\begin{center}
\epsfig{file=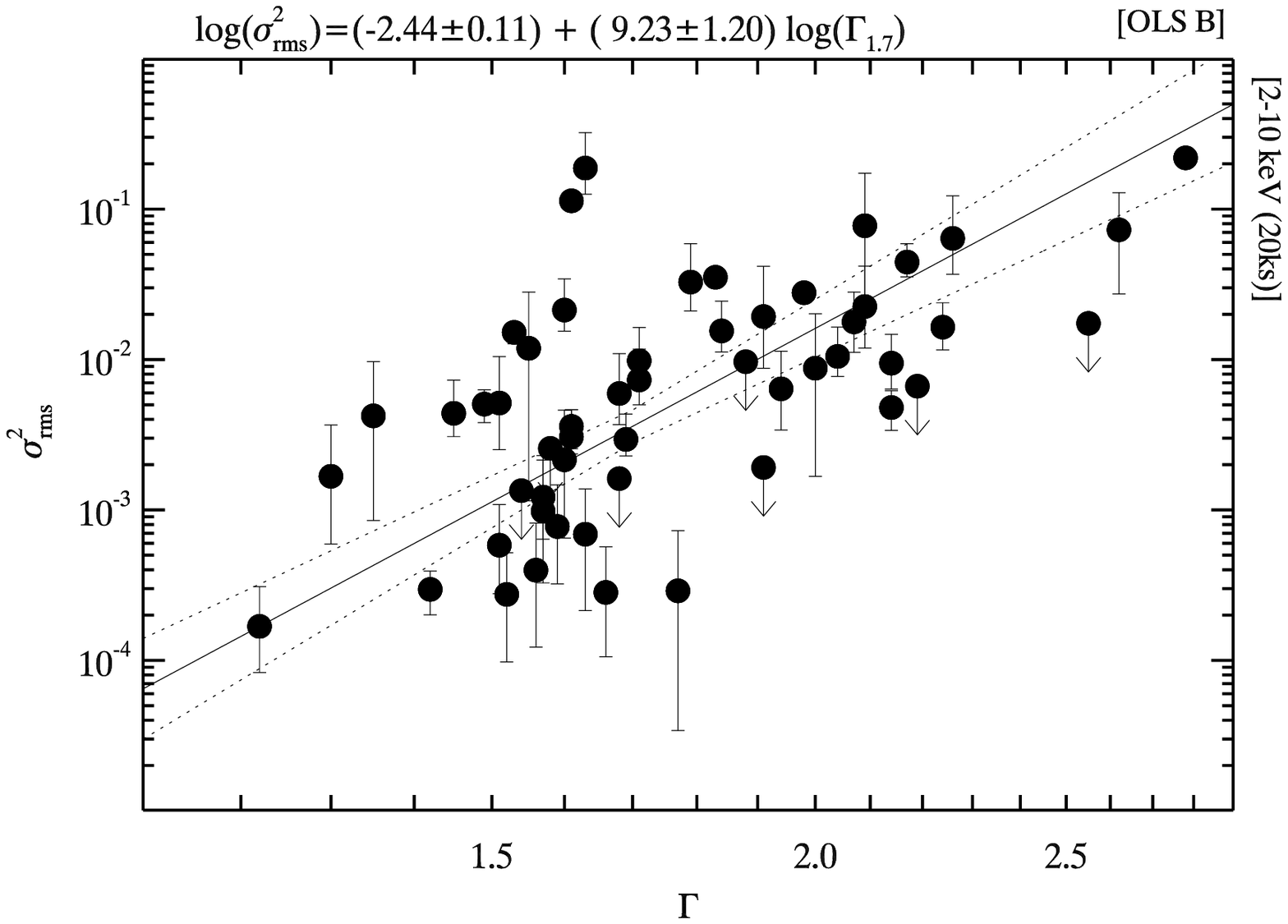,width=0.49\textwidth}
\epsfig{file=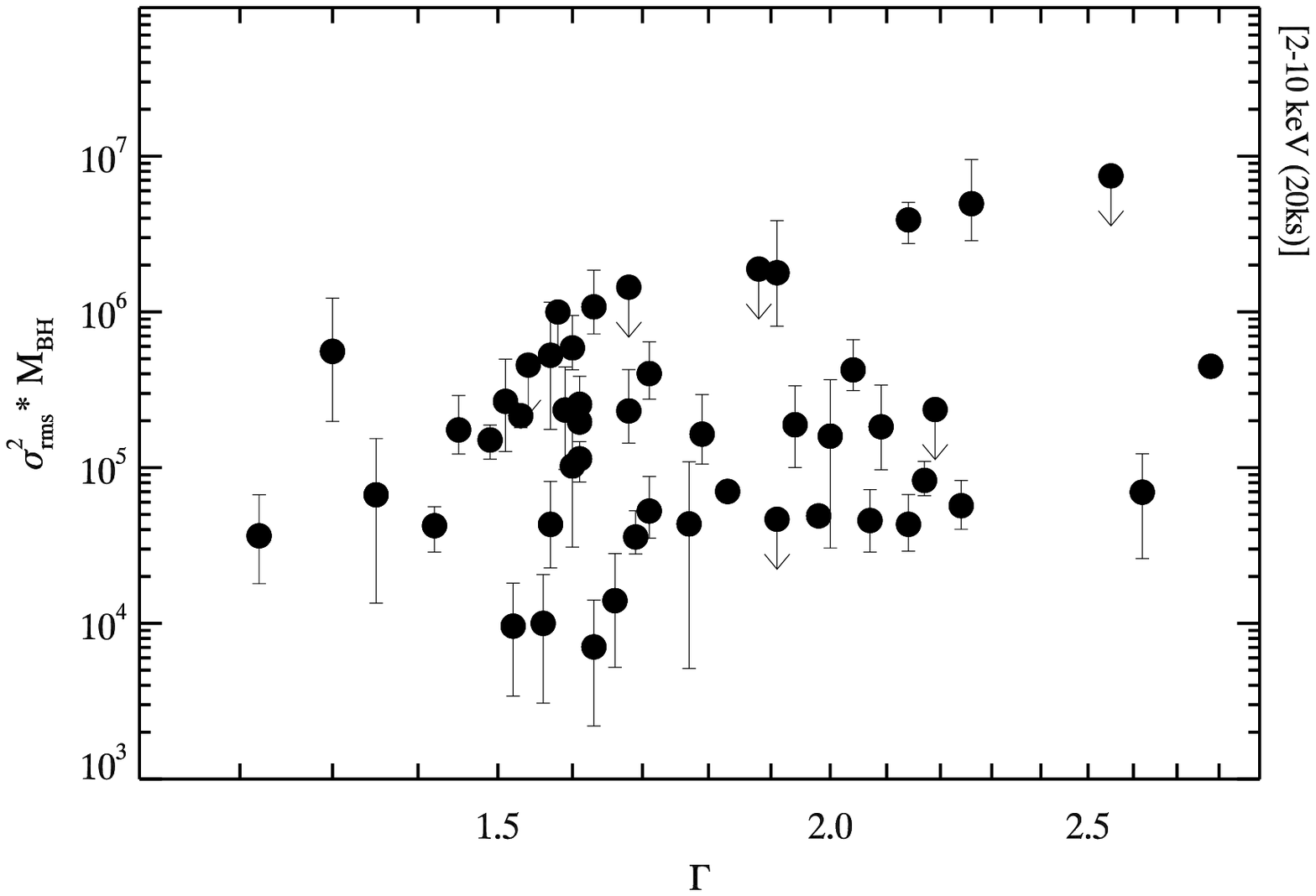,width=0.49\textwidth}
\end{center}
\caption{The $\sigma^2_{\rm rms,20}$ and $\sigma^2_{\rm rms,20}\times{\rm M_{\rm BH}}$ vs. 2-10 keV spectral index correlations (left and right panel). $\Gamma_{1.7}$ indicates $\Gamma$ in units of 1.7.}
\label{Gamma} 
\end{figure*}

The left panel of Fig. \ref{Gamma} shows the $\sigma^2_{\rm rms,20}$ vs. the 2-10 keV spectral index relation for the CAIXAvar sample (see Bianchi et al. 2009 for a description of how the spectral index is calculated). The variability is strongly correlated with $\Gamma$. This is also the case for $\sigma^2_{\rm rms,80}$, $\sigma^2_{\rm rms,40}$, $\sigma^2_{\rm rms,20}$ and the  $\sigma^2_{\rm rms,10}$ vs. $\Gamma$ relations (see Tab. \ref{relation}). In fact, the probability that these correlations are significant are all above 98.5~\% and reaches values larger than 99.99~\%.

This correlation (see also Turner et al. 1999) is rather surprising and not easy to understand. As before, we also investigated the $\sigma^2_{\rm rms,80,40,20,10}\times$M$_{\rm BH}$ vs. $\Gamma$ relations. When the variability amplitude is ``normalized" to M$_{\rm BH}$, the correlation disappears (see right panel of Fig. \ref{Gamma}  and Tab. \ref{relation}). However, even this result is not easy to understand, as there are no ``a priory" direct physical connection between the M$_{\rm BH}$ and the 2-10 keV spectral index. In fact, although some authors claim the presence of a correlation between spectral index and M$_{\rm BH}$, only a marginally significant correlation  between these quantities is observed in CAIXA (Bianchi et al. 2009).  To be scrupulous, we repeated the study of this correlation in the case of  the CAIXAvar sample (due to the slightly different definition of the CAIXAvar sample compared to CAIXA). Again, only a marginal correlation is present. 

Several authors have argued in the past that the 2-10 keV spectral index  is not correlated with the M$_{\rm BH}$, but with the accretion rate (see e.g. Porquet et al. 2004; Piconcelli et al. 2005; Shemmer et al. 2006; Saez et al. 2008; Papadakis et al. 2009; Sobolewska \& Papadakis 2009; Risaliti et al. 2009; Scott et al. 2011). This argument is strengthened by the fact that the same spectral index vs. accretion rate correlation is observed in BHB as well (see e.g. Wu \& Gu 2008). If this is true, then we can explain the $\sigma^2_{\rm rms}$ vs. $\Gamma$ relation as follows: Sobolewska \& Papadakis (2009) have argued that $\Gamma \propto \dot{m}^{0.1}$, while we observe that $\sigma^2_{\rm rms} \propto \dot{m}$ (see Table \ref{relation}). Therefore, we would expect to observe $\sigma^2_{\rm rms} \propto \Gamma^{10}$, perfectly consistent with our observations (see Table \ref{relation}).  

\begin{figure}
\begin{center}
\epsfig{file=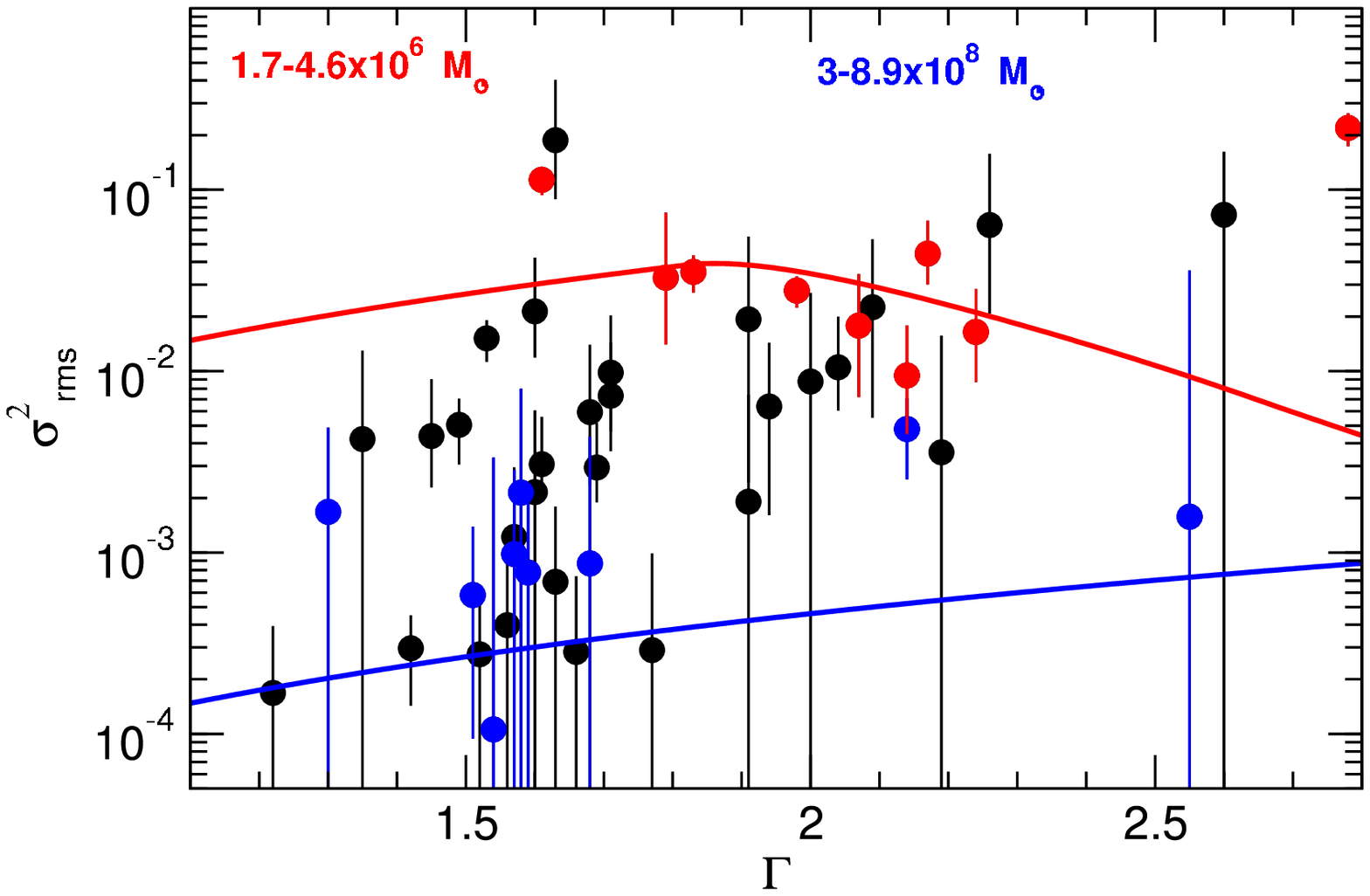,width=0.49\textwidth,bb=20 430 535 770, clip=}
\end{center}
\caption{The $\sigma^2_{\rm rms,20}$ vs. 2-10 keV spectral index correlation. Solid line show Model B predictions for different M$_{\rm BH}$.}
\label{GammaTheo} 
\end{figure}

In Fig. \ref{GammaTheo} we plot again the $\sigma^2_{\rm rms}$ vs. $\Gamma$ data for all the CAIXAvar sources  (irrespective of the magnitude of their error), in the log--log space. Red and blue circles indicate the data for AGN with $1.7\times10^6$~M$_{\odot} <$~M$_{\rm BH}<4.6\times10^6$~M$_{\odot}$ and $3\times10^8$~M$_{\odot} <$~M$_{\rm BH}<8.9\times10^8$~M$_{\odot}$, respectively, exactly like in Fig. \ref{mbhTheo}. The red and blue lines indicate the Model B $\sigma^2_{\rm rms}-\Gamma$ predictions for an AGN with M$_{\rm BH}=2.5\times10^6$ and $5\times10^8$~M$_{\odot}$ (i.e. the mean M$_{\rm BH}$ of the points indicated with the red and blue circles). For each of the two M$_{\rm BH}$ values we have estimated the excess variance for various accretion rates assuming that PSD$_{\rm amp}\propto \dot{m}^{-0.8}$, as explained in Section 6.1.3. Then, instead of plotting $\sigma^2_{\rm rms}$ as a function of $\dot m$ (as we did in Fig. \ref{accr_all}), we plot it as function of $\Gamma$, using the Sobolewska \& Papadakis (2009) relation mentioned above. The agreement between the data and the model predictions is rather good. This result suggests that the $\sigma^2_{\rm rms} - \Gamma$ relation we observe (Fig. \ref{Gamma}) is due to the $\sigma^2_{\rm rms}$ vs. $\dot m$ and $\Gamma$ vs. $\dot m$ relations. When "normalized" to M$_{\rm BH}$, the excess variance is no more clearly correlated with $\Gamma$, because, according the Model B predictions, it is expected to be weak (with a slope of just $\sim 0.2$), and with a considerable scatter at high accretion rates, where the excess variance may increase with increasing $\dot m$ for AGN with large M$_{\rm BH}$, but it may also decrease with increasing $\dot m$ for lower M$_{\rm BH}$ AGN.

\section{Are NLS1 intrinsically more variable than other AGN?}
\label{nls1}

\begin{figure*}
\begin{center}
\epsfig{file=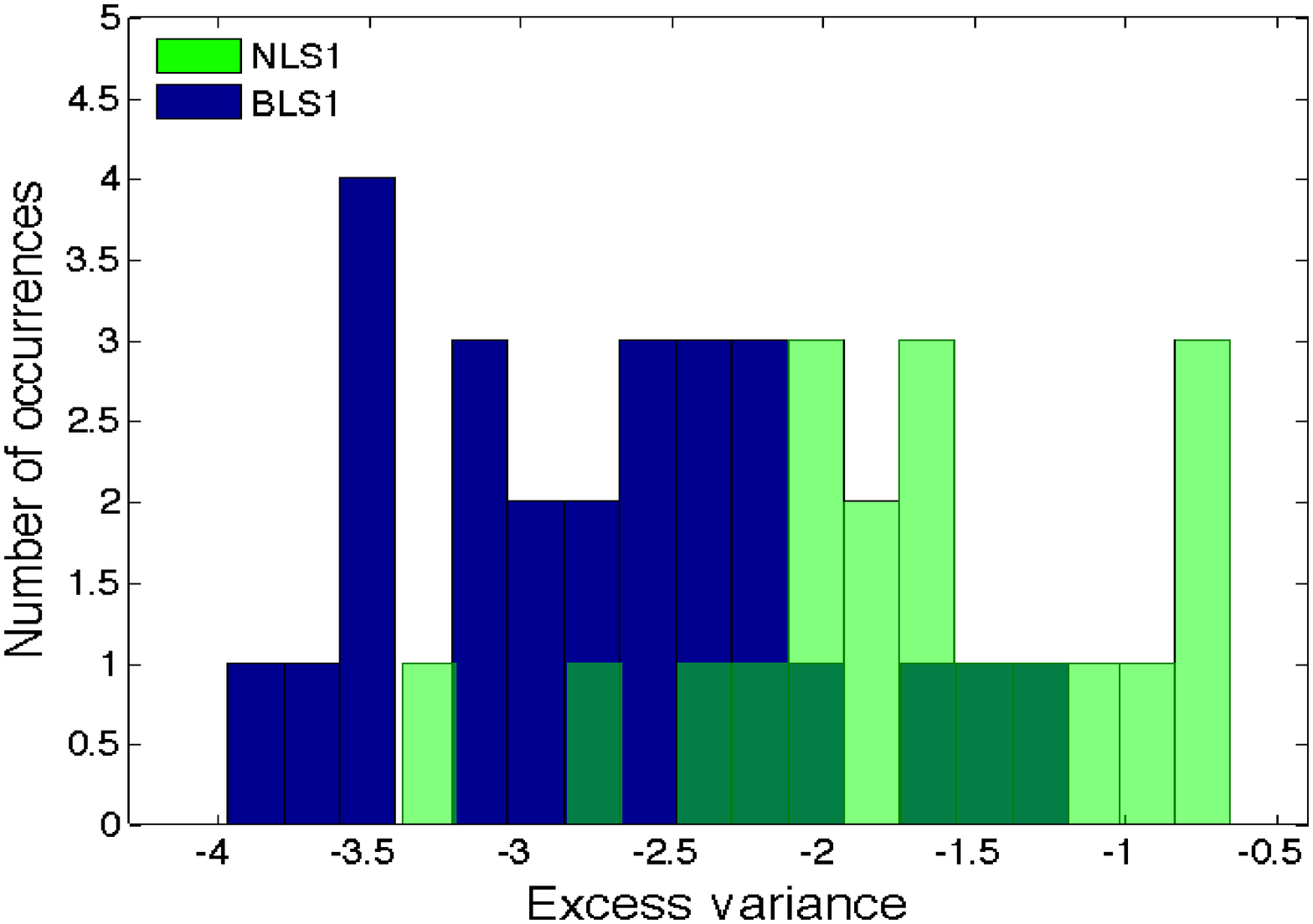,width=0.23\textwidth,angle=-90}
\epsfig{file=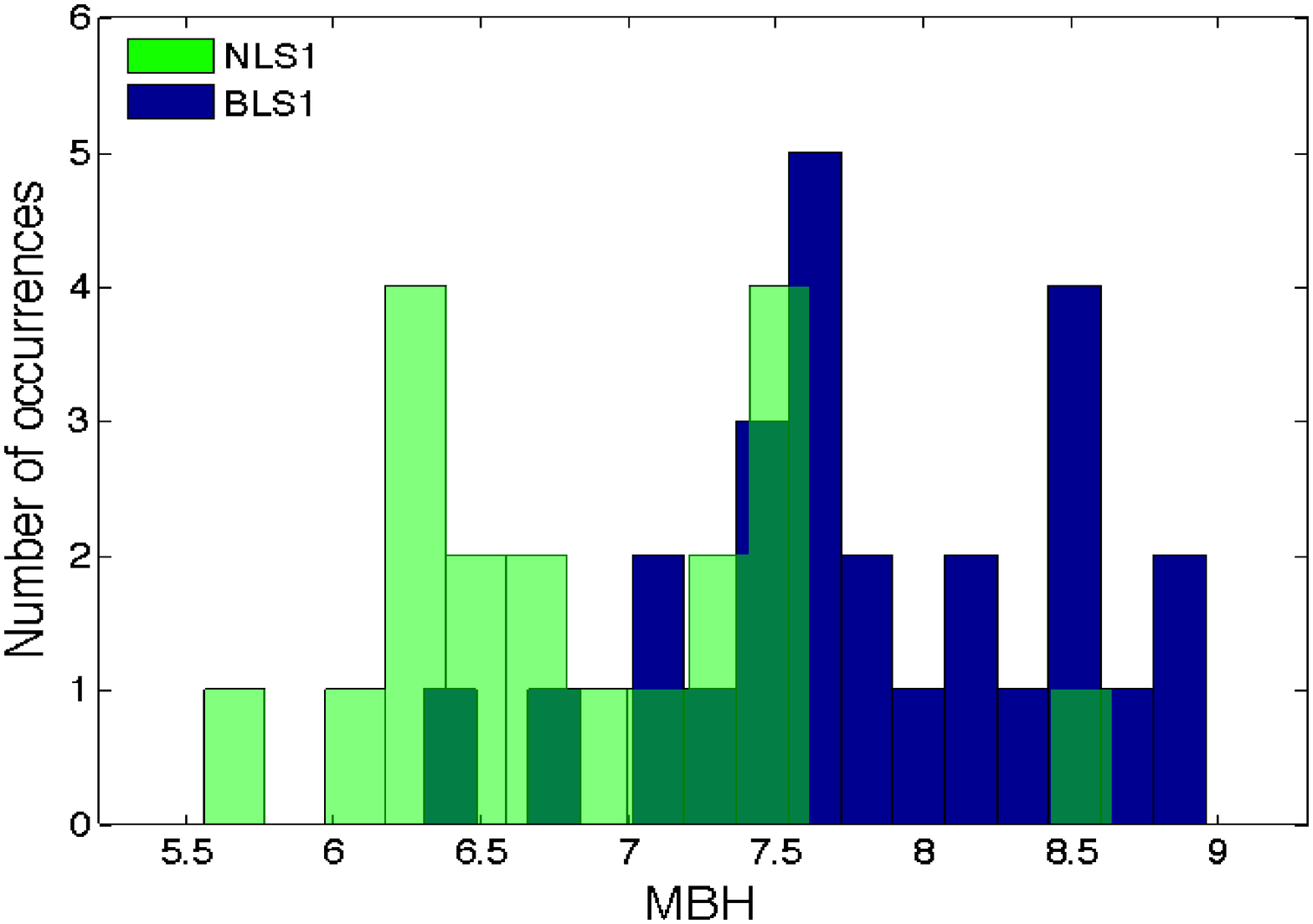,width=0.23\textwidth,angle=-90}
\epsfig{file=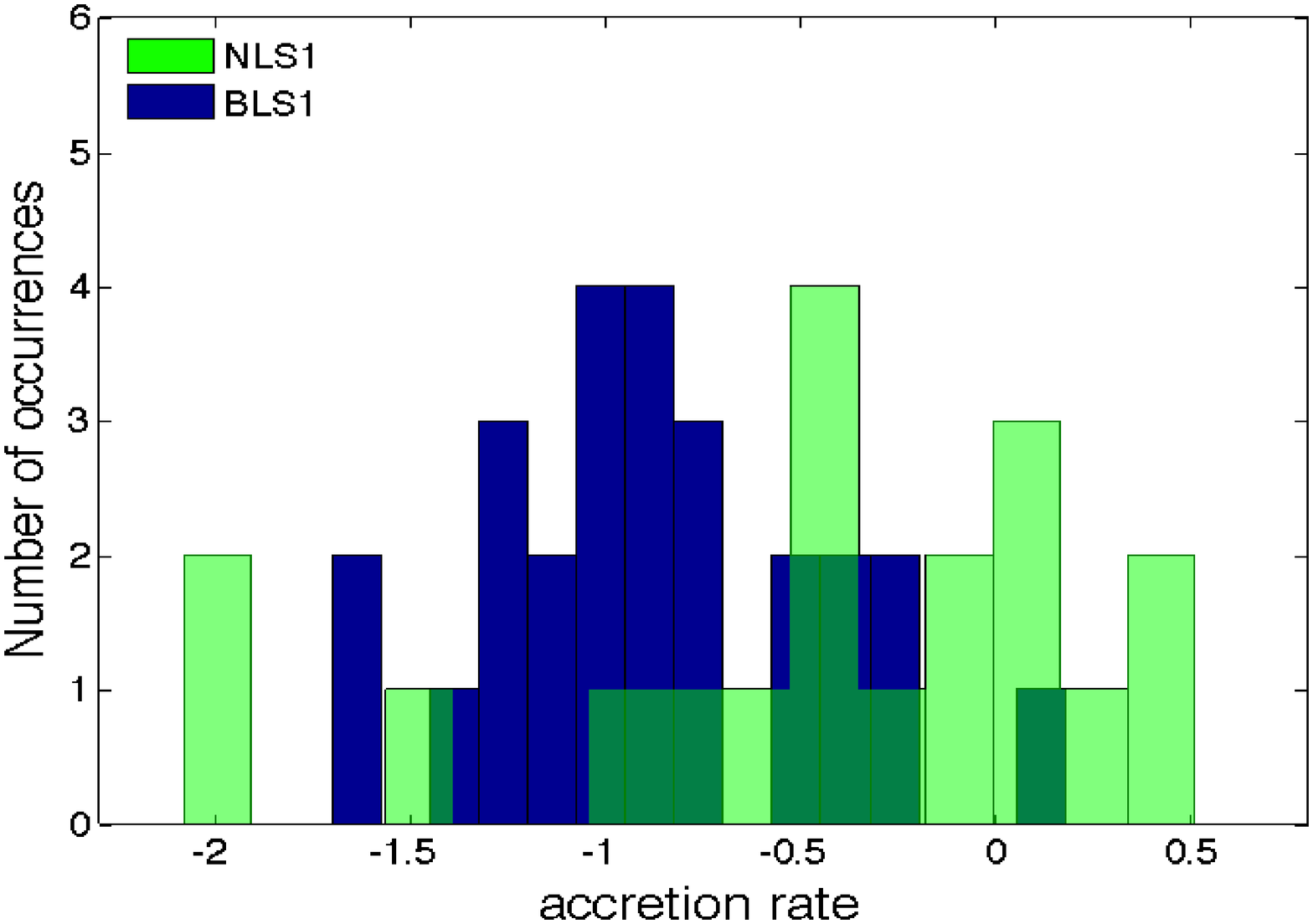,width=0.23\textwidth,angle=-90}
\epsfig{file=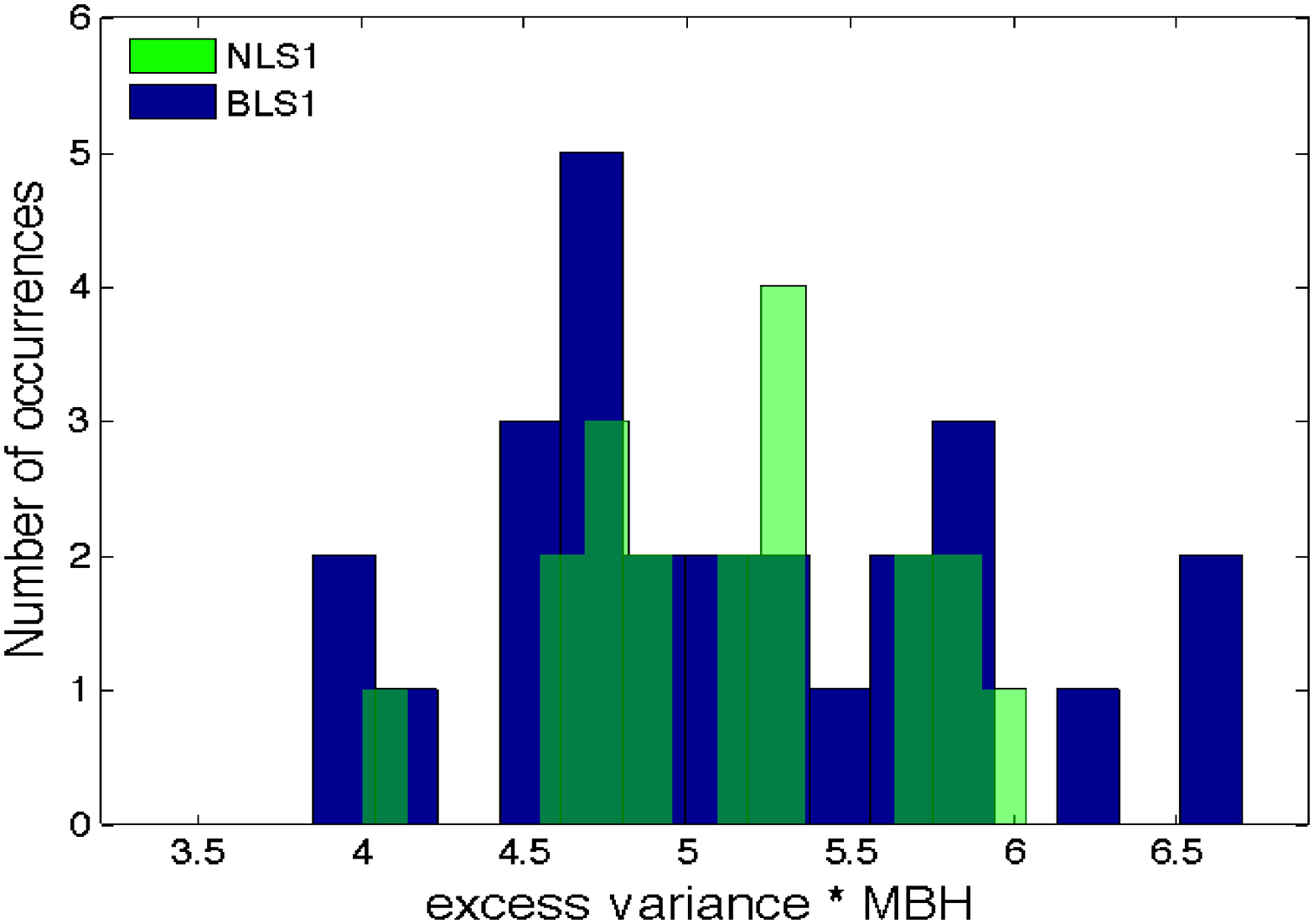,width=0.23\textwidth,angle=-90}
\epsfig{file=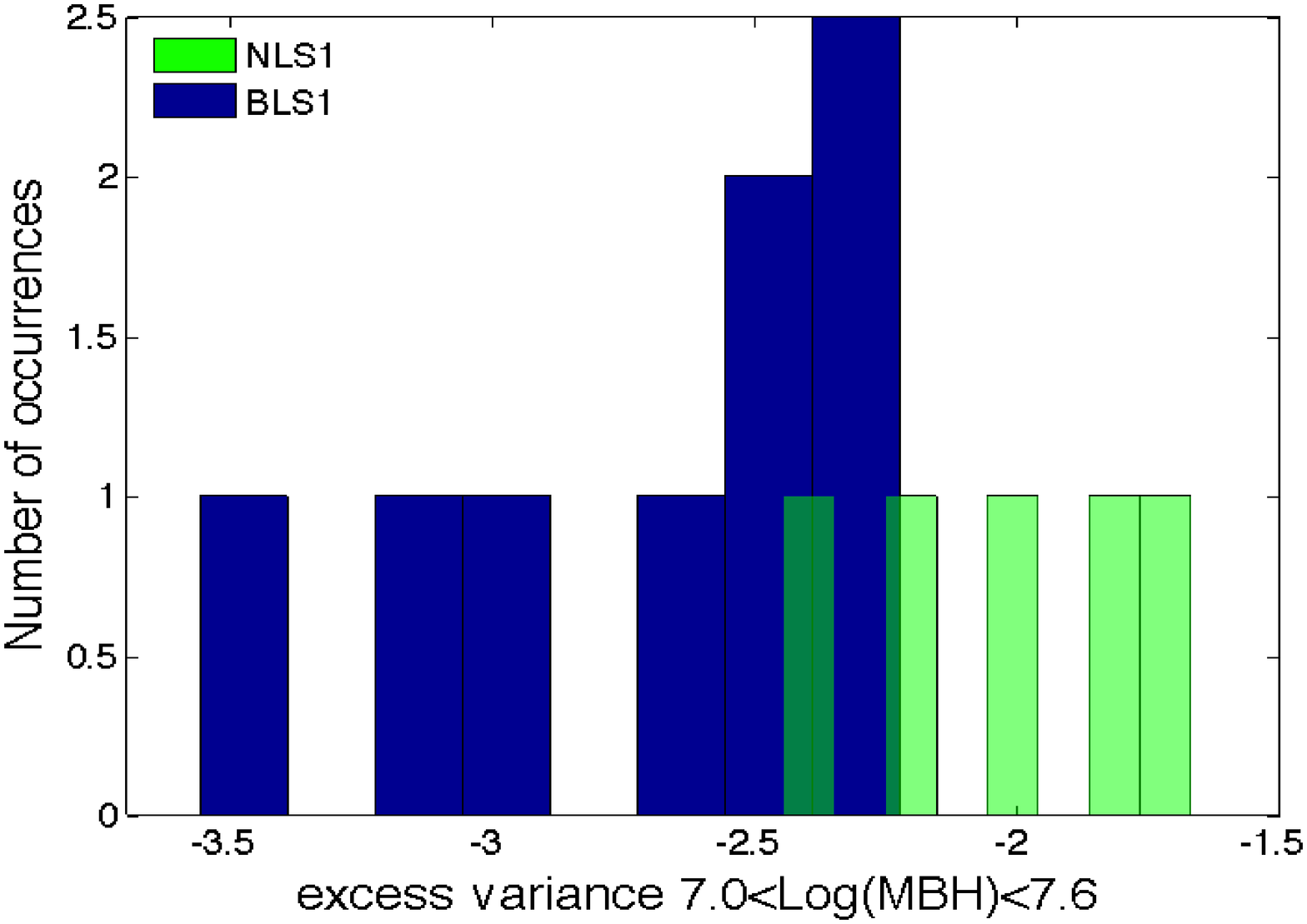,width=0.23\textwidth,angle=-90}
\epsfig{file=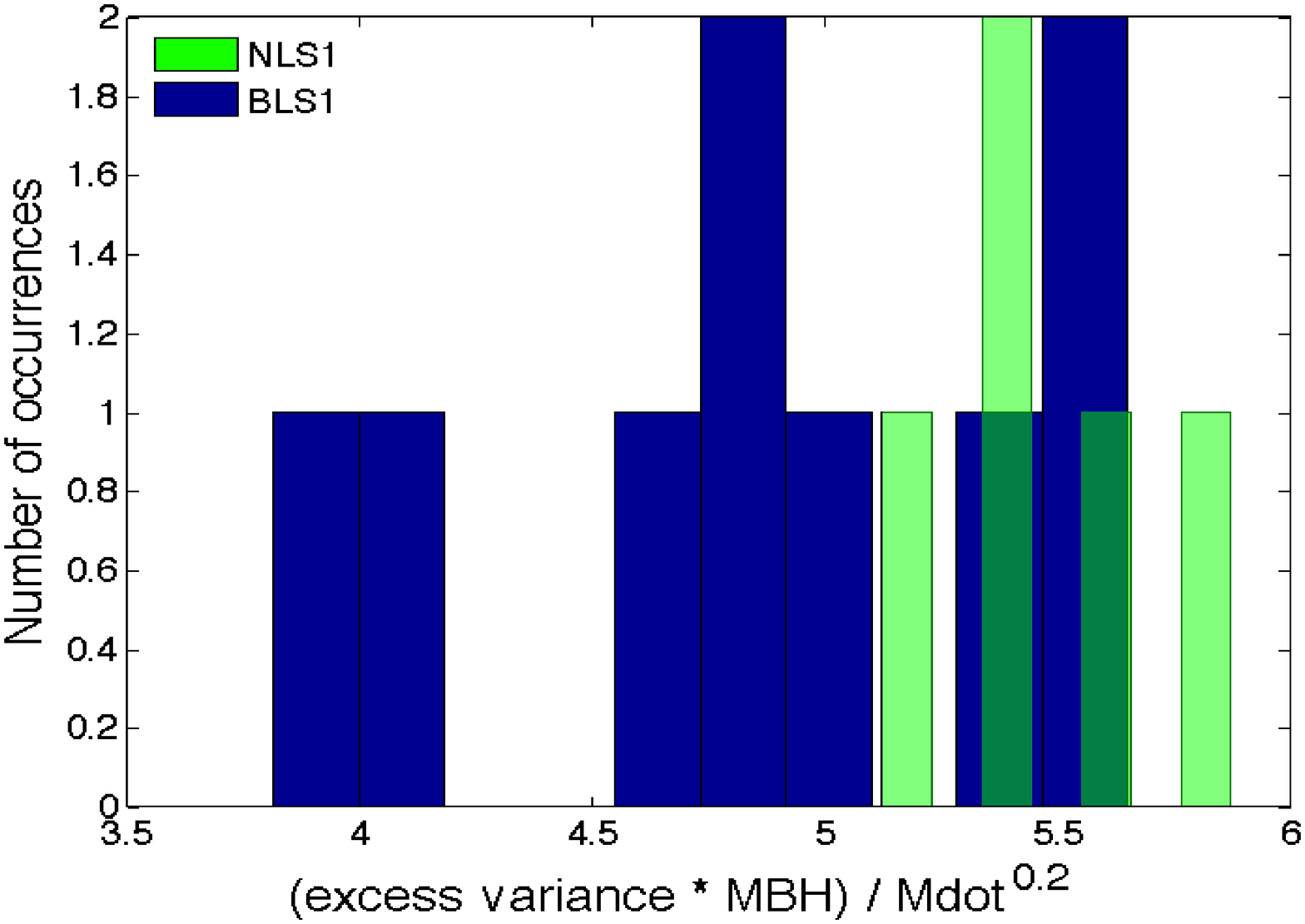,width=0.23\textwidth,angle=-90}
\end{center}
\caption{{\it (Upper left panel)} The light green histogram shows the 
excess variance distribution of the NLS1 which are variable in 20 ks segments and present in CAIXAvar. The blue histogram shows the 
excess variance distribution for the other (broad line Sayfert 1) 
sources in CAIXAvar. The NLS1 are significantly (KS test) 
more variable than BLS1. {\it (Upper middle panel)} M$_{\rm BH}$ 
distribution (colour code as before). The NLS1 have smaller BH 
masses than BLS1. {\it (Upper right panel)} Accretion rate distribution. 
The NLS1 are accreting at significantly higher accretion rates than BLS1. 
{\it (Lower left panel)} $\sigma^2_{\rm rms,20}\times$M$_{\rm BH}$
distribution. Once that the M$_{\rm BH}$ dependence is taken into 
account the NLS1 are not anymore distinguishable from BLS1.
{\it (Lower middle panel)} Excess variance distribution of the 
AGN with 7.0$<$Log(MBH)$<$7.6. At the same M$_{\rm BH}$, NLS1 
are more variable than BLS1 objects. {\it (Lower right panel)}
$\sigma^2_{\rm rms,20}\times$M$_{\rm BH}$/$\dot{m}^{0.2}$ distribution 
for the AGN with 7.0$<$Log(MBH)$<$7.6. The higher variability of the NLS1 
objects can be due to their higher accretion rate.
}
\label{histonls1} 
\end{figure*}

Several works in the past have shown that NLS1 appear to have higher variability than normally expected for their M$_{\rm BH}$ (see e.g. Leighly 1999; Turner et al. 1999; Nikolajuk et al. 2004; 2009). To investigate this issue, we compared the excess variance of the NLS1 and BLS1 in CAIXAvar. The upper left panel of Fig. \ref{histonls1} shows the distribution of Log$(\sigma^2_{\rm rms,20})$ for the NLS1 (light green) and the other AGN (BLS1, in blue) in the CAIXAvar sample. Clearly, the NLS1 appear to be more variable than BLS1. The mean value of the distributions are ${\rm <Log(}\sigma^2_{\rm rms,20}){\rm >} = -1.69$ and  ${\rm <Log(}\sigma^2_{\rm rms,20}){\rm >} = -2.70$ for the NLS1 and BLS1, respectively.  Application of the Student's t-test indicates that they are significantly different (null hypothesis probability, NHP$ = 3\times10^{-5}$). At the same time, the upper middle and upper right panels of Fig. \ref{histonls1} shows that NLS1 have significantly smaller M$_{\rm BH}$ ( $<$Log(M$_{\rm BH, NLS1}) >= 6.86$ and $<$Log(M$_{\rm BH, BLS1}) > = 7.85$; NHP$ = 2\times10^{-5}$). They also appear to have higher accretion rates (upper right panel in Fig. \ref{histonls1}), although the difference between NLS1 and BLS1 $\dot{m}$ is less significant ( $ <$Log($\dot {m}_{\rm NLS1}) > = -0.47$ and $<$Log($\dot {m}_{\rm BLS1}) >= -0.89$; NHP$ = 2.2\times10^{-2}$), just a $\sim$ 1-sigma effect. 

The lower left panel of Fig. \ref{histonls1} shows the $\sigma^2_{\rm rms,20}\times$M$_{\rm BH}$ distribution for the NLS1 and BLS1 as well. They look very similar and in fact the application of the t-test indicates no significant difference between the mean of the two distributions (NHP $> 0.05$). We also test that, given the number of points in each sample, we would be able to detect with high significance (more than $3 \sigma$) a difference in the mean of the two distributions of the order of 4-5. This indicates that, once the M$_{\rm BH}$ dependence is taken into account, NLS1 have variability amplitudes, to first approximation indistinguishable from the ones of BLS1. Thus the difference in the variability properties of NLS and BLS1 is mainly due to the fact that NLS1 host a smaller M$_{\rm BH}$, on average, when compared to BLS1. 

In order to investigate whether the accretion rate also contributes to the difference in the variability amplitudes of NLS1 and BLS1 we selected a subsample of NLS1 and BLS1 which have the same M$_{\rm BH}$, on average. The upper middle panel of Fig. \ref{histonls1} indicates that several NLS1 and BLS1 have Log(M$_{\rm BH}$) between 7 and 7.6. We note that MRK110 belongs to this M$_{\rm BH}$ range. The classification of this AGN is quite uncertain. In fact, according to the FWHM of H$_{\rm \beta}=1760$ km s$^{-1}$, it would be classified as a NLS1, however it lacks all the other features typical of NLS1. For this reason, detailed studies of its optical spectrum suggest that MRK110 is a BLS1 with relatively "narrow" broad lines (Veron-Cetty et al. 2007). Thus, we decide to exclude MRK110 from the following analysis. Both the NLS1 and BLS1 distributions in this narrow M$_{\rm BH}$ range have mean Log(M$_{\rm BH}$) of 7.37. 

The lower middle panel of Fig. \ref{histonls1} shows the excess variance distribution of the NLS1 and BLS1 within the selected M$_{\rm BH}$ bin mentioned above. Despite the small number of objects considered here, the distributions suggest that NLS1 show higher variability (KS and t-test more than 98.7 \% probability; NHP $= 1.3\times10^{-2}$; the mean excess variance being ${\rm <Log(}\sigma^2_{\rm rms,20}){\rm >} = -2.04$ and $-2.70$ for NLS1 and BLS1, respectively). This difference cannot be ascribed to the M$_{\rm BH}$, thus either the NLS1 are different from the other AGN or the difference is due to their higher accretion rate ($<\dot{m}>_{\rm NLS1}=-0.65$ and $<\dot{m}>_{\rm BLS1}=-1.03$, although not significantly different, NHP$>0.05$). Thus, to test if the enhanced variability is due to the accretion rate dependence, we multiply the excess variance for M$_{\rm BH}$ and divide for the accretion rate to the expected power ($\dot{m}^{1}$ as in Model A and $\dot{m}^{0.2}$ as in Model B). The right lower panel of Fig. \ref{histonls1} shows that, once that the accretion rate is taken into account (showing the case B of $\dot{m}^{0.2}$), the distribution and the mean of the variability of NLS1 and BLS1 are perfectly consistent (NHP $> 0.05$ in both cases). 

\section{Discussion}

We have studied the X-ray variability properties of a large (161 AGN) sample of radio quiet, X-ray unobscured, bright AGN. We estimated their excess variance on various time scales (80, 40, 20 and 10 ks) and various energy bands (0.3-0.7, 0.7-2 and 2-10 keV), using \xmm\ data and taking full advantage of the high sensitivity, long orbit and wide energy range of \xmm. 
We have systematically investigated the relation between the variability amplitude of these objects (as parametrized by $\sigma^2_{\rm rms}$) and other source parameters like, M$_{\rm BH}$, accretion rate, optical emission line width, and X-ray spectral slope. The AGN of the CAIXAvar sample, with well measured $\sigma^2_{\rm rms}$, mainly belong to the local universe (z$<0.2$, with just 3 objects at intermediate redshift, see Fig. 1) and have relatively modest bolometric luminosity (L$_{\rm Bol}\sim 10^{41-47}$ erg s$^{-1}$). Thus, the observed relations and variability properties are, strictly speaking, characteristic of the AGN of the local Universe. A larger sample of higher redshift and luminosity AGN is needed in order to confidently extrapolate the AGN variability properties observed in the local universe to higher redshift. We discuss below some of the implications of our results. 

\subsection{$\sigma^2_{\rm rms}$ in various energy bands}

One of the main results of our work is that the variability amplitude is comparable in the hard, medium and soft energy bands, at least on time scales $\le 40$ ks. Typically, different spectral components contribute to the AGN X--ray spectra at different energy bands. At energies above 2 keV, the spectrum is generally dominated by the continuum power-law component. At lower energies a ``soft excess"  component appears in many objects, on top of the extrapolation of the power law to lower energies. If this component is constant (or less variable than the continuum), then we should expect $\sigma^2_{\rm rms, soft} < \sigma^2_{\rm rms, 2-10 keV}$, but this is not the case. Furthermore an absorbing material (either neutral or ionised) is also present in many radio quiet AGN. This material acts as a filter, absorbing mainly the soft and medium energy band photons, and may vary in ionisation, responding to the continuum variations, and/or in its column and covering fraction of the source. If these variations were to happen on time scales less than $\sim$ half a day then we would expect to measure a larger variability amplitude in the medium and soft energy bands, since they are more affected by absorption, compared to the hard band. This is contrary to what we observe. Finally, if there were significant, large amplitude intrinsic spectral slope variations with a pivot point at energies lower than 0.3 keV, then we would expect $\sigma^2_{\rm rms, 2-10 keV}>\sigma^2_{\rm rms, soft,~or~medium}$, and the opposite trend if the pivot point were at energies higher than 10 keV. Again, this is contrary to our results. Consequently, our results strongly indicate that, on time-scales less than 40 ks, intrinsic spectral slope variations or absorption variations, must be of small amplitude so that they do not contribute significantly to the observed {\it flux} variability of the sources in the different energy bands. Of course we know that the rms-spectra of AGN are {\it not} flat (e.g. Ponti et al. 2004; 2006; 2007; 2010; Gallo et al. 2004; 2007; Gierli{\'n}ski \& Done 2006; Goosman et al. 2006; Petrucci et al. 2007; Larsson et al. 2008; Zoghbi et al. 2010). In other words, the components mentioned above, do affect the observed variability at different energy bands to some extent. However, our results show that, {\it to first order}, the main driver for the observed {\it flux} variations in AGN at all bands, between 0.3--10 keV, is the {\it continuum normalisation variability}.  

On 80 ks time-scales a deviation, although not more significant than $\sim2~\sigma$, from the 1-to-1 correlation appears. 
In particular, the more variable (i.e. the smaller M$_{\rm BH}$) AGN appear to have a {\it higher} soft and medium variability, when compared to the hard band. This trend had already been noticed in the past as well (e.g. Nandra et al. 1997; Leighly 1999; Markowitz \& Edelson 2001). If real, this result is most probably due to a complex variation of the PSD normalisation and slope with energy (McHardy et al. 2004; Uttley \& McHardy 2005). Alternatively, this trend might indicate some sort of triggering of the soft/medium band variability occurring only on time-scales longer than a given time-scale and-or characteristic length. 

If this deviation from the 1-to-1 correlation in the $\sigma^2_{\rm rms,80}$ plots is real, then we would expect it to appear even stronger at longer time-scales. Indeed, Markowitz \& Edelson (2001) measured the $\sigma^2_{\rm rms}$ on time-scales of 300 days ($\sim26$ Ms) with time bins of roughly 5-13 days ($\sim 800$ ks) for 9 Seyfert 1 galaxies. On such long timescales, the authors find that all the sources exhibit stronger variability in the softer X-ray (the 2--4 keV band, in their case) than in the hard one (7--10 keV band). When the $\sigma^2_{\rm rms,7-10~keV}$ is plotted vs. the $\sigma^2_{\rm rms,2-4~keV}$ a highly significant deviation from the 1-to-1 correlation is observed (best fit slope being $0.724\pm0.041$; Markowitz \& Edelson 2001). This result strongly supports the idea of a triggering of the spectral variability at a timescale of about 80 ks (at least for the low M$_{\rm BH}$ AGN). For this reason we discuss below some ideas which could explain such a behaviour. 

Absorbers change their ionisation balance (i.e. change their opacity at a given energy), responding to the illuminating sources,  only on time-scales longer than their recombination time (on shorter time-scales the absorber has not enough time to react to the variation of the illuminating source). The recombination time depends critically on the density of the absorber, being shorter for denser clouds. Interestingly accretion disc theory suggests that smaller mass BH should have denser discs and environment. If this is the case it would indicate that the absorber's recombination time scale is of the order of 40-80 ks for small M$_{\rm BH}$ AGN (and longer for larger M$_{\rm BH}$ AGN). A time scale of 80 ks corresponds to about 1 light day (roughly the distance of the BLR in a small mass BH) or a light crossing time of $2\times10^4$ r$_g$, for a BH of $10^6$ M$_{\odot}$ (r$_g=\frac{GM}{c^2}$). Such radius is comparable to the Compton radius where a disc wind might be formed (Begelman et al. 1983a,b; Woods et al. 1996; Ponti et al. 2012). Assuming that the excess of soft and medium variability is due to absorption located either in the BLR (see also Risaliti et al. 2011) or at the Compton radius, we would naively then expect the absorber to be variable on time-scales larger than about several $10^3\times$M$_{\rm BH,7}$ s (see Krongold et al. 2005; 2007; Kaastra et al. 2011; for more accurate computations). 

Alternatively the spectral variability might be intrinsic to the source, with the spectral slope being correlated with flux, having a pivot point above 10 keV, thus producing higher soft variability. Such a phenomenon has been envisaged by Comptonisation models (Haardt et al. 1991; 1993; 1994; 1997) in which variations of the coronal optical depth, size and-or temperature drive spectral changes, with the spectral index steepening with flux. If the time scale for significant spectral slope variations corresponds to the thermal and-or sound crossing time of a standard accretion disc at a few r$_g$, this could be of the right order of $\sim40-100$ ks for small M$_{\rm BH}$ AGN (where the thermal and sound crossing time should be $\sim30-150$ ks for an alpha disc with $\alpha=0.01$, H/R=0.01 at 7 $r_g$ for M$_{\rm BH}=10^6$ M$_{\odot}$; Treves et al. 1988; Edelson \& Nandra 1999).  

\subsection{$\sigma^2_{\rm rms}$ vs. M$_{\rm BH}$, bolometric luminosity and H$\beta$}

The CAIXAvar sample shows statistically significant correlations between $\sigma^2_{\rm rms}$ and M$_{\rm BH}$, bolometric luminosity and H$\beta$ (McHardy et al. 2006). We believe that the latter correlations are a by-products of the $\sigma^2_{\rm rms}$ vs M$_{\rm BH}$ relation, which is probably the fundamental correlation in AGN (Papadakis 2004). There are two reasons why this should be the case: a) the scatter of the $\sigma^2_{\rm rms}$ vs. M$_{\rm BH}$ is much smaller than the correlation of the $\sigma^2_{\rm rms}$ with L$_{\rm Bol}$ or with the FWHM of H$\beta$; b) standard accretion disc theory (Shakura \& Sunyaev 1973) predicts that all the time-scales in an $\alpha-$disc (like the light crossing, orbiting, thermal, sound crossing, radial drift, and viscous time scale), which should affect the variability amplitude of an object, scale linearly with M$_{\rm BH}$ (Molendi et al. 1992; Treves et al. 1988).

Our best-fit slope and normalisation of the $\sigma^2_{\rm rms,40}$ vs. M$_{\rm BH}$ relation are in agreement with the results of Zhou et al. (2010), who considered a sample of AGN with M$_{\rm BH}$ from reverberation mapping only. Their best-fit slope of $\alpha_{\rm 40}=-1.05\pm0.08$ for the $\sigma^2_{\rm rms}-$M$_{\rm BH}$ relation is in agreement with our best-fit slopes of both the CAIXAvar and Rev samples. O'Neill et al. (2005) computed the $\sigma^2_{\rm rms,40}$ for a large ASCA AGN sample. They also found a significant correlation between normalized excess variance and M$_{\rm BH}$, and estimated a slope of $\alpha_{\rm 40}=-0.57$, steepening to $\alpha_{\rm 40}=-0.84$ (roughly in agreement with the slope obtained with CAIXAvar) once a broken power law was considered. 

One of our major results is that the scatter in the $\sigma^2_{\rm rms}$ vs. M$_{\rm BH}$ relation of the Rev sample is very small. It corresponds to a scatter of 2.7-3 in linear space, which is comparable to the scatter of 2.6--2.9 that Peterson et al. (2004) measured around the AGN M$_{\rm BH}$ vs. $\sigma_*$ relation. This result implies that there may be no intrinsic scatter in the $\sigma^2_{\rm rms}$ vs. M$_{\rm BH}$ relation, and the scatter we observe may be entirely due to the uncertainty on the M$_{\rm BH}$ values from reverberation. Zhou et al. (2010) used a Nuker's linear regression method to estimate the intrinsic scatter of this relation. They fitted the relation taking into account the uncertainties associated with both quantities ($\sigma^2_{\rm rms}$ and M$_{\rm BH}$), then they added a component (that would measure the intrinsic scatter) to the $\sigma^2_{\rm rms}$ uncertainties until reaching a $\chi^2=1$. In this way they estimated an intrinsic scatter of 0.2 dex (in agreement with the Rev sample results that show that intrinsic scatter on $\sigma^2_{\rm rms}<0.4$ dex). 

A scatter around the best fit relation of less than a factor of 3 is challenging for the current variability models. We know that the reverberation sample spans a wide range of accretion rates from below 1\% up to the Eddington limit. If the break frequency scales linearly with accretion rate (McHardy et al. 2006; Koerding et al. 2007), and the PSD amplitude is constant in all AGN, we would expect a scatter of more than $\sim 10-20$ for the 20 and 40 ks relations at M$_{\rm BH}>10^7$M$_{\odot}$  (see Fig. \ref{mbhTheo}). We found that the tight correlation between variability and M$_{\rm BH}$ can be explained if the PSD normalisation decreases with increasing accretion rate as PSD$_{\rm amp}\propto \dot{m}^{-0.8}$. In this case, the combined variations of the $\nu_{\rm br}$ and PSD normalisation with accretion rate results in the PSD integral above the longest frequency sampled in the light curves (i.e. $\sigma^2_{\rm rms}$) being only weakly dependent on accretion rate. Thus, the expected $\sigma^2_{\rm rms}$ at a given M$_{\rm BH}$ is almost the same even for objects with a large difference in accretion rates. We therefore conclude that, the presence of the $\sigma^2_{\rm rms}$ vs. M$_{\rm BH}$ correlation, its best-fit slope and normalisation, {\it and} its small scatter, are in agreement with the idea that every object in CAIXAvar has the same PSD with low frequency slope of $-1$ breaking to a slope of $-2$ at a break frequency that decreases with M$_{\rm BH}$ and increases with $\dot{m}$ (McHardy et al. 2006; Koerding et al. 2007), but only if the PSD normalization also decreases with increasing $\dot m$. This is a new result, which has never been observed before. Investigation of the published PSD results indicates that this is indeed the case (Papadakis et al., in preparation). 

\subsection{The X--ray variability M$_{\rm BH}$ estimation in AGN}

The use of the $\sigma^2_{\rm rms}$ as a tool to measure the M$_{\rm BH}$ has already been proposed in the past. Nikolajuk et al. (2004) provided the first recipe to measure AGN M$_{\rm BH}$ from $\sigma^2_{\rm rms}$ measurements. Their recipe could work for both BLS1 and NLS1, but only if a shift of a factor of 20 in M$_{\rm BH}$ was used in the case of NLS1. With the introduction of the correction factor, the authors were in effect assuming a dependence on accretion rate (NLS1 have in general higher accretion rates than BL AGN, see i.e. \S \ref{secGamma} and \S \ref{nls1}), but rather than a continuous scaling with $\dot{m}$ (such as i.e. McHardy et al. 2006; Koerding et al. 2007) they were assuming a bimodal dependence in the form of a correction factor for NLS1. A few years later, Gierlinski et al. (2008) showed that the PSD integral above $\nu_{\rm br}$ remains approximately constant for a given AGN and BHB (in the hard state), regardless of the source luminosity (thus accretion rate), unlike $\nu_{\rm br}$ which correlates strongly with the latter. They argued that it is for this reason that the M$_{\rm BH}$ estimates from $\sigma^2_{\rm rms}$ measurements of light curve segments which are shorter than $1/\nu_{\rm br}$ are accurate. To further investigate the difference between NLS1 and BL AGN, Nikolajuk et al. (2009) computed the $\sigma^2_{\rm rms}$ for a large sample (21) of NLS1 and confirmed that, if the recipe to estimate M$_{\rm BH}$ from $\sigma^2_{\rm rms}$ was deduced from the PSD scaling with M$_{\rm BH}$ only (without containing a term related to $\dot{m}$), a shift of a factor of 20 had to be used for NLS1. However, the authors also showed that this factor is consistent with being just the product of a weak dependence of $\sigma^2_{\rm rms}$ with $\dot{m}$. Interestingly, the measured dependence of the variability with $\dot{m}$ was weaker (with a slope flatter $a=0.79$) than the one expected assuming the McHardy et al. (2006) relation (with $a=1.4$). This finding further supports the idea that the PSD$_{\rm amp}$ lowers with $\dot{m}$, as proposed in this work. 

Given the small scatter in the $\sigma^2_{\rm rms}$ vs. M$_{\rm BH}$ relation for the Rev sample objects, we argue that the best-fit results that we have listed in Tables \ref{relationRev}, can be used to measure the M$_{\rm BH}$ for an AGN if any of the $\sigma^2_{\rm rms,80, 40, 20,10}$ values is available.  If the $\sigma^2_{\rm rms}$ is measured with an uncertainty comparable to the uncertainty of $\sigma^2_{\rm rms}$ for the AGN in the Rev sample, this method can potentially yield M$_{\rm BH}$ estimates as accurate as the ones from the reverberation mapping. In fact, we followed this approach in this work, and we provide new M$_{\rm BH}$ estimates for 55 AGN in CAIXAvar, of which 6 had no previous M$_{\rm BH}$ estimates. 47 (1 and 7) of our estimates were based on $\sigma^2_{\rm rms}$ measurements from 20 ks (40 and 10 ks) long light curves. These estimates should be on average as accurate as the M$_{\rm BH}$ estimates from reverberation mapping, and most probably more accurate than the M$_{\rm BH}$ estimated from single epoch spectra. Moreover, we find that this recipe can be applied to both BLS1 and NLS1. 

\begin{figure}
\begin{center}
\epsfig{file=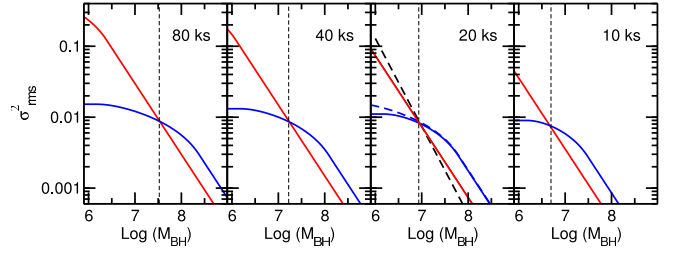,width=0.49\textwidth,bb=15 250 742 590, clip=}
\end{center}
\caption{{\it Left panel:} Expected $\sigma^2_{\rm rms,80}$ vs. M$_{\rm BH}$ relations for AGN with an accretion rate of 0.01 (red lines) and 1 (blue lines) in Model B scenario and $\sigma^2_{\rm rms,80}$ computed with 250 s bins and 80 ks intervals. {\it Middle left:} Same as before but for 40 ks intervals. {Middle right:} Same as before but for 20 ks intervals. Dashed lines show the expected $\sigma^2_{\rm rms,20}$ computed with time bins of 50 s. {\it Right panel:} 10 ks intervals.}
\label{Theo} 
\end{figure}

Since it is easier to estimate the $\sigma^2_{\rm rms}$ than to perform a series of reverberation mapping measurements for an AGN, X--ray variability based methods should be useful in determining M$_{\rm BH}$ of many objects.  However, we stress that one has to be careful when using the $\sigma^2_{\rm rms}$ vs. M$_{\rm BH}$ best fit results listed in Table \ref{relationRev} (and in fact, any X--ray variability based method to estimate the M$_{\rm BH}$ in an AGN). Figure \ref{Theo} shows the expected $\sigma^2_{\rm rms}$ vs. M$_{\rm BH}$ relations according to our Model B prescription, for AGN with an accretion rate of 0.01 (red lines) and 1 (blue lines) and for light curves which are 80, 40, 20 and 10 ks long (from the left to the right panel), binned at 250 s. A shorter bin size does not affect the expected relation significantly, except for the smallest M$_{\rm BH}$ and highest accretion rate sources. To illustrate this point, the dashed lines in the 3rd panel of Fig. \ref{Theo} indicate the predicted $\sigma^2_{\rm rms}$ vs. M$_{\rm BH}$ relation for a light curve with a bin size of 50 s (as opposed to the solid lines which corresponds to a 20 ks long, 250 s binned light curve). The relations are identical in the case of the low accretion rate objects (hence it is not possible to separate the dashed and solid red lines in this panel). On the other hand, a shorter bin size results in a different variability vs. M$_{\rm BH}$ relation for the high accretion rate objects, but only when M$_{\rm BH}\lesssim 5\times 10^6$ (see blue solid and dashed lines in the same panel). Even in this case though, the difference between the two relations is really small. 

The important thing to notice in all panels in Fig. \ref{Theo} is that Model B predicts a small scatter in the variability vs. M$_{\rm BH}$ relation but only above a certain M$_{\rm BH}$ which is equal to $3\times 10^7, 1.8\times 10^7, 10^7,$ and $5\times 10^6$ M$_{\odot}$, in the case of the 80, 40, 20 and 10 ks long light curves. This effect is equivalent to what the previous studies have argued: the M$_{\rm BH}$ estimates which are based on the use of $\sigma^2_{\rm rms}$ measurements are ``valid" as long as the length light curve used to measure the $\sigma^2_{\rm rms}$ is shorter than $1/\nu_{\rm br}$. For example, if we use an $\sigma^2_{\rm rms}$ estimate from a 80 ks long light curve for an object which has a M$_{\rm BH}$ {\it smaller} than $\sim 3\times 10^7$ M$_{\odot}$, and an accretion rate {\it larger} than $\dot{m} \sim 0.6-0.7$, the break time scale will be {\it shorter} than 80 ks, hence, due to the decrease of PSD$_{\rm amp}$ with increasing accretion rate, the expected excess variance will be {\it smaller} than the excess variance of objects which accrete at a smaller rate. In this case, the use of the best fit results we list in Table \ref{relationRev} (which indicate {\it no} break of the variability vs. M$_{\rm BH}$ relation at low M$_{\rm BH}$ objects) may result in an inaccurate M$_{\rm BH}$ estimate. One should be careful if the use of the Rev sample best fit results yields M$_{\rm BH}$ estimates significantly smaller than the masses mentioned above. In this case, an accurate estimate of M$_{\rm BH}$ requires knowledge of the L$_{\rm Bol}/$L$_{\rm Edd}$ ratio for the object, which may not be possible (even if L$_{\rm Bol}$ is known, L$_{\rm Edd}$ requires an apriori knowledge of M$_{\rm BH}$). Obviously, it is preferable to use 10 ks light curves to estimate the M$_{\rm BH}$, however, such a short light curve may not result in a positive excess variance measurement, at least for large M$_{\rm BH}$ objects. 

Given the discussion above, it seems rather peculiar that a simple straight line provides a good fit to the Rev 40 variability vs. M$_{\rm BH}$ relation, even when we use 80 and 40 ks long segments and for objects with M$_{\odot}$ less than $10^7$ M$_{\odot}$. In principle, we would expect a higher spread around the best fit line at low M$_{\rm BH}$. There are 5 objects with M$_{\rm BH}$ less than $10^7$ M$_{\odot}$ in the 20 ksec Rev relation. Three of these AGN (namely NGC4395, NGC4051 and NGC3227) actually have a low $\dot{m}$ (ranging from about 0.1 to 2 \%). The other two AGN (MRK766 and NGC4593) are accreting at about 60 and 10 \% Eddington and, in fact, they lie below the best fit relation. Although this is in line with the Model B predictions, the presence of only 5 AGN in this M$_{\rm BH}$ range is hampering us from deriving any strong conclusion. The CAIXAvar (20~ks) has more objects in this M$_{\rm BH}$ range. As expected, most of the objects do lie below the best fit line, suggesting a flattening of the relation at low M$_{\rm BH}$. We also observe that most of the objects with low $\sigma^2_{\rm rms}$ have high $\dot{m}$, while the two object with higher variability (1H0707-495 and IRAS13224-3809) have quite uncertain M$_{\rm BH}$ and L$_{\rm Bol}$. Obviously, we need to observe more low mass objects (M$_{\rm BH} < 10^7$), with a high accretion rate, to determine whether the variability vs. M$_{\rm BH}$ relation stays linear (in log--log space) at low M$_{\rm BH}$ or not. If it does, our assumptions on the PSD need to be modified. On the other hand, if our PSD ``results" are correct, then the relation should ``break" and one could not use the best-fit results we give to estimate M$_{\rm BH}$ below the numbers we give above.

\subsection{$\sigma^2_{\rm rms}$ vs. accretion rate}

It is not easy to determine observationally if the variability amplitude depends on accretion rate as well. Although reverberation mapping can deliver M$_{\rm BH}$ estimates with uncertainties of a factor three, to measure the accretion rate, one requires not only a reliable measurement of M$_{\rm BH}$, but of L$_{\rm Bol}$ as well. However, there are larger uncertainties associated with the L$_{\rm Bol}$ estimates in most objects, which introduce a large scatter in the variability vs. accretion rates plots. A second source of significant scatter in such plots is the fact that, for a given light curve length, $\sigma^2_{\rm rms}$ will be different for different M$_{\rm BH}$ objects. Consequently, the variability vs. $\dot m$ relation will also depend on M$_{\rm BH}$.  

Early studies already suggested a weak dependence of $\sigma^2_{\rm rms}$ on accretion rate (Bian \& Zhao 2003; Lu \& Yu 2001). O'Neill et al. (2005) showed that even this weak dependence disappears when $\sigma^2_{\rm rms}$ is ``normalized" to M$_{\rm BH}$. Zhou et al. (2010) confirmed the O'Neill et al. (2005) result and they found that to reproduce the small scatter in the $\sigma^2_{\rm rms}$ vs. M$_{\rm BH}$ relation, the slope of the variability -- accretion rate relation has to be smaller than 0.2. We found a strong, almost linear correlation between $\sigma^2_{\rm rms}$ and accretion rate in CAIXAvar. However, this is mainly introduced by the large M$_{\rm BH}$ range of the AGN present in this sample, as clearly suggested by the fact that, in agreement with the previous studies, this correlation is disappeared when we multiply the normalised excess variance values with the M$_{\rm BH}$ of the objects. 

This result is at odds with the McHardy et al. (2006) and Koerding et al. (2007) scaling laws for the PSD break time scale, {\it if} the PSD amplitude was the same in all objects. We found that the same PSD model which can explain the small scatter in the variability vs. M$_{\rm BH}$ relations (Model B), also predicts a much shallower variability vs. accretion rate relation. If the PSD amplitude decreases with $\dot m$ as PSD$_{\rm amp}\propto \dot{m}^{-0.8}$, {\it and} the PSD break frequencies scale with M$_{\rm BH}$ and $\dot m$ according to McHardy et al. (2006) and Koerding et al. (2007), then $\sigma^2_{\rm rms}\times$M$_{\rm BH}\propto \dot{m}^{0.2}$, in agreement with the Zhou et al. (2010) results, but also with our data (solid lines in the bottom panel of Fig. \ref{accr_all})

Clearly, it is difficult to detect such a shallow relation. As we discussed above, the most obvious difficulty is the scatter introduced in the data plots due to the large uncertainties of the accretion rate measurements when we go beyond the 10-15 nearby better studied AGN. Another important effect is that, in data plots with objects which have significantly different M$_{\rm BH}$, even the ``normalized"  estimates $\sigma^2_{\rm rms}\times$M$_{\rm BH}$ are {\it not} expected to strictly follow the same relation at all $\dot{m}$ (an effect that was not realised by O'Neill et al. 2005, and Zhou et al 2010). For a given light curve length of say $T$, and for sufficiently small M$_{\rm BH}$ objects, we expect that $\nu_{\rm br}> 1/T$ above a certain accretion rate. At higher rates, the $\sigma^2_{\rm rms}\times$M$_{\rm BH}$ vs. accretion rate relation is not linear any more (in the log-log space), as the blue and red, solid lines in the bottom panel of Fig. \ref{accr_all} clearly demonstrate. This effect, is expected to increase even further the scatter in these plots, rendering even more difficult the detection of the intrinsic variability -- accretion rate relation in AGN. 

In any case, the Model B curves shown in both panels of Fig. \ref{accr_all} indicate that our results are consistent with the McHardy et al. (2006) and Koerding et al. (2007) scaling laws, but only if PSD$_{\rm amp}\propto \dot m^{-0.8}$. This relation introduces a dependence of variability on accretion rate, which although weak, is necessary to explain other observational results, such as the variability properties of the NLS1. Previous work on large samples of NLS1 do find a significant difference between the variability of broad and narrow line objects that can not be justified by the difference in their M$_{\rm BH}$, only (Nikolajuk et al. 2009). 
We also observe indications for higher variability amplitude in NLS1, even when we consider NLS1 and BLS1 of comparable M$_{\rm BH}$. But we showed that, once the difference in their accretion rate is taken into account properly, i.e. when we consider the Model B PSD$_{\rm amp}$ dependence on accretion rate, then both NLS1 and BLS1 show comparable variability amplitudes. This result indicates that the variability mechanism is the {\it same} in both classes objects, and the differences in their variability amplitudes can be fully understood once the differences in their average M$_{\rm BH}$ and accretion rate is taken into account properly. 

\subsection{$\sigma^2_{\rm rms}$ vs. $\Gamma$}

The AGN in CAIXAvar show a strong correlation between $\sigma^2_{\rm rms}$ and the 2-10 keV spectral index $\Gamma$. Early works on this relations (Turner et al. 1999; Bian \& Zhao 2003) already found indications for these two quantities to be correlated ($\sim99$ \% confidence), with the more variable NLS1 having steeper spectral indexes than broad line AGN. However, larger samples indicated this correlation to be rather weak ($\sim96$ \% confidence; O'Neill et al. 2005). Using a larger sample than ever before (to study AGN X-ray variability on time scales less than a day) we estimate the significance of the correlation to be as high as $99.99$ \% with a steep slope of $\sim9.2\pm1.2$. 

The physical origin of this correlation is not easy to understand, but we argued that it is the product of two underlying correlations: the one between variability and accretion rate, and the relation between $\Gamma$ and $\dot m$. The latter correlation has been suggested by many AGN studies in the past  (e.g. Porquet et al. 2004; Piconcelli et al. 2005; Shemmer et al. 2006; Saez et al. 2008; Papadakis et al. 2009; Sobolewska \& Papadakis 2009; Risaliti et al. 2009; Scott et al. 2011) and has also been observed in BHB (see e.g. Wu \& Gu 2008). We note that this correlation is not significant in CAIXA (Bianchi et al. 2009) but this may be due to the flatness of the correlation and the large scatter of the data points (both in accretion rate and $\Gamma$) in CAIXA. If $\Gamma \propto \dot{m}^{0.1}$, as Sobolewska \& Papadakis (2009) suggested, then since $\sigma^2_{\rm rms} \sim \dot{m}$ (see Table 2) in CAIXAvar (a relation which is mainly driven by the presence of different M$_{\rm BH}$ objects in this sample, as we discussed in the previous section), we would expect to observe a $\sigma^2_{\rm rms} \sim\Gamma^{10}$ relation, which is entirely consistent with our observations. 

If we multiply $\sigma^2_{\rm rms}$ with M$_{\rm BH}$, then the variability vs. spectral index  correlation disappears. But in this case, it is difficult to argue that this is due to the fact that both $\sigma^2_{\rm rms}$ and $\Gamma$ correlate with M$_{\rm BH}$. In fact, both in CAIXA and in CAIXAvar a $\Gamma$ vs. M$_{\rm BH}$ correlation is not significant. Instead, when normalized to M$_{\rm BH}$, the expected variability vs. accretion rate relation is much flatter (as we discussed above) and hence much more difficult to measure accurately. Therefore, just like with the $\sigma^2_{\rm rms}\times$M$_{\rm BH}$ vs $\dot m$ plot, the $\sigma^2_{\rm rms}\times$M$_{\rm BH}$ vs $\Gamma$ plot is flat, but nevertheless, the predicted relations, assuming Model B and the $\Gamma-\dot m$ relation of Sobolewska \& Papadakis, are entirely consistent with the data (see Fig. \ref{GammaTheo}).

\section{Conclusions}

We studied the variability properties, computing the excess variance, of all the radio quiet, X-ray unobscured, AGN observed by \xmm\ for more than 10 ks in pointed observations. The \xmm\ high sensitivity and long orbit allows us to measure the variability amplitude of 161 AGN and probe the variability on shorter timescales down to 10 ks. We sample timescales going from 80 to 10 ks, with the latter being significantly smaller than the time scales considered in all previous works (like e.g. O'Neill et al. 2005; Zhou et al. 2010). CAIXAvar is the largest sample in which short time-scale variability (on time-scales smaller than about a day) has been studied and is composed primarily by local AGN. Thus the relations and variability properties investigated in this work are representative of a population of AGN characteristic of the local universe. 

\begin{itemize}

\item{} We systematically explore the energy dependence of the excess variance, comparing the variability of the hard (2-10 keV) with the soft (0.3-0.7 keV) and medium (0.7-2 keV) bands. Surprisingly, although different spectral components, such as the soft excess and warm absorbers ``dominate" the emission at softer energies, the soft band variability amplitude  is very well correlated with the hard band variability. In particular, no deviations from the linear one-to-one correlations are observed, suggesting that these components add a minor contribution to the total variability.

\item{} We observe a highly significant and tight (0.4 dex) correlation between $\sigma^2_{\rm rms}$ and M$_{\rm BH}$ for the reverberation sample. A larger scatter (0.7 dex) is observed for CAIXAvar. This is due to the larger uncertainties on the M$_{\rm BH}$ of the non-reverberation estimates. This result implies that our best-fit Rev sample results could be used to measure M$_{\rm BH}$ from $\sigma^2_{\rm rms}$ measurements. Provided that these estimates are as accurate as the variability estimates for the objects in our Rev sample, the M$_{\rm BH}$ estimates should be more accurate than the ones based on single epoch spectra. Particular care though should be given to the results from the use of this method if the M$_{\rm BH}$ turns out to be smaller than $\sim 5\times 10^{6}-5\times 10^{7}$ M$_{\odot}$ (depending on the length of the light curve in use), as in this regime the variability - M$_{\rm BH}$ relation may have a larger intrinsic scatter. 

\item{} The $\sigma^2_{\rm rms}$ vs. M$_{\rm BH}$ correlation has a small scatter (smaller than expected, see i.e. McHardy et al. 2006; Koerding et al. 2007) because the PSD normalisation decreases with accretion rate as PSD$_{\rm amp}\propto \dot {m}^{-0.8}$. The combination of this relation with the PSD break frequency scaling relations of McHardy et al. (2006) and Koerding et al. (2007) results in the PSD high frequency integral being weakly dependent on accretion rate, being thus a good tool to estimate M$_{\rm BH}$ (Nikolajuk et al. 2004; 2006; 2009). 

We note that the universal constancy of the high frequency tail (rather than the break frequency) of the PSD in {\it all} BH accreting sources, seems to suggest a deep link with some common physical process experienced by all accreting BH. 

\item{} We measure M$_{\rm BH}$ or provide stringent upper limits for 65 AGN, the majority of which (44) have no either reverberation mapping or stellar velocity dispersion M$_{\rm BH}$ estimates and for 6 of which we did not find any M$_{\rm BH}$ estimate in literature. We also estimate lower limits to the M$_{\rm BH}$ for the remaining 96 AGN.

\item{} We observe a significant correlation between  $\sigma^2_{\rm rms}$ and $\dot{m}$, which dissapears when we ``normalize" $\sigma^2_{\rm rms}$ to the M$_{\rm BH}$ of the objects, in agreement with the previous results of O'Neill et al. (2005) and Zhou et al. (2010). We argue that, the lack of significant intrinsic scatter in the $\sigma^2_{\rm rms}$ vs. M$_{\rm BH}$ relation, and the lack of a significant correlation between variability (normalised to M$_{\rm BH}$) and $\dot{m}$ are in agreement with the hypothesis that: all AGN have the same PSD shape, with the break frequencies scaling with M$_{\rm BH}$ and $\dot m$ as in McHardy et al. (2006) and Koerding et al. (2007), but only {\it if} PSD$_{\rm amp}\propto \dot {m}^{-0.8}$, in a fashion similar to what is observed in BH binaries (Axelsson et al. 2005; Gierlinski et al. 2008). 

\item{} We observe that NLS1 are more variable than broad line AGN. This is due to their smaller M$_{\rm BH}$ and higher accretion rate. Once rescaled, no difference from broad line AGN is required to explain their different variability properties.

\item{} We observe significant correlations between $\sigma^2_{\rm rms}$ and both the bolometric luminosity and the FWHM of the H$\beta$. Both these correlations disappears when the $\sigma^2_{\rm rms}$ vs. M$_{\rm BH}$ is considered, suggesting that the formers are just by-products of the variability vs. M$_{\rm BH}$ dependence.

\item{} We observe, for the first time at very high significance ($>99.99$ \%), a correlation between $\sigma^2_{\rm rms}$ and 2-10 keV spectral index. This correlation provides indirect support to previous suggestions that the spectral index in AGN depends on accretion rate. In particular, the CAIXAvar $\sigma^2_{\rm rms}-\Gamma$ relation can be fully explained if $\Gamma \propto \dot {m}^{0.1}$, as suggested by Sobolewska \& Papadakis (2009).

\end{itemize}

In summary, all our results are in agreement with a picture where, to first approximation, all AGN have {\it the same} variability properties once rescaled properly for  M$_{\rm BH}$ and accretion rate. This confirms and extends the work of McHardy et al. (2006) and Koerding et al. (2007) who studied the PSD of less than two dozen sources. In our case, this picture is based on the study of a significantly larger number of objects. Our work demonstrates the significance of studies which are based on the estimation of the excess variance, which although as a statistic carries far less information than a full PSD analysis, its use allows the consideration of large AGN samples, hence establishing the validity of the current paradigm for the X-ray variability of AGN, and even allowing the unexpected detection of important ``complications", such as the PSD amplitude dependence on accretion rate. 

\section*{Acknowledgments}

The work reported here is based on observations obtained with XMM-Newton, an ESA science mission with instruments and contributions directly funded by ESA Member States and NASA. We thank the anonymous referee for very helpful comments.  GP thanks Rob Fender, Simona Soldi and Massimo Cappi for many useful comments and suggestions. GP acknowledges support via an EU Marie Curie Intra-European Fellowship under contract no. FP7-PEOPLE-2009-IEF-254279. SB and GM from ASI (grant I/088/06/0).


\newpage
\begin{landscape}
\begin{table}
\begin{center}
\begin{tabular}{ |l |  ccl |  ccl  | ccl | ccl |}
\hline                      
Relation & & 80 ks & & & 40 ks & & & 20 ks & & & 10 ks & \\
              & Norm & Slope & prob & Norm & Slope & prob & Norm & Slope & prob & Norm & Slope & prob\\
\hline
CAIXAvar & & & & & & & & & & & & \\
$\sigma^2_{\rm rms0.3-0.7}$ vs. $\sigma^2_{\rm rms2-10}$ & $0.51\pm0.25$ & $1.25\pm0.13$ & 99.999 & $-0.03\pm0.19$ & $1.02\pm0.09$ & 99.999 & $0.13\pm0.24$ & $1.08\pm0.12$ & 99.999 & $0.13\pm0.21$ & $1.08\pm0.09$ & 99.999  \\
$\sigma^2_{\rm rms0.7-2}$ vs. $\sigma^2_{\rm rms2-10}$    & $0.42\pm0.19$ & $1.19\pm0.10$ & 99.999 & $0.10\pm0.14$ & $1.05\pm0.07$ & 99.999 & $0.16\pm0.20$ & $1.08\pm0.10$ & 99.999 & $0.21\pm0.15$ & $1.10\pm0.07$ & 99.999  \\
\hline
$\sigma^2_{\rm rms}$ vs. $M_{BH}$                                  & $-1.68\pm0.11$ & $-0.92\pm0.12$ & 99.995 & $-1.60\pm0.12$ & $-1.10\pm0.12$ & 99.997 & $-1.83\pm0.10$ & $-1.04\pm0.09$ & 99.999 & $-2.09\pm0.10$ & $-1.03\pm0.10$ & 99.999\\
$\sigma^2_{\rm rms}$ vs. accre                  & $-0.95\pm0.17$ & $1.19\pm0.15$ & 82.6     & $-1.26\pm0.16$ & $1.09\pm0.18$ & 67.1     & $-1.50\pm0.13$ & $1.11\pm0.13$ & 86.9 & $-1.54\pm0.13$ & $1.11\pm0.11$ & 90.4 \\
$\sigma^2_{\rm rms}\times M_{BH}$ vs. accre & ... & ... & 45.8 & ... & ... & 63.3 & $4.38\pm0.05$ & $-1.06\pm0.03$ & 81.6 & $4.22\pm0.12$ & $-1.00\pm0.07$ & 86.2 \\
\hline
$\sigma^2_{\rm rms}$ vs. L$_{Bol}$           & $-1.49\pm0.19$ & $-0.92\pm0.18$ & 99.69    & $-1.37\pm0.17$ & $-1.03\pm0.13$ & 99.93  & $-1.44\pm0.15$ & $-1.04\pm0.12$ & 99.95 & $-1.66\pm0.14$ & $-1.04\pm0.13$ & 99.96 \\
$\sigma^2_{\rm rms}\times M_{BH}$ vs. L$_{Bol}$          & ... & ... & 63.1 & ... & ... & 61.2 & $4.37\pm0.18$ & $0.93\pm0.08$ & 85.2 & ... & ... & 64.7\\ 
\hline
$\sigma^2_{\rm rms}$ vs. H$_{\rm \beta}$ & $-1.70\pm0.14$ & $-2.87\pm0.39$ & 99.97    & $-1.76\pm0.12$ & $-3.25\pm0.35$ & 99.998 & $-2.03\pm0.10$ & $-3.01\pm0.27$ & 99.999 & $-2.28\pm0.11$ & $-2.38\pm0.35$ & 99.999 \\
$\sigma^2_{\rm rms}\times M_{BH}$ vs. H$_{\rm \beta}$& ... & ... & 37.0 & ... & ... & 36.3 & ... & ... & 60.3 & ... & ... & 34.3 \\ 
\hline
$\sigma^2_{\rm rms}$ vs. $\Gamma$         & $-1.98\pm0.14$ & $  7.33\pm1.31$ & 98.5     & $-2.15\pm0.13$ & $  7.93\pm1.90$ & 99.94   & $-2.44\pm0.11$ & $  9.22\pm1.20$ & 99.992 & $-2.54\pm0.13$ & $9.02\pm1.38$ & 99.97 \\
$\sigma^2_{\rm rms}\times M_{BH}$vs. $\Gamma$         & ... & ... & 27.9 & ... & ... & 53.3 & ... & ... & 64.3 & ... & ... & 44.2 \\ 
\hline                      
\end{tabular}
\small
\caption{List of all best fit relations of the CAIXAvar sample as well as their probabilities. For the correlations with significance higher than 65 \% we also report the best fit slope and normalisation as well as the associated 1-$\sigma$ errors. We do not report the best fit values of the less significant correlations because of their loose physical meaning.}
\label{relation}
\end{center}
\end{table} 
\end{landscape}

\appendix

\section{Estimating the excess variance uncertainty}

The uncertainty on the excess variance depends both on the measurement
uncertainties (e.g. Poisson noise) in the light curve and on the stochastic
nature of the variability. As shown by Vaughan et al. (2003) through
MonteCarlo simulations, or Ponti et al. (2004) in the particular case of large
number of photons, the former can be approximated by the formula (see eq. 11 of Vaughan
et al. 2003): 

\begin{equation}
\left(\Delta \sigma^2_\mathrm{rms}\right)_\mathrm{meas} = \sqrt{ \left(\sqrt{\frac{2}{N}} \frac{ <\sigma^2_i> }{\mu^2}\right)^2 + \left( \sqrt{\frac{<\sigma^2_i>}{N} \frac{2F_\mathrm{var}}{\mu}} \right)^2 }
\label{Poiss}
\end{equation}
where $<\sigma^2_i>$ is the mean of the uncertainties squared and $F_\mathrm{var}$ is the fractional variability. 

The uncertainty owing to the stochastic nature of the process is more difficult
to estimate. Vaughan et al. (2003) showed that the uncertainty increases with
the steepness of the power spectrum slope and it is large and highly
non-Gaussian for steep Power Spectral Densities (PSD). A simple
estimate of the uncertainty cannot be analytically derived, however two 
approaches can be pursued. 
\begin{figure} [t]
\begin{center}
\epsfig{file=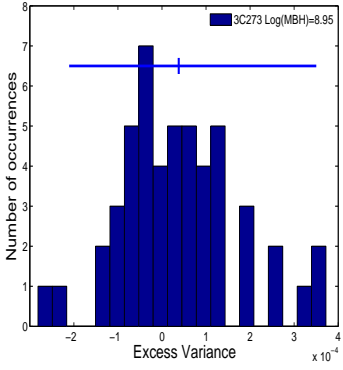, width=4.5cm,height=4.9cm}
\end{center}
\caption{Distribution of the observed excess variances computed for 
10 ks intervals for 3C273. The distribution of each of the other 6 sources 
with longer \xmm\ exposures are displayed on the on-line version 
of the paper. The points above each distribution indicate the average excess 
variance and the ``error bars" around this mean indicate the mean 
upper and lower 90~\% confidence limits on $\sigma^2_{\rm rms}$, estimated 
as described in the text. They are compatible with 
the region where $\sim 90$~\% of the points are located (see text for more details).
}
\label{ErrRms} 
\end{figure}

\subsection{Estimating the "red noise" scatter: more than 10 intervals}

The first approach is to estimate the uncertainty directly from the data,
deriving the scatter from the distribution of the measured excess variances of
the different intervals. Every excess variance measurement is an independent
variable with identical distribution (unless the process is non-stationary). This is only approximately valid, because successive light curve parts may not be truly independent, if the ``memory" of the system is longer than the segment's length. Nevertheless, every time that more than 10 valid segments are available for the same source, we estimate the uncertainty on the mean excess variance computing the sigma of the mean of the different measurements. Furthermore, due to the central limit theorem, the distribution of the mean of all
the excess variance estimates  will become normally distributed for a large
number of measurements. Obviously it is not easy to estimate how ``large" the
number of $\sigma^2_{\rm rms}$ estimates should be but, following this line of
reasoning, in the case of sources with more than 10 light curve intervals, we consider the error 
on the mean to be symmetric (as if its distribution were normal), and in order to estimate the
90 \% confidence level, this value is then multiplied by 1.6. Despite the numerous limitations, 
 at least this approach allows us to estimate the uncertainty of our final estimate without
the need to adopt any assumptions about the underlying variability process. 

\subsection{Estimating the "red noise" scatter: 1 interval}

When the number of valid segments is just one, we estimated the uncertainty on
$\sigma^2_{\rm rms}$ using  the results from the Monte Carlo simulations of Vaughan et al (2003)
and assuming a PSD shape for each source. Detailed studies of
PSD in AGN have shown an almost ubiquitous PSD shape characterised by a steep
power law shape ($\alpha\sim2$) above a special frequency $\nu_\mathrm{br}$ and
a flatter ($\alpha\sim1$) power law slope below (Uttley \& McHardy 2005; but see
also the case of ARK564; McHardy et al. 2007). In particular it has been
observed that the characteristic break frequency ($\nu_\mathrm{br}$) scales
primarily with mass and  with accretion rate (McHardy et al. 2006). We thus
expect that the slope of the PSD, within the frequency range probed by our light
curves (from $4\times10^{-3}$ to either $10^{-4}$, $5\times10^{-5}$ or
$2.5\times10^{-5}$ Hz, for the 10, 20 and 40 ks segments, respectively), may be
different  for the sources in the sample, depending on their  M$_{\rm BH}$ and
accretion rate. Assuming that all AGN show the same PSD and that the PSD shape scales with mass and accretion rate as measured by McHardy et al. (2006), 
we can predict the position of the break frequency and the slope of the PSD 
in the frequency band which corresponds to 250 s binned light curves, 
of duration equal to 10, 20, 40 and 80 ks.  

Using then the results listed in Table 1 of Vaughan et al. (2003), we assume a
$\Delta log(S^2)=+0.45$ and $-0.71$, for the positive and negative  error on
$\sigma^2_{\rm rms~}$, respectively,  if the break time-scale is longer than
the length of the interval (this means that in the frequency window on which
the excess variance is computed, the PSD has a slope of -2). We assumed $\Delta
log(S^2)=+0.28$ and $-0.36$ if the break time-scale is shorter than the light
curve time bin (in which case we would expect a PSD of a -1 slope in the
frequency range sampled by each light curve segment). If the break time-scale
falls within the frequency window, then we combine the errors with the formula:
$0.28\times\frac{(Log(\nu_{br})-Log(\nu_{min}))}{Log(\nu_{max})-Log(\nu_{min})}
+ 0.45\times\frac{(Log(\nu_{max})-Log(\nu_{br}))}{Log(\nu_{max})-
Log(\nu_{min})}$ (and similarly for the negative error). When no M$_{\rm BH}$
estimates are available, we are conservative and assume the largest
uncertainties, associated with a PSD slope of -2. As computed by Vaughan et al.
(2003) this scatter estimates the 90 \% confidence interval. In this way we are
able to conservatively estimate the scatter in the excess variance measurements
introduced by the red noise, in the case when we have just a single excess
variance value. 

\subsection{Estimating the "red noise" scatter: checking the approximations}

In order to judge how accurate are our excess variance
uncertainties, given the various assumptions that underlie the methods we described above, we performed the following test. We considered  the 7 sources
with the longest XMM-Newton observations in CAIXAvar. The number of 10 ksec
segments for each one of these objects is significantly larger than 10 (so the
10 ksec excess variance measurement we list in Table \ref{exCAIXA} for these objects
is the mean of all the individual measurements, and its error is based on the
true scatter of the points around their mean, as explained above). Figure
\ref{ErrRms} shows the distribution of the observed 10 ksec excess variance values,
 for these objects.  As expected, whenever we have
a ``signal" (i.e. the $\sigma^2_{\rm rms}$ values are positive) the
distributions are asymmetric with a tail at large values. In the case
of 3C273, due to its large M$_{\rm BH}$,  we do not expect large amplitude variations
on such short time scales. As a result, the distribution of the 10ks
$\sigma^2_{\rm rms}$ values is dominated by the uncertainties associated with
the Poisson noise of the light curves, hence it is more symmetric. 

We then used the method described above to obtain the uncertainty of each
individual $\sigma^2_{\rm rms}$ value, and we calculated the mean upper and
lower 90\% confidence limits on $\sigma^2_{\rm rms}$, for each object. The 
points on top of each sample distribution in Fig. \ref{ErrRms} indicate the
average excess variance, and the ``error bar" around this mean indicate the mean
upper and lower 90\% confidence limits on $\sigma^2_{\rm rms}$. 
A visual inspection shows that these confidence limits are
compatible with the area where $\sim 90$\% of the points are. 
In fact, we measure that the average 90 \% uncertainty estimated with our 
method actually contains 84, 92, 80, 93, 87, 97, 92 \% of the measured values 
of NGC4051, 1H0707-495, MGC-6-30-15, MRK766, MRK335, NGC3516 and 3C273, 
respectively. Thus the typical difference between the estimated 90 \% 
uncertainty and the one measured from the observed distribution is less 
than 10 \%. 

The 1-$\sigma$ ``error" has been estimated simply dividing the 
90 \% uncertainties by 1.6. This approximation is valid for normal 
distributions, only. However, in the cases of NGC4051, 1H0707-495,  
MGC-6-30-15, MRK766, MRK335, NGC3516 and 3C273, using the observed distribution, we can compare the 1-$\sigma$ uncertainty computed in this 
way with the corresponding probabilities that are 73, 69, 51, 62, 58, 54 
and 82 \%, respectively. We, thus, measure that the typical difference between the 1-$\sigma$ computed in this way and the 1-$\sigma$ measured from the observed distribution, is of the order of 10-15~\%, reaching maximum values of 25~\% for MCG-6-30-15.

Fig. \ref{ErrRms}, suggests that the method we have adopted to estimate the uncertainty on the excess variance in the case when there is just a single interval available results in an acceptable-conservative estimate of the true scatter of $\sigma^2_{\rm rms}$. The same figure also suggests that this result should be valid for AGN with different M$_{\rm BH}$ and intrinsic excess variance values which spans almost four orders of magnitude.

\subsection{Estimating the "red noise" scatter: between 2 and 10 intervals}

When the number of valid segments is higher than one and lower than 10, we estimated the stochastic scatter for each segment as detailed above, in the case of sources with just a single interval available. Then, following O'Neill et al. (2005), our final estimate of the mean excess variance is equal to the square root of the sum of the squared ``error" of the individual segments, divided by the number of intervals. Finally, both in the case of single and less than 10 intervals, we combined in quadrature the stochastic scatter and the one associated with the Poissonian noise (see eq. \ref{Poiss}). 

\section{$\sigma^2_{\rm rms}$  computed in fixed length intervals and time dilation with redshift}

Since CAIXAvar is composed mainly of local AGN (see Fig. \ref{z}), the impact of the differences in red-shift on the excess variance, if we estimate it using intervals with a fixed length in the observers frame, should be minimal. This is even more obvious in the case of the sources with at least one variable segment detected, whose redshift is less than 0.3 in all (but one) cases. For these objects, segments of fixed length in the observer's frame should imply differences of less than $\sim 30$\% in the rest frame segment's length. Since the excess variance does depend on the maximum frequency sampled, we expect that the intrinsic excess variance should be different, depending on the source's redshift. However we verified that, for power-spectra like the ones typically observed in AGN (see discussion in Section 6.1.3 below), the resulting differences in the intrinsic variance of the sources should be less than $\sim 20$\% (for a large range in BH mass and accretion rates). This {\it maximum} difference is much smaller than the observed scatter in all the variability plots we study below. For this reason, we decided to work with the excess variance measurements that we estimated from the intervals we mentioned above, whose length is fixed, irrespective of the sources redshift. Regarding the highest redshift source (at z=0.9) in the sample of "variable" objects, we always checked that its presence does not affect in anyway our results from the study of the correlations between the variability amplitude and the various physical parameters we present in this paper. 

\end{document}